\documentclass[nohyper,notoc,a4]{article} 
\usepackage{axodraw2}
\usepackage{epsfig}
\usepackage{bm}
\usepackage{color}
\usepackage[utf8]{inputenc} 
\usepackage[T1]{fontenc}

\usepackage{graphicx} 
\usepackage{amssymb}
\usepackage{amsmath}
\usepackage{xfrac}
\usepackage{booktabs} 
\usepackage{array} 
\usepackage{paralist} 
\usepackage{verbatim} 
\usepackage{subfig} 
\usepackage{braket}
\usepackage{slashed}

\numberwithin{equation}{section}

\newcommand{\brackets}[1]{\left ( #1 \right )}
\newcommand{\sbrackets}[1]{\left [ #1 \right ]}

\def\bit{\begin{itemize}}
\def\eit{\end{itemize}}
\def\ben{\begin{enumerate}}
\def\een{\end{enumerate}}
\def\beq{\begin{equation}}
\def\eeq{\end{equation}}
\def\bea{\begin{eqnarray}}
\def\eea{\end{eqnarray}}
\def\bq{\begin{quote}}
\def\eq{\end{quote}}

\def\gappeq{\mathrel{\rlap {\raise.5ex\hbox{$>$}}
{\lower.5ex\hbox{$\sim$}}}}
\def\lappeq{\mathrel{\rlap{\raise.5ex\hbox{$<$}}
{\lower.5ex\hbox{$\sim$}}}}

\def\Dslash{ \, D  \! \! \! \! / ~ }
\def\Wslash{ \, W  \! \! \! \! / ~ }

\def\mec{\mu \! \to \! e~ {\rm conversion}}
\def\teg{\tau \to e \gamma}
\def\meg{\mu \to e \gamma}
\def\tmg{\tau \to \mu \gamma}

\def\meee{\mu \to e \bar{e} e}

\def\a{\alpha}
\def\b{\beta}
\def\g{\gamma}
\def\d{\delta}

\def\m{\mu}

\def\r{\rho}
\def\s{\sigma}

\begin{document}
\vspace*{-1in}
\renewcommand{\thefootnote}{\fnsymbol{footnote}}
\begin{center}
{\Large {\bf 
Majorana Neutrino Masses in the RGEs for Lepton Flavour Violation}}
\vskip 25pt
{\bf   Sacha Davidson $^{1,}$\footnote{E-mail address:
    s.davidson@lupm.in2p3.fr},
Martin Gorbahn $^{2,}$\footnote{E-mail address: martin.gorbahn@liverpool.ac.uk
}
  and Matthew Leak  $^{2,}$}\footnote{E-mail address: matt.leak@hotmail.co.uk}
 \vskip 10pt  

$^1${\it LUPM, CNRS,
Université Montpellier
Place Eugene Bataillon, F-34095 Montpellier, Cedex 5, France
}\\
$^{2}${\it Theoretical Physics Division,
Department of Mathematical Sciences,
University of Liverpool,\\
 Liverpool L69 3BX, United Kingdom
}\\

\end{center}

\vskip 10pt

\begin{abstract}
\noindent
We suppose that  the observed neutrino masses
can be parametrised by  a lepton number violating  dimension-five operator, 
and  calculate the mixing of double insertions of this
operator 
 into lepton flavour changing
dimension-six operators of the standard model effective theory. This allows to  predict the log-enhanced, but $m_\nu^2$-suppressed  lepton flavour violation  that is generic to  high-scale  Majorana
neutrino mass models.
We also consider the Two Higgs Doublet Model,
where the second Higgs  allows the construction of three additional dimension-five operators, and  evaluate the corresponding anomalous dimensions. The sensitivity of current searches for lepton flavour violation to  these
additional Wilson coefficients is then examined.

\end{abstract}

\setcounter{footnote}{0}
\renewcommand{\thefootnote}{\arabic{footnote}}


\section{Introduction}
\label{intro}

Neutrinos are  elusive and enigmatic particles: uncoloured, uncharged, and
very light. Nonetheless, their observed masses and mixing angles~\cite{PDB}
imply that   Lepton Flavour Violation (LFV) must occur, where
we define LFV as flavour-changing contact interactions of charged leptons
(for a review, see {\it e.g.}~\cite{KO}).
Since these do not occur  in the Standard Model~(SM), LFV is considered
to be ``New Physics'', and
searched for in a  wide variety of  experiments~\cite{PDB,TheMEG:2016wtm,Aad:2014bca,Abreu:1996mj,Akers:1995gz,Khachatryan:2016rke,Khachatryan:2015kon,Hayasaka:2010np,Bellgardt:1987du,Aubert:2009ag,Hayasaka:2007vc}.
Neutrinos could also induce another kind of New Physics:
if their small masses are ``Majorana'', they are
Lepton Number Violating~(LNV), and could for instance
mediate neutrinoless double-$\beta$-decay~\cite{0nu2B}.
Below the weak scale, such masses appear as  renormalisable
terms in the Lagrangian,  
but in the full  SU(2) gauge invariant Standard Model,  they
arise as a non-renormalisable, dimension-five operator. 

  In this paper, we will  assume  that neutrino masses
are Majorana, and that the scale  $\Lambda$ of New Physics  in the
lepton sector is large. We focus on the theory
at scales above $m_W$ but below $\Lambda$,
where it can be described in the framework of the
Standard Model Effective Field Theory\footnote{For
  an introduction to EFT, see {\it e.g.}~\cite{Georgi,burashouches}.} (SMEFT). 
The neutrino masses can be  parametrised  by 
operators of dimension  five, and LFV is parametrised
by  operators  of dimension-six. 
Our aim is to  obtain the  log-enhanced loop contributions
of   two LNV  operators
to LFV processes. These can be calculated via
renormalisation group equations (RGEs), and in particular
we aim to   calculate  the anomalous dimensions 
that mix two dimension-five operators into a dimension-six operator.
The renormalisation group running of the dimension-five operators has been extensively studied in the literature~\cite{Babu,Antusch:2001vn,Grimus:2004yh},
and the mixing of the dimension-six operators among themselves have been evaluated at one-loop~\cite{JMT} in the  ``Warsaw''-basis~\cite{polonais} of SMEFT operators.
The mixing of two dimension-five operators into
dimension-six operators was calculated in~\cite{BGJ},
using  the Buchmuller-Wyler~\cite{Buchmuller:1985jz} basis at dimension-six. 
We perform this calculation using the ``Warsaw''-basis,
and our results appear to disagree with~\cite{BGJ}.

The mixing  of neutrino masses into LFV amplitudes
is ${\cal O}(m_\nu/m_W)^2 \ln(\Lambda/m_W)$, so negligibly small,
but completes the anomalous  dimensions required  to perform
a one-loop renormalisation-group anlysis of the SMEFT at dimension-six.
In addition, we explore an extention of the SMEFT with two
Higgs doublets~\cite{2HDM},  where the second Higgs doublet  lives at
a scale $m_{22}$ between $m_W$ and significantly below the lepton number/flavour-changing scale $\Lambda$, and we impose that
   LFV at the weak scale is
still described by the dimension-six operators of the SMEFT.
In this scenario, there are four LNV dimension-five  operators above
$m_{22}$, but only one combination of coefficients
contributes to neutrino masses.
We calculate the mixing of these LNV operators into the
LFV operators of the SMEFT, and estimate the sensitivity of
current LFV  experiments to their coefficients.

The paper is organised as follows. In Section 2 we introduce the notation of our standard model and two-Higgs doublet model calculation. The main results are presented in Section 3, where we discuss the general structure of our calculation and give the relevant counterterms, anomalous dimensions and renormalisation group equations. Section 4 discusses the phenomenological implications of both results before we conclude. We provide the relevant Feynman rules, further details of the calculation (including a careful treatment of the flavour structures), and the renormalisation group in the Appendices A -- C. The LFV operators of the SMEFT are recalled in  Appendix D,
and Appendix E  gives  the current experimental constraints on
some LFV  coefficients of  the SMEFT at the weak scale. Appendix F provides a comparison with the  previous calculation of ~\cite{BGJ}
and Appendix G presents the lepton conserving contributions to the anomalous dimensions.


\section{ Notation and Review }
\label{sec:notn}

The SM Lagrangian for leptons can be written as 
\begin{equation}
\mathcal{L}_{lep} =  i     \overline{\ell}_\a \, 
\gamma^\mu D_\mu \, \ell_{\a}
+ i \overline{e}_\a 
 \gamma^\mu D_\mu  \,e_{ \a}
- {\Big (} [Y_e]_{\alpha \beta} \bar{\ell}_{\alpha} H e_{\beta} 
+ \mbox{h.c.}  {\Big )}
\label{L1}
\end{equation}
where Greek letters represent lepton generation indices
in the charged-lepton mass eigenstate basis, $[Y_e]$
is the  diagonal charged-lepton Yukawa matrix, $\ell$ is a doublet of left-handed leptons, and
  $e$ is a right-handed charged-lepton singlet. The explicit form of the lepton and Higgs doublets is
\begin{equation}
\ell = \left(
\begin{array}{c}
\nu_L\\
e_L
\end{array}
\right)
,~~~
{H} = \left(
\begin{array}{c}
H^+\\
H_0
\end{array}
\right)
,
\end{equation}
which have hypercharge $y_{\ell} = - 1/2$ and $y_H = 1/2$ respectively.
The covariant derivative for a lepton doublet is
\begin{equation}
(D_\mu \ell)^{i}_{\alpha} = \left( \delta_{ij}\partial_\mu + i \frac{g}{2} \tau^a_{ij} W^a_\mu 
+ i \delta_{ij} g' y_{\ell}
  B_\mu \right) \ell^{j}_{\alpha},
\end{equation}
where $\tau^a$ are the Pauli matrices. 
This sign convention for the covariant derivative agrees with~\cite{JMT}.


Heavy New Physics can be parameterised
by adding   non-renormalisable operators to the
SM Lagrangian that respect the SM gauge symmetries~\cite{Buchmuller:1985jz}.
There is only a single operator at dimension-five in the SM, which is the Lepton Number Violating
``Weinberg'' operator~\cite{WeinbergOp} 
which is responsible for Majorana masses of left-handed neutrinos.
The resulting effective Lagrangian at dimension-five is
\begin{equation}
\delta \mathcal{L}_5 = 
\frac{C_5^{\a \b}}{2 \Lambda} 
 (\overline{\ell_{\alpha}} \varepsilon H^{*})(\ell^c_{\beta} \varepsilon H^{*})
+\frac{C_5^{\a \b *}}{2 \Lambda} 
 (\overline{\ell_\beta ^c} \varepsilon {H}) (\ell_\alpha \varepsilon H) 
 \,,
\label{d5Weinberg}
\end{equation}
where $\varepsilon$ is the totally antisymmetric rank-2 Levi-Civita symbol with $\varepsilon_{12} = +1$,
all implicit $\rm SU(2)$ indices inside brackets are
contracted,
and the charge conjugation acts on the $\rm SU(2)$ component $\ell^{i}$ of the lepton doublet  
as $\left (\ell^{i}\right)^c = C  \overline{\ell^{i}}^T$.
The charge conjugation matrix $C$
fulfils the properties of the charge-conjugation matrix used in
\cite{Denner}\footnote{Note that this definition of the dimension-five operator is the hermitian conjugate of the one used in~\cite{polonais} where $C=i \gamma^2 \gamma^0$
  in the Dirac representation, since in this representation
  $C^{-1} = -C$.
}.
The coefficient $C^{\a\b}_5$ is symmetric
under the interchange of the generation indices $\a,\b$,
 the  New Physics scale  $\Lambda$ is assumed $\gg m_W$, 
 and the second term is the hermitian conjugate of the first.

In the broken theory, with  $ H_0 = v + (h/\sqrt{2})$,  $v\simeq m_t$,
this gives   a  Majorana neutrino mass matrix
\begin{equation}
  \delta {\cal L} = - \frac{1}{2} [m_\nu]_{\a \b}\overline{\nu_{\a}} \nu^c_\b + h.c
~~~~~~~~~~~~~~~~~[m_\nu]_{\a \b} = - \frac{v^2}{\Lambda} C_5^{\alpha\beta}
\end{equation}
In the charged lepton mass  eigenstate basis,
this mass matrix is diagonalised by the PMNS matrix
 $ [m_\nu]_{\a \b}  =U_{\a i} m_{\nu i} U_{\b  i} $. 

At dimension-six, we will be interested in  SM-gauge invariant operators that violate lepton flavour; a  complete list is given in Appendix \ref{app:ops}.
Following the conventions of~\cite{polonais,JMT}, they
are  added to the Lagrangian as:
\begin{equation}
\delta {\cal L}_6 = \sum_{X,\zeta}  \frac{C^\zeta_X}{\Lambda^2} {\cal O}^\zeta_X + {\rm h.c.} ~~~
\label{L3}
\end{equation}
where $X$  is an operator label and $\zeta$ represents all required
generation indices which are summed over all generations.
Of particular interest are
the operators that can be generated at one-loop with
two insertions  of  dimension-five operators,
as illustrated in figure \ref{fig:2}.
With  SM particle content,
these operators 
involve two Higgs doublets and two lepton doublets, four lepton doublets, 
or three Higgs doublets and leptons
of both chiralities.  In the ``Warsaw'' basis~\cite{polonais}, the possibilities at dimension-six are:
\begin{align}
\mathcal{O}_{H \ell (1)}^{\alpha \beta} =& \frac{i}{2}(H^\dagger \overset{\leftrightarrow}{ D_\m }  H )
(\overline{\ell_\alpha} \gamma^\m \ell_\beta) &
\mathcal{O}_{H \ell (3)}^{\alpha \beta} =& \frac{i}{2}(H^\dagger \overset{\leftrightarrow}{D_{\mu}^{a}}  H )
(\overline{\ell_\alpha} \gamma^\m \tau^{a} \ell_\beta) \nonumber \\
\mathcal{O}_{eH}^{\alpha \beta} =& (H^\dagger  H) \overline{\ell}_\alpha H e_\beta &
\mathcal{O}_{\ell \ell}^{\alpha \beta \gamma \delta} =& \frac{1}{2} (\bar{\ell}_\alpha \gamma_{\mu} \ell_\beta) (\bar{\ell}_\gamma \gamma^{\mu} \ell_\delta)
\label{LFVWarsaw}
\end{align}
where we normalise the ``Hermitian''  operators
 with a factor of 1/2 (see appendix
 \ref{app:ops} for  a discussion) in
 order to agree with~\cite{polonais,JMT}, and
\begin{align}
i(H^\dagger \stackrel{ \leftrightarrow}{  D_\m } H) &\equiv
i(H^\dagger   D_\m H) - i( D_\m H)^\dagger H \nonumber \\
&= H^\dagger (i\partial_\m H) -i (\partial_\m H )^\dagger  H
- gH^\dagger \tau^a W_{ \m}^a H   - 2y_{H} g'H^\dagger B_{ \m} H  \,,   \nonumber \\
i(H^\dagger \overset{\leftrightarrow}{D_\m^a } H) &\equiv
i(H^\dagger  \tau^a D_\m H) - i( D_\m H)^\dagger \tau^a H \,.
\label{doubleD}
\end{align}

The choice of operator basis implies a choice of operators that vanish
by the Equations of Motion (EOMs). For example $i\Dslash \ell_\a -
[Y_e]^{\a\s} He_\s = 0$ implies that the following operators
\begin{eqnarray}
{\cal O}^{\alpha\beta}_{v(1)}& =& (H^\dagger H)
i(\overline{\ell}_\alpha  \overset{\leftrightarrow}{ \Dslash } \ell_\beta ) 
- (H^\dagger H)(\overline{\ell}_\alpha H e_\sigma [Y_e^T]_{\sigma\beta} + 
[Y_e^*]_{\alpha\sigma} \overline{e}_\s H^\dagger \ell _\beta) \,,  \nonumber\\
{\cal O}^{\alpha\beta}_{v(3)} &= &i(H^\dagger \tau^{a} H )
(\overline{\ell}_\alpha  \overset{\leftrightarrow}{  \Dslash^{a} } \ell_\beta) - 
(H^\dagger H)( 
\overline{\ell}_\alpha He_\sigma [Y_e^T]_{\sigma\beta} + 
[Y_e^*]_{\alpha\sigma} \overline{e}_\sigma H^\dagger\ell _\beta) \,,
\label{vanish}
\end{eqnarray}
are EOM-vanishing operators. The role of these operators becomes clear
by noting that in intermediate steps of our off-shell calculations,
additional structures appear that can conveniently be matched onto
combinations of EOM-vanishing operators and operators of the Warsaw
basis. For example the structures involving two Higgs fields and a
covariant derivative of a lepton doublet are expressed in terms of
the above operators as:
\begin{eqnarray}
{\cal S}^{\alpha\beta}_{HD\ell(1)} &=&
(H^\dagger H) i(\overline{\ell}_\alpha  \stackrel{ \leftrightarrow}{ \Dslash } \ell_\beta ) 
=  {\cal O}^{\alpha\beta}_{v(1)} +
 {\cal O}^{\alpha\s}_{eH}[Y_e^T]_{\s\beta}
+ [Y_e^*]_{\alpha \s}{\cal O}^{\dagger \s\beta}_{eH} \,,
\nonumber\\
{\cal S}^{\alpha\beta}_{HD\ell(3)}& =&
i(H^\dagger \tau^{a} H )
(\overline{\ell}_\alpha  \overset{\leftrightarrow}{  \Dslash^a } \ell_\beta)
=  {\cal O}^{\alpha\beta}_{v(3)}  +
 {\cal O}^{\alpha\s}_{eH}[Y_e^T]_{\s\beta}
+ [Y_e^*]_{\alpha \s}{\cal O}^{\dagger \s\beta}_{eH} \,.
\label{EoMOEH}
\end{eqnarray}
In practice, if the coefficients $C^{\b\a}_{HD\ell(1)}$
and $ C^{\b\a}_{HD\ell(3)}$ of these structures are present, 
they are equivalent to 
$C^{\b\s}_{eH} =  C^{\b\a}_{HD\ell(1)} [Y_e]^{\a \s}
+  C^{\b\a}_{HD\ell(3)} [Y_e]^{\a \s}$
(and the hermitian conjugate relation).

\subsection{In the case of the 2HDM}
\label{ssec:2HDML}

In this section, we
consider  the addition of  a second  Higgs doublet  $H_2$ to the SM,
of the  same hypercharge as  the SM Higgs (which we relabel $H_1$). 
The   LFV induced by double-insertions
of dimension-five operators could be   more significant
in this model, because there
are several dimension-five operators,  so neutrino masses cannot
constrain them all.  However, a complete analysis of LFV in the
2HDM  would require extending the operator basis at dimension-six
and calculating the additional  terms in the RGEs, which is beyond the
scope of this work.  So for simplicity, we  make
three restrictions:
\ben
\item First, we  consider only the
  dimension-six LFV operators of the SMEFT. This is the
  appropriate  set of dimension-six operators
 just  above $m_W$, provided that  $H_2$  has  no vev, and that
 the  mass $m_{22}$ of the additional Higgses is
 sufficiently high: $m_W^2 \ll m_{22}^2 \ll \Lambda^2$.
 In our phenomenological analysis we extend this range to the scenario $m_W^2 \lesssim m_{22}^2 \ll \Lambda^2$, by considering a Higgs potential where the additional Higgses are not directly observable at the LHC, and where the Yukawa couplings of $H_2$ are vanishing. Such a scenario would for example be realised in the inert two Higgs doublet model~\cite{Deshpande:1977rw,Ma:2006km,Barbieri:2006dq,Belyaev:2016lok} and setting the scale $m_{22}$ close to the electroweak scale will not require the consideration of additional renormalisation group effects in the SMEFT.

\item Second, we suppose that at the high scale $\Lambda$ 
  no dimension-six LFV operators are generated.  This is
  unrealistic, but allows us to focus on the LFV
  generated by   double-insertions of the
  dimension-five operators. 
\item Third, we suppose
there is no LFV in the renormalisable couplings of the
2HDM (in particular, in the lepton  Yukawas), so that 
when matching the 2HDM + dimension-five operators
onto the SMEFT at the intermediate scale $m_{22}$,
no additional LFV operators are generated.
\een

Consider first the renormalisable Lagrangian. 
The Yukawa couplings can be written~\cite{howie}:
\bea
\delta {\cal L}_{2HDM} = 
  - ( \overline{\nu},  \overline{e_{ L}} )     [Y^{(1)}]
\left( \begin{array}{c} H_1^+ \\
 H_1^{0}
\end{array} \right)  e
 -  \overline{e}   [Y^{(1)}]^\dagger H_1^\dagger  \ell
-    ( \overline{\nu},  \overline{e_{ L}} )     [Y^{(2)}]
\left( \begin{array}{c} H_2^+ \\
 H_2^{0}
\end{array} \right)  e 
 - \overline{e}   [Y^{(2)}]^\dagger  H_2^\dagger  \ell \,,
\eea
where   the
flavour indices are implicit, and the basis in $(H_1, H_2)$ space is taken to be the ``Higgs basis'' where $\langle H_2\rangle = 0$. We suppose that
$ [Y^{(1)}]$ and $ [Y^{(2)}]$ 
are simultaneously diagonalisable on
their lepton flavour indices.

The  second Yukawa coupling
changes the Equations of Motion for the leptons, so
the  2HDM version of  the equation-of-motion vanishing
operators  (given in  eqn (\ref{vanish}) for the single Higgs model)
should be modified.
As a result,  the operators  ${\cal O}_{HD\ell(1)}$ and ${\cal O}_{HD\ell(3)}$
should  not be replaced only  by the SMEFT operator ${\cal O}_{eH}$,
as given in  eqns (\ref{EoMOEH}), but also by an operator
with an external $H_2$ leg. However, since  we neglect
dimension-six  operators
with external $H_2$,  we  use the
relations (\ref{vanish}) and  (\ref{EoMOEH})
also in the 2HDM case.

In this ``Higgs'' basis,
the most general Higgs potential is 
\bea
\label{pothbasis}
{V}&=& m_{11}^2 H_1^\dagger H_1+m_{22}^2 H_2^\dagger H_2
-[m_{12}^2 H_1^\dagger H_2+{\rm h.c.}]\nonumber\\[4pt]
&&\quad\!\!\!\! +\frac{1}{2}\lambda_1(H_1^\dagger H_1)^2
+\frac{1}{2}\lambda_2(H_2^\dagger H_2)^2
+\lambda_3(H_1^\dagger H_1)(H_2^\dagger H_2)
+\lambda_4(H_1^\dagger H_2)(H_2^\dagger H_1)
\nonumber\\[4pt]
&&\quad\!\!\!\! +\left\{\frac{1}{2}\lambda_5(H_1^\dagger H_2)^2
+\big[\lambda_6\,(H_1^\dagger H_1)
+\lambda_7(H_2^\dagger H_2)\big]
H_1^\dagger H_2+{\rm h.c.}\right\}\,.
\eea
In order to decouple
  the additional Higgses, we can, for instance,
 set  $m_{12}^2 = 0$ and assume
 $m_{22}^2 \gg m_W^2$, or  leave   $m_{22}^2 $
 free, and  impose  $m_{12}^2  = \lambda_6  = \lambda_7= [Y^{(2)}] =0$.

At dimension-five in the 2HDM, there are  four  operators~\cite{Babu}:
\begin{eqnarray}
  \label{eq:10}
\delta {\cal L} &=&
+ \frac{C_{5}^{\a \b }}{2\Lambda} (\overline{\ell_\alpha } \varepsilon H_1^{*})
(\ell^c_\beta  \varepsilon H_1^{*})
+ \frac{C_{5}^{\a \b *}}{2\Lambda} (\overline{\ell_\beta ^c}\varepsilon H_1)
(\ell_\alpha \varepsilon {H_1})
\nonumber\\ &&
+\frac{C_{21}^{\a \b }}{2\Lambda} {\Big (} (\overline{\ell_\alpha } \varepsilon H_2^{*})
(\ell^c_\beta  \varepsilon H_1^{*})
+(\overline{\ell_\beta } \varepsilon H_1^{*}) (\ell^c_\alpha \varepsilon H_2^{*}) {\Big )}
+\frac{C_{21}^{\a \b *}}{2\Lambda}
 {\Big (} (\overline{\ell_\beta ^c} \varepsilon {H_2})
(\ell_\alpha \varepsilon {H_1}) +(\overline{\ell_\alpha ^c} \varepsilon {H_1}) (\ell_\beta \varepsilon  {H_2}){\Big )}
 \nonumber\\ &&
 +\frac{C_{22}^{\a \b }}{2\Lambda} (\overline{\ell_\alpha }   \varepsilon H_2^{*})
 (\ell^c_\beta   \varepsilon H_2^{*})
 +\frac{C_{22}^{\a \b *}}{2\Lambda} (\overline{\ell_\beta ^c} \varepsilon {H_2})
 (\ell_\alpha \varepsilon {H_2})  
\nonumber\\ &&
-\frac{C_{A}^{\a \b }}{2\Lambda} (\overline{\ell_\alpha } \varepsilon \ell^c_\beta )
 (H_1^\dagger \varepsilon {H}_2^*)
-\frac{C_{A}^{\a\b *}}{2\Lambda} (\overline{\ell^c_\beta } \varepsilon \ell_\alpha)
 (H_2 \varepsilon {H_1} )  \,,
\label{d52HDM}  
\end{eqnarray}
where  $\{ C_{5},  C_{22},  C_{21}\}$ are symmetric on
flavour indices (so can contribute to
neutrino masses). In the ${\cal O}_{21}$ operator,
$(\overline{\ell_\alpha} \varepsilon H_2^{*}) (\ell_\beta^c \varepsilon H_1^{*}) = 
 (\overline{\ell_\beta} \varepsilon H_1^{*}) (\ell_\alpha^c \varepsilon H_2^{*})$,
but   both terms  are retained here because
they are convenient  in our Feynman rule conventions
\footnote{ The  operator
${\cal O}_{21}$ can also be written as
$2(\overline{\ell_\b } \epsilon H_1^{*}) (\ell^c_\a \epsilon H_2^{*})$
  $+(\overline{\ell_\b }\epsilon \ell^c_\a ) (H_1^{*} \epsilon H_2^{*})$
  using the identity (\ref{anotherid}), as done in  the first
  reference of~\cite{Babu}.
}.

Tree-level LFV is often avoided in the 2HDM by imposing a $Z_2$
symmetry on the renormalisable Lagrangian: if  under the
$Z_2$ transformation, $H_1\to H_1$ and  $H_2 \to -H_2$, then
$[Y_2], \, \lambda_6$ and $\lambda_7$ are forbidden. We will
later discuss this case, but do not  impose the $Z_2$ symmetry
from the beginning,
because it also forbids  the $C_{21}, C_{12} $ and $C_A$ coefficients
at dimension-five.


\section{The EFT Calculation}
\label{sec:caln}

\subsection{Diagrams and Divergences}

\begin{figure}[ht]
\unitlength.15mm
\SetScale{.4254}
\centering
  \begin{center}
\begin{picture}(244,220)(0,0)
\ArrowLine(25,125)(55,160)
\DashArrowLine(25,195)(55,160){3}
\DashArrowArc(110,160)(45,0,180){3}
\ArrowArc(110,160)(45,180,0)
\ArrowLine(165,160)(195,125)
\DashArrowLine(165,160)(195,195){3}
\GCirc(160,160){7}{.7}
\GCirc(60,160){7}{.7}
\put(-64,116){$\ell_{\a}^{n}(p_i)$} 
\put(100,86){$(\ell^c)^{l}_ {\r}$}
\put(200,126){$\ell^{i}_ {\b}(p_f)$}
\put(-70,196){$H^{M }(q_i)$} 
\put(200,196){$H^{J }(q_f)$}
\put(100,219){$H^{K }$}
\end{picture}
\hspace{2cm}
\begin{picture}(200,100)(0,0)
\ArrowLine(10,150)(40,150)
\ArrowLine(40,150)(100,150)
\DashArrowLine(150,150)(190,190){3}
\DashArrowLine(75,100)(100,150){3}
\DashArrowLine(110,100)(100,150){3}
\ArrowLine(100,150)(150,150)
\ArrowLine(150,150)(190,150)
\DashArrowArc(100,150)(50,0,180){3}
\GCirc(150,150){7}{.7}
\GCirc(100,150){7}{.7}
\put(-29,150){$e_{ \a}$}
\put(50,125){$\ell_{\a}^{o}$}
\put(120,122){$(\ell^c)^{l}_{\r}$} 
\put(200,150){$\ell^{n}_ { \b}$}
\put(60,84){$H^{K }$} 
\put(180,200){$H^ {I }$}
\put(110,90){$H^ {J }$}
\put(80,215){$H^ {M}$}
\end{picture}
\hspace{2cm}
\begin{picture}(244,220)(0,0)
\ArrowLine(25,125)(60,155)
\ArrowLine(60,165)(25,195)
\DashArrowArc(110,160)(50,0,180){3}
\DashArrowArcn(110,160)(50,0,180){3}
\ArrowLine(195,125)(160,155)
\ArrowLine(160,165)(195,195)
\GCirc(60,160){7}{.7}
\GCirc(160,160){7}{.7}
\put(-16,116){$\ell^{j}_{ \a}$} 
\put(100,220){$H^M$}
\put(200,126){$(\ell^c)_ { \r}^{l}$}
\put(-21,196){$(\ell^c)^{i}_ {\s}$} 
\put(200,196){$\ell^{k}_ { \b} $}
\put(100,86){$H^ {N }$}
\end{picture}
\vspace{-1.5cm}
 \end{center}
\caption{Diagrams involving two insertions of  dimension-five operators,
that can contribute to dimension-six lepton-flavour-violating operators.
SU(2) indices run from $I,...,O$ and $i,...,o$, lepton flavour indices
are $\a,\b,\r,\s$.
\label{fig:2}}
\end{figure}
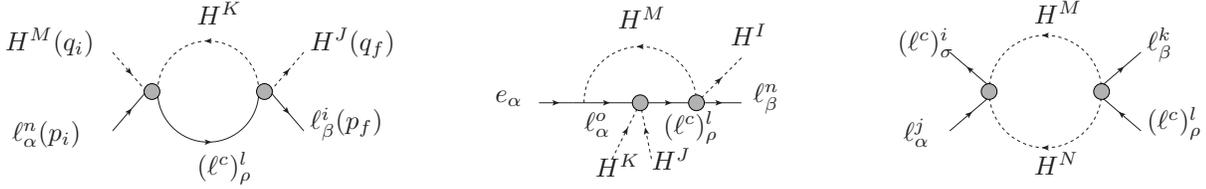

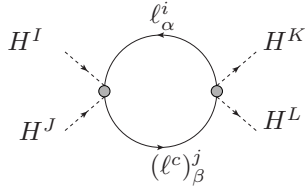
\begin{figure}[ht]
\unitlength.15mm
\SetScale{.4254}
  \begin{center}
\begin{picture}(244,220)(0,0)
\DashArrowLine(25,125)(60,155){3}
\DashArrowLine(25,195)(60,165){3}
\ArrowArc(110,160)(50,180,0)
\ArrowArc(110,160)(50,0,180)
\DashArrowLine(160,155)(195,125){3}
\DashArrowLine(160,165)(195,195){3}
\GCirc(160,160){5}{.7}
\GCirc(60,160){5}{.7}
\put(-16,116){$H^J$}
\put(100,220){$\ell^{i}_ { \a}$}
\put(200,126){$H^ {L}$}
\put(-26,196){$H^ {I}$} 
\put(200,196){$H^ {K} $}
\put(100,86){$(\ell^c)^{j}_{ \b}$}
\end{picture}
\vspace{-1.5cm}
 \end{center}
\caption{Two insertions of  dimension-five operators can also 
  contribute to 
  dimension-six operators involving four Higgses via this diagram.
\label{fig:2b}}
\end{figure}

Diagrams with  two insertions of the dimension-five operators are illustrated in 
figures \ref{fig:2} and  \ref{fig:2b}. We focus on the
lepton flavour violating  diagrams of figure \ref{fig:2},
and discuss the
  four-Higgs operators generated by figure \ref{fig:2b} in Appendix
  \ref{sec:lept-cons-contr}, because
four-Higgs interactions are flavour conserving and arise in the SM.

The Feynman rules arising from the
(tree-level) Lagrangian of equations
(\ref{L1}, \ref{d5Weinberg}, \ref{L3})
are given in Appendix \ref{app:caln}.
We use them to  evaluate,  using   dimensional
regularisation  in $4-2\epsilon$ dimensions in $\rm \overline{MS}$,
the coefficient of the  $1/\epsilon$ divergence
of each diagram  of  figure \ref{fig:2}.
These coefficients can be expressed as
a sum of numerical factors multiplying the Feynman
rules for the dimension-six operators of equations
(\ref{LFVWarsaw}) and (\ref{EoMOEH}) (these Feynman
rules are given in Appendix \ref{app:caln}), and
then the EOMs  are  used to
transform  the operators of eqn~(\ref{EoMOEH})
to ${\cal O}_{eH}$ and  ${\cal O}^\dagger_{eH}$.
The   required counterterm $\Delta C_O$
for each of  the  dimension-six  operators 
given in 
eqn~(\ref{LFVWarsaw}) can be identified 
 as  $(-1)\times$ the numerical factor that multiplies
 its Feynman rule. This counterterm 
 is added in the Lagrangian to the
 operator  coefficient  $C_{\cal O}$,
 resulting  in a ``bare'' coefficient
 $
 C_{{\cal O},bare} = \mu^{2 \epsilon} (C_{\cal O} + \Delta C_{\cal O})
$
 that should
 be independent of the $\overline{\mathrm{MS}}$ renormalisation scale $\mu$.
 Note that the factor $\mu^{2 \epsilon}$ is chosen such that bare Lagrangian remains $d$-dimensional.

 A more complete and rigorous presentation
 will be required in the next section,
 in order to derive the
 RGEs, so  let us replace counterterms
 by $Z$ factors in order to minimise notation and 
introduce the necessary factors of $\mu^{2\epsilon}$ 
to obtain the correct dimensions.
 More details of the formalism and
calculations are given in Appendix
\ref{app:RGEs}.

 We allow for multiple operators at both
 dimension-six and -five,
   and align the dimension-six coefficients in a row vector
 $\tilde{C}$, and
  the dimension-five coefficients in a row  vector $\vec{C}$.
  Then the bare coefficients can
 be written
\begin{equation}
\label{Cbare2}
\vec{C}_{bare} = \mu^{2\epsilon} \vec{C} [Z] ~~~,
~~~\tilde{C}_{bare} =  \mu^{2\epsilon}\big[ \tilde{C} \hat{Z} +  \vec{C} [\tilde{Z}] \vec{C}^\dagger \, \big]
\end{equation}
where   matrices  wearing a hat act
on the space of dimension-six
coefficients, and 
matrices in square brackets act in the dimension-five   space,
so  $\hat{Z}$ represents the renormalisation of dimension-six
coefficients amongst themselves, and $[Z]$ represents the renormalisation of dimension-five coefficients.
The quantity $[\tilde{Z}]$
renormalises insertions of two dimension-five operators;
$[\tilde{Z}]_k^{ij}$ is
  a vector in the  dimension-six space with index $k$, and  a matrix in
  the dimension-five space with indices $i,j$. In the single
  Higgs model, $i,j$ correspond to the flavour indices
  of the Weinberg operator, e.g. $i = \a\b$,
  $j = \r\s$. The index $k$ corresponds to the
  operator labels and flavour indices of
  dimension-six  operators. The counterterms
  that renormalise the diagrams of figure~\ref{fig:2}
  are then components of the vector $\vec{C} [\tilde{Z}] \vec{C}^\dagger$.
  All terms in the above expressions assume an implicit sum over flavour indices; the explicit flavour dependence is presented in Appendix~\ref{app:renormalisation}.

The first diagram of figure \ref{fig:2} has two Higgs
and two doublet-lepton legs and so must be renormalised by the operators
${\cal O}_{H \ell(3)}$ and 
${\cal O}_{H \ell(1)}$, and the structures
${\cal S}_{HD \ell(1)}$
and ${\cal S}_{HD \ell(3)}$. Since these all
involve a derivative, the diagram is calculated
for finite external momenta. 
The  counterterms that we obtain
from this diagram  differ
from those given  in~\cite{BGJ};
as discussed in Appendix \ref{sec-1-3}, it
appears that the authors of~\cite{BGJ} dropped  one
of the terms multiplying the $1/\epsilon$
divergence. We check our result by
attaching an external $B_\mu$ or
$W^a_\mu$ boson, in all possible
ways, to the first diagram of figure \ref{fig:2},
and verify that our counterterms also
cancel the divergences of the
2-Higgs-2-lepton-gauge boson vertices
generated by two insertations of
the Weinberg operator (this is outlined in Appendix \ref{app:W}). This diagram
can be renormalised using the following counterterms:
\begin{eqnarray}
(\vec{C} [\tilde{Z}] \vec{C}^\dagger)
^{\b \a}_{H\ell (1)} &=& -\frac{3}{4} \frac{1}{16 \pi^2 \epsilon} [C_5C_5^*]^{\b \a} \,,
\label{ct0}\\
(\vec{C} [\tilde{Z}] \vec{C}^\dagger)^{\b \a}_{H\ell (3)} &=& + \frac{2}{4} \frac{1}{16\pi^2 \epsilon} [C_5C_5^*]^{\b \a} \,,
\label{ct1} \\
(\vec{C} [\tilde{Z}] \vec{C}^\dagger)^{\b \a}_{HD\ell (1)} &=& - \frac{3}{4} \frac{1}{16 \pi^2 \epsilon} [C_5C_5^*]^{\b \a} \,,
\label{adD1} \\
(\vec{C} [\tilde{Z}] \vec{C}^\dagger)^{\b \a}_{HD\ell (3)} &=& + \frac{2}{4} \frac{1}{16 \pi^2 \epsilon} [C_5C_5^*]^{\b \a} \,,
\label{adD3}
\end{eqnarray}
where the last two counterterms represent divergences proportional to the structures $\mathcal{S}_{HD \ell (1)}$ and $\mathcal{S}_{HD \ell (3)}$, which contribute to the renormalisation of $C_{eH}$ through the linear combination given in eqn~\eqref{EoMOEH}.

The middle  diagram of figure \ref{fig:2}
contributes to ${\cal O}^{ \b\a}_{eH} $,
and  the divergence
it  induces can be removed by the counterterm
$(16\pi^2 \epsilon)^{-1} [{C_5 C_5^*} Y]^{ \b\a}$
(where the flavour index order 
is doublet-singlet).
Including also
 the counterterms   for  ${\cal S}_{HD\ell (1)}^{\b\a}$ and
 ${\cal S}_{HD\ell (3)}^{\b\a}$ (eqns (\ref{adD1},\ref{adD3}))
gives
\begin{equation}
\label{ct2}
(\vec{C}  [\tilde{Z}] \vec{C}^\dagger)_{eH}^{\b \a } = +
\frac{3}{4} \frac{1}{16 \pi^2 \epsilon} [ C_5C_5^* Y]^{\b \a}
\end{equation}
Since the structures ${\cal S}_{HD \ell (3)}$ and
  ${\cal S}_{HD \ell (1)}$ are hermitian, they contribute to the  renormalisation  of
both ${\cal O}_{eH} $ and ${\cal O}^\dagger_{eH} $ (see eqn(\ref{EoMOEH})).
Only the contribution to  ${\cal O}_{eH} $ is included in (\ref{ct2}),
because 
the hermitian conjugate in \eqref{L3} generates a counterterm proportional to $\mathcal{O}_{eH}^{\dagger}$ that absorbs the divergence of the the ``conjugate'' process of figure \ref{fig:2}.

The  third diagram of figure \ref{fig:2}
contributes to the four-lepton operator $ {\cal O}^{\r\a\b\s}_{\ell \ell}$, and  the divergence it induces can be removed by the counterterm
\begin{equation}
\label{ct3}
(\vec{C} [\tilde{Z}] \vec{C}^\dagger)^{\r\a\b\s}_{\ell\ell} = -\frac{1}{4} \frac{1}{16\pi^2 \epsilon} C_5^{\rho \b}C^{* \s \a}_5 \,.
\end{equation}

\subsection{The 2HDM}

In the 2HDM, we consider diagrams analogous
to figure  \ref{fig:2},  but  with  insertions
of any of the dimension-five
operators given in eqn (\ref{d52HDM}). 
The external Higgs lines are required to be
$H_1$, but the  internal  Higgs
lines can be either doublet.  The counterterms required to 
cancel double-insertions of  the ${\cal O}_5$ operator,
discussed in the previous section,  also  arise  in the 2HDM. In this
section, we only list the additional contributions to the counterterms.

We start again with the first diagram of figure \ref{fig:2}, with
${\cal O}_{21}$ or ${\cal O}_{A}$ at the vertices. Since by construction,
the Feynman rule for ${\cal O}_{21}$ is identical to the
rule for ${\cal O}_{5}$, double-insertions of ${\cal O}_{21}$ require
the  same counterterms as given in eqns (\ref{ct0}) to (\ref{adD3}), but with
${C}_{5}, {C}^{*}_{5}$ replaced by  ${C}_{21}, {C}^*_{21}$.
Double insertions of the antisymmtric operator
${\cal O}_{A}$ require the counterterms:
\bea
\Delta (\vec{C}  [\tilde{Z}] \vec{C}^\dagger)
^{\b \a}_{H \ell (1)} = \frac{1}{4} \frac{1}{16\pi^{2}\epsilon} [C_AC_A^*]^{\b \a} \, ,\label{ctAA1}\\
\Delta (\vec{C}  [\tilde{Z}] \vec{C}^\dagger)^{\b \a}_{HD \ell (1)} =\frac{1}{4}\frac{1}{16\pi^{2}\epsilon} [C_AC_A^*]^{\b \a} \, .
\label{ctAAD1}
\eea
Finally, ${\cal O}_{A}$ at one vertex and ${\cal O}_{21}$ at
the other require the  contributions to the counterterms:
\bea
\Delta (\vec{C}  [\tilde{Z}] \vec{C}^\dagger)
^{\b \a}_{H \ell (3)} =\frac{1}{4}\frac{1}{16\pi^{2}\epsilon} [C_AC_{21}^* - C_{21}C_A^*]^{\b \a}\,  ,\label{ctA12a}\\
\Delta (\vec{C}  [\tilde{Z}] \vec{C}^\dagger)^{\b \a}_{HD \ell (3)} =\frac{1}{4}\frac{1}{16\pi^{2}\epsilon} 
[C_AC_{21}^* - C_{21}C_A^*]^{\b \a}  \, .
\label{ctA12b}
\eea
It is straightforward to check, using respectively  the antisymmetry and
symmetry of
$C_A$  and $C_{21}$ on flavour indices,
that the combination $[C_AC_{21}^* - C_{21}C_A^*]$ is hermitian,
as expected for the coefficients of ${\cal O}_{H\ell(3)}$
and ${\cal O}_{HD\ell(3)}$. 

Consider next the middle diagram of figure \ref{fig:2}. Only the
internal Higgs lines can be $H_2$, so the additional
divergences in the 2HDM will arise from ${\cal O}_A$
or ${\cal O}_{21}$ at the  vertex farthest from the Yukawa
coupling, which can be cancelled by the counterterms
$(16\pi^2 \epsilon)^{-1} [C_{21} C_5^* Y_2]^{ \b\a}$
  and $
- [C_{A} C_5^* Y_2]^{ \b\a}/( 16\pi^2 \epsilon)$.
Including also
 the additional  counterterms   for  ${\cal O}_{HD\ell (1)}^{\b\a}$ and
 ${\cal O}_{HD\ell (3)}^{\b\a}$  in the 2HDM
gives
\begin{equation}
  \Delta (\vec{C}  [\tilde{Z}] \vec{C}^\dagger)_{eH}^{\b \a } = \frac{1}{4}\frac{1}{16\pi^{2}\epsilon} \left(
4[ (C_{21} -C_A)C_5^* Y_2]^{\b \a}
+ [ ( C_{A}C_A^*  + C_{A}C_{21}^* - C_{21}C_{A}^* - C_{21}C_{21}^*) Y_1]^{\b \a}
\right) \,.
\label{ct2THDM}
\end{equation}

Finally, for the four-lepton operator, there are additional
counterterms in the 2HDM to cancel the divergences induced by
double-insertions
of ${\cal O}_{22}$, of ${\cal O}_{21}$, and  of ${\cal O}_{A}$.
(The possible  diagrams with an insertion of both
 ${\cal O}_{21}$ and  ${\cal O}_{A}$ vanish due to anti-symmetry.)
We obtain:
\begin{equation}
\label{eq:1}
\Delta (\vec{C}  [\tilde{Z}] \vec{C}^\dagger)^{\r\s\b\a}_{\ell\ell}
=  -\frac{1}{4}\frac{1}{16\pi^{2}\epsilon}C_{22}^{\rho \b}C_{22}^{* \a \s}
 -\frac{1}{2}\frac{1}{16\pi^{2}\epsilon}C_{21}^{\rho \b}C_{21}^{* \a \s}
 +\frac{1}{2}\frac{1}{16\pi^{2}\epsilon}C_{A}^{\rho \b}C_{A}^{* \a \s} \,.
\end{equation}

\subsection{The Renormalisation Group Equations}
\label{ssec:RGE}

The  contribution  of dimension-five
 operators  to the 
 Renormalisation Group Equations of dimension-six operators,
   due to double insertions, can be obtained
 following the discussion of Herrlich and
 Nierste \cite{Herrlich:1994kh}. The derivation is
 presented 
  in Appendix \ref{app:RGEs}.  Here we schematically
  outline the result.

 The bare Lagrangian coefficients are defined at one loop
 as  in eqn (\ref{Cbare2}), where the counterterm
   for  one operator can depend on the coefficients of
   other operators.    Recall that the bare coefficients
   are  independent of the dimensionful parameter $\mu$,
   and that the renormalised $C$s are dimensionless.
   Using $ \vec{C}=\mu ^{-2 \epsilon} \vec{C}_{bare} [Z^{-1}]$
   allows one to  obtain, in $4-2\epsilon$ dimensions: 
   \beq
   (16 \pi^2) \mu \frac{d}{d \mu}  \vec{C} = 
- \vec{C} \left\{ 2 \epsilon (16 \pi^2) + (16 \pi^2) \left[ \mu \frac{d}{d \mu} Z \right][ Z^{-1}] \right\} \equiv 
\vec{C}  [\gamma] -  2 \epsilon (16 \pi^2) \vec{C} \,
\label{eps}
\eeq
where $[\gamma]$ denotes the 4-dimensional anomalous dimension matrix, 
and we   (unconventionally)\footnote{The usual definition \cite{burashouches} is
  $\mu \frac{d}{d \mu}  C = C \gamma$, then $\gamma$ is expanded in loops:
  $\gamma = \frac{\alpha_s}{4\pi} \gamma_0 + ...$. However, here we
  only work at one loop,  have other subscripts  on our $\g$s and
  the one loop mixing of dimension-five-squared into dimension-six
  is not induced by a renormalisable coupling. So we factor
out the $16\pi^2$.} 
 factor the  $16\pi^2$ out of  the  anomalous dimension matrices. 
While the $- 2\epsilon$ term does not contribute in $d=4$ dimensions to the mixing of the dimension-five operators,
it plays an essential role in the renormalisation group equations of the dimension-six operators.

  For the dimension-six coefficients, it is straightforward to obtain
 from eqn (\ref{Cbare2}):
\begin{equation}
\label{RGE1}
\begin{split}
  \mu \frac{d}{d \mu}\tilde{C} =& -\tilde{C} \cdot \left\{ \mu \frac{d}{d \mu}  \hat{Z}\right\}
  \hat{Z}^{-1}
+ 2 \epsilon \, \vec{C} \cdot \tilde{Z} \cdot \vec{C}^\dagger   \hat{Z}^{-1}
\\ 
&
- \vec{C} \cdot \left[ \mu \frac{d}{d \mu} \tilde{Z} \right] \cdot \vec{C}^\dagger   \hat{Z}^{-1}
-   \vec{C} \cdot [Z] \left[ \mu \frac{d}{d \mu} Z^{-1} \right]
\cdot [\tilde{Z}] \cdot \vec{C}^\dagger  \hat{Z}^{-1}
-   \vec{C} 
\cdot [\tilde{Z}] \cdot  \left[ \mu \frac{d}{d \mu} Z^{-1} \right]^{\dagger}
 [Z]^{\dagger}\vec{C}^\dagger  \hat{Z}^{-1} \,,
\end{split}
\end{equation}
  where terms of  ${\cal O}(\epsilon)$ that vanish in 4 dimensions are neglected, and  the summation over flavour and operator indices is indicated with a dot.
  The second line can be dropped, because the first term vanishes at one loop,
  and   the remaining terms are of two-loop order because  both $[\tilde{Z}]$ and $  d [Z^{-1}]/d\mu$ arise at one-loop.   So the renormalisation group equations
  for the dimension-six coefficients can be written
\begin{equation}
(16 \pi^2) \mu \frac{d}{d \mu}\tilde{C} =
\tilde{C}  \hat{\gamma} 
+ \vec{C} [ \tilde{\gamma}]  \vec{C}^\dagger \,,
\end{equation}
where $\hat{\gamma}$  is  the  one-loop anomalous dimension
matrix  for dimension-six operators \cite{JMT} and $[\tilde{\gamma}] =
2 (16\pi^2) \epsilon [ \tilde{Z} ]$ is the anomalous dimension tensor.

We give below the anomalous dimensions describing the
one-loop mixing of double-insertions of  dimension-five operators
into  LFV dimension-six operators, in the 2HDM.  The
single Higgs model can be easily retrieved by
setting $C_{21} = C_A = C_{22} = 0$ in the equations below.
The anomalous dimension tensor mixing  a pair
of dimension-five operators into a dimension-six operator
is neccessarily a three-index object; below we sum over
the two  dimension-five indices, and give these
summed components of the tensor as elements of a vector
in the dimension-six operator space.
These anomalous dimensions  parametrise
the mixing  of  figure   \ref{fig:2}  in the 2HDM
(recall that a factor $1/16\pi^2$ is scaled out of our anomalous dimensions):
\bea
(\vec{C}  [\tilde{\g}] \vec{C}^\dagger)^{\b\a}_{H\ell(1)} &= & -C_5^{\b\r}\frac{3\delta_{\r\s}}{2} C_5^{*\s\a} \nonumber\\
&& 
-C_{21}^{\b\r}\frac{3\delta_{\r\s}}{2} C_{21}^{*\s\a} +C_{A}^{\b\r}\frac{\delta_{\r\s}}{2} C_{A}^{*\s\a}
\label{AD1}\\
(\vec{C}  [\tilde{\g}] \vec{C}^\dagger)^{\b\a}_{H\ell(3)} &= & C_5^{\b\r}\delta_{\r\s} C_5^{*\s\a}\nonumber\\
&& +  C_{21}^{\b\r}\delta_{\r\s} C_{21}^{*\s\a}
+ C_A^{\b\r}\frac{\delta_{\r\s}}{2} C_{21}^{*\s \a} - C_{21}^{\b\r} \frac{\delta_{\r\s}}{2}C_A^{*\s \a} 
\label{AD2} \\
(\vec{C}[ \tilde{\g}] \vec{C}^\dagger)^{\b\a}_{eH} &= & C_5^{\b\r}\frac{3 [Y_1]_{\eta\a} \delta_{\r\s}}{2} C_5^{*\s\eta} \nonumber \\
& &+
 2[(C_{21} -C_A) C_5^{*} Y_2]^{\b\a} 
+ \frac{1}{2}[ ( C_{A}C_A^*  + C_{A}C_{21}^* - C_{21}C_{A}^* - C_{21}C_{21}^*) Y_1]^{\b \a}
  \label{AD3} \\
(\vec{C} [\tilde{\g} ]\vec{C}^\dagger)^{\r\s\b\a}_{\ell\ell} &=& 
-C_5^{\b\r}\frac{ 1 }{2} C_5^{*\s\a} \nonumber \\
& & -C_{22}^{\b\r}\frac{1}{2} C_{22}^{*\s\a}  
    - C_{21}^{\b\r} C_{21}^{*\s\a}
    + C_{A}^{\b\r} C_{A}^{*\s\a}
\label{AD5}
\eea
where the operator label and flavour indices  on the left-hand-side 
refer to  the dimension-six operator (the dimension-five indices are summed).

In the next section, we will need
  the RGEs for dimension-five operators.  Recall that
in the single Higgs model,   $ [\gamma]$ is
in principle a 9$\times 9$ matrix (or 6$\times 6$, if
one uses the symmetry of $C_5^{\a\b}$),
mixing the  elements of $C_5$ among themselves.
However, in  the  basis
where the charged leptons are diagonal,
 $[\gamma]$ is diagonal, and 
the anomalous dimension   for the coefficient
$C_5^{\alpha \beta}$ of the Weinberg operator is \cite{Babu}:
\bea
16\pi^2   \gamma &=& - \frac{3}{2} ([Y_e]_{\a\a}^2 + [Y_e]_{\b\b}^2) +
    (\lambda - 3g_2 + 2 {\rm Tr}(3[Y_u]^\dagger [Y_u] + 3[Y_d]^\dagger [Y_d]
+[Y_e]^\dagger [Y_e]))
\eea
where the Higgs self-interaction in the SM Lagrangian
is $\frac{\lambda}{4} (H^\dagger H)^2$, and $[Y_f]$ are
the fermion Yukawa matrices.


\section{Phenomenology}
\label{sec:pheno}


In order to solve the RGEs,
it is convenient to define 
$t = \frac{1}{16\pi^2}  \ln \frac{\mu}{m_W}$,  in which case the  one-loop RGEs
for dimension-five and -six operator coefficients can be written as
\bea
\frac{d}{d t}\tilde{C} &=&
\tilde{C} \cdot \hat{\gamma} 
+   \vec{C} \cdot [\tilde{\g}] \cdot \vec{C}^\dagger
\nonumber \\
\frac{d}{d t}\vec{C} &=& \vec{C} \cdot [{\gamma}] ~~~.
\eea
These are among
the most familiar of differential equations,
whose  solutions
have the form
\bea
\vec{C}(t_f)  &=& \vec{C}(0) \exp\{ \g t_f \} \simeq \vec{C}(0) {\Big [} 1 + \g
  \frac{1}{16\pi^2} \ln \left( \frac{\Lambda}{m_W} \right) +... {\Big ]}
\label{diffequ1}
\\
\tilde{C}(t_f)  &=&  {\Big [}
  \int _{0}^{t_f} d\tau    \vec{C}(0) e^{\g\tau}   [\tilde{\g}]  [e^{\g\tau}]^T
       \vec{C}^\dagger(0)
 e^{ -\hat{\g} \tau }
  + \tilde{C}(0)
  {\Big ]}  e^{ \hat{\g} t_f }
\label{diffequ}
\eea
where $ 16 \pi^2 t_f = \ln \left( \frac{\Lambda}{m_W} \right)$. In these
solutions, the anomalous dimension
matrices were  approximated  as constant;
this is not a good approximation, because the
anomalous dimensions 
depend on running  coupling constants,
in particular the Yukawa couplings  can evolve  significantly
above $m_W$. 

A simple solution to eqn (\ref{diffequ}) can be obtained by expanding
the exponentials under the integral, as in eqn (\ref{diffequ1}):
\bea
\tilde{C}(m_W)  &=&
 \tilde{C}(\Lambda) -  \tilde{C}(\Lambda)  \hat{\g} \frac{1}{16\pi^2}\ln \frac{\Lambda}{m_W}  -
  \vec{C}(\Lambda)    [\tilde{\g}] 
       \vec{C}^\dagger(\Lambda)   \frac{1}{16\pi^2} \ln\frac{\Lambda}{m_W} + ...
\label{diffequ3}
\eea

\subsection{The single Higgs model}
\label{ssec:sHm}

In the SM case where there is only one Higgs doublet, there is
only the Weinberg operator at dimension-five:  a symmetric $3\times 3$
matrix, whose entries 
are  determined by neutrino masses and mixing angles
(in the mass basis of charged leptons). We now want to
estimate the contribution of  double-insertions of this
dimension-five operator to  lepton-flavour violating
processes. 

We neglect the
``Majorana phases'',  suppose that the lightest neutrino
mass is negligible, and  and neglect
the lepton Yukawas in the RGEs. 
Then the RG running 
of $C_5^{\a\b}$ between $m_W$ and $\Lambda$
can be approximated as a rescaling, with
$\gamma \approx \lambda -3 g_2+6 y_t^2\approx 3.5$:
\beq
{C}_5^{\a\b}(\Lambda)  = {C}_5^{\a\b}(m_W)  {\Big [} 1 + 3.5
  \frac{1}{16\pi^2} \ln \frac{\Lambda}{m_W} +... {\Big ]}
\eeq
For $\Lambda\leq 10^{16}$ GeV, the log is $\leq  32$.

We can now estimate the contribution of the neutrino
mass operator to lepton flavour violating processes from eqn
(\ref{diffequ3}).  We neglect $\tilde{C}(\Lambda)$
and find that the  contribution is $\frac{1}{16 \pi^2} \ln \frac{\Lambda}{m_W}\times$ the  coefficients of eqns (\ref{AD1}) to (\ref{AD5}),
that is, of order 
\beq
\tilde{C}(m_W) \sim \frac{C_5^2}{16\pi^2} \ln  \frac{\Lambda}{m_W}~~.
\label{slooppy}
\eeq
As expected, this is negligibly small, because 
$   {C_5^2}/{\Lambda^2} \sim {m_\nu^2}/{v^4} \,.$

\subsection{The two Higgs doublet model}
\label{sec:two-higgs-doublet}

Experimental Neutrino data constrain the dimension-five operator in the one Higgs doublet model, so the
lepton flavour violating effects estimated in eqn (\ref{slooppy}) are suppressed by the smallness of the neutrino masses.
The situation changes in an extended Higgs sector, where more than one dimension-five operator is present.
The operator $\mathcal{O}_A$ cannot contribute to neutrino masses as it is anti-symmetric in flavour space and is hence unconstrained.
In addition, the neutrino mass contribution of operators $\mathcal{O}_{21}$ and $\mathcal{O}_{22}$ is suppressed if the vacuum expectation value of the second Higgs doublet is small.
Renormalisation group effects \cite{Babu,Antusch:2001vn,Grimus:2004yh} will in general mix all operators, which could lift these suppression mechanisms at loop level.
However the mixing factorises in the limit where $\lambda_6, \lambda_7$ and $Y^{(2)}$ tend to zero: then the operators $\mathcal{O}_{21}$ and $\mathcal{O}_A$ will not mix into $\mathcal{O}_5$ and $\mathcal{O}_{22}$ and are hence not constrained by the observed neutrino masses.
Furthermore, the mixing of $\mathcal{O}_{22}$ into $\mathcal{O}_5$ vanishes in the limit where in addition $\lambda_5$ tends to zero (see~\cite{Gorbahn:2009pp} for a symmetry argument).

In the following we will study the sensitivity of lepton-flavour violating decays to these additional operators.
We assume that the Wilson coefficients of the dimension-five operators are generated at $\Lambda = 10 \mathrm{TeV}$, while all other dimension-six Wilson coefficients are zero at this scale.
To avoid constraints from the observed neutrino masses we consider the scenario where the second Higgs doublet has a negligible vacuum expectation value and a mass at the weak scale.
The Higgs sector could be assumed to be close to that of an inert two-Higgs doublet model \cite{Deshpande:1977rw,Ma:2006km,Barbieri:2006dq,Belyaev:2016lok} and the dangerous couplings $\lambda_6, \lambda_7$ and $Y^{(2)}$ are not generated radiatively. 
Renormalisation group running will then generate non-zero Wilson coefficients of several dimension-six operators at $\mu \sim v$.
Only those dimension-six operators that involve standard model particles are of interest to us, since the vanishing vacuum expectation value of the second Higgs doublet will suppress the contribution of the other operators after spontaneous symmetry breaking.
Applying the constraints of Table~\ref{tab:bds} of the Wilson coefficients evaluated using eqn~\eqref{diffequ3} neglecting the small log $\ln ({m_{22}}/m_W)$, we find the following: 
the $\mu \to 3 e$ decays provide the greatest sensitivity to the additional dimension-five Wilson coefficients. In particular the left-handed contribution implies
\begin{equation}
  \label{eq:mu3e1}
  \bigg|  
  C_{21}^{ee} C_{21}^{e\mu *} +
  0.5 C_{22}^{ee} C_{22}^{e\mu *} +
  0.1 \sum_{\sigma} 
   \left( C_A^{e\sigma } - C_{21}^{e\sigma } \right) \left( C_A^{\sigma \mu *} + C_{21}^{\sigma \mu *}  \right)
  \bigg| < 
  \frac{1}{5.2 \ln \left( \Lambda/m_{22} \right)} 
  \left( \frac{\Lambda}{10 \mathrm{TeV}} \right)^2
 \,,
\end{equation}
where we neglected the mixing of the dimension-five operators amongst themselves, as this would contribute at two-loop order to the lepton flavour violating processes. 
For the right-handed contribution we find
\begin{equation}
  \label{eq:mu3e2}
  \bigg|  
  \sum_{\sigma} 
   \left( C_A^{e\sigma } - C_{21}^{e\sigma } \right) \left( C_A^{\sigma \mu *} + C_{21}^{\sigma \mu *}  \right)
  \bigg| < 
  \frac{1.6}{\ln \left( \Lambda/m_{22} \right)} 
  \left( \frac{\Lambda}{10 \mathrm{TeV}} \right)^2
 \,,
\end{equation}
which exhibits a weaker sensitivity.
It is interesting to note that current experimental data is already sensitive to this parameter space of Wilson coefficients.
The contribution to the $\mu \to e \gamma$ is further suppressed by the smallness of the Yukawa couplings
which puts these beyond current experimental sensitivity. We also checked that the current experimental situation for $\tau$ decays does not lead to significant constraints.


\section{Summary}
\label{sec:summ}

Motivated by neutrino masses and
the  expected progress in  searches for lepton flavour violation,
we calculated the leading one-loop contribution
of a pair of lepton number violating dimension-five operators
to  the coefficients
of lepton flavour violating dimension-six operators.
The diagrams are given in  figure \ref{fig:2}.
The dimension-five operators that we considered are
the Weinberg operator, constructed out of SM fields and
given in eqn~(\ref{d5Weinberg}), and 
three additional  dimension-five
operators that can be constructed in the Two Higgs Doublet Model,
given in eqn~(\ref{eq:10}).
The dimension-six,  lepton flavour violating
operators  of the SMEFT are listed in  Appendix \ref{app:ops},
in the ``Warsaw'' basis,
and the subset of  these  operators
relevant for our calculation is given
in eqn (\ref{LFVWarsaw}). 
A selection   of constraints on  their
coefficients, evaluated at the weak scale,
is given in Appendix \ref{app:bds}.

In section \ref{sec:caln},  we obtain the anomalous
dimensions mixing two dimension-five operators into
the lepton flavour violating operators of eqn~(\ref{LFVWarsaw}). 
The required  counterterms are given in
eqns~(\ref{ct0}-\ref{ct3}) for the Standard Model with a single Higgs,
and in eqns~(\ref{ctAA1}-\ref{eq:1}) for the
case of the Two Higgs doublet model.
Then in section \ref{ssec:RGE},  we  outline
the derivation of the renormalisation group equations
(Appendices \ref{app:renormalisation} and \ref{app:RGEs}  present
our calculation and the flavour dependence of our result in more detail),
and the resulting  anomalous  dimensions
are listed in eqns~(\ref{AD1}-\ref{AD5}).
The mixing of two
dimension-five operators into the lepton flavour
conserving four-Higgs operator, via the diagram
of figure \ref{fig:2b} is given in Appendix
 \ref{sec:lept-cons-contr}; however, we do not consider mixing
into dimension six operators constructed
with the second Higgs of the 2 Higgs Doublet Model.
This  completes the one loop renormalisation group equations
of the standard model effective theory, up to
operators of dimension six.
  It is amusing that the insignificant effect we calculate
  does not involve Standard Model couplings, so, in
  an expansion in terms of SM couplings, our result is the
  ``leading'' contribution to the one-loop RGEs of
  the  dimension-six SMEFT \footnote {Mixing among dimension
  six operators  occurs via the exchange of a SM particle, so 
is $\propto$  [SM  coupling$]^2$).}.

In  the effective field theory constructed
with Standard Model fields,  the
coefficient of the Weinberg operator
is proportional to the neutrino mass
matrix.  So the    lepton flavour changing amplitudes
induced by  double insertions of the Weinberg operator
are  $\propto (m_\nu/m_W)^2 \ln \Lambda/m_W$,
and far below current sensitivities.
This is outlined in section \ref{ssec:sHm}. However, the
situation is different in the
2 Higgs doublet model, as
discussed in section \ref{sec:two-higgs-doublet}: there
are four operators at dimension five,  and  the neutrino
mass matrix  only constrains one  combination. 
 We evaluated the mixing of  the four operators into
lepton flavour violating operators of the standard model effective
theory, and for a lepton number violating scale of
$10~\mathrm{TeV}$
we found that the current experimental value of $\mu \to 3 e$
is sensitive to the Wilson coefficients of these additional operators.

\subsubsection*{Acknowledgements}

The research of MG was supported in part by U.K. Science and
Technology Facilities Council (STFC) through the Consolidated Grant
ST/L000431/1. MG thanks the MITP workshop on ``Low-Energy Probes of
New Physics'' for hospitality and a stimulating work environment.
ML would like to thank the Les Houches Summer School Session
CVIII for much useful and interesting discussion, and a wonderful
environment in which part of this work was done.
We thank Rupert  Coy for drawing our attention to reference
\cite{BGJ}. 

\appendix


\section{ Feynman rules and Identities}
\label{app:caln}

\subsection{ Feynman rules}

We use Feynman rules of reference~\cite{Denner},  in order  that the fermion
traces  in loops multiply spinors in the correct order. 
The Feynman rule for the Weinberg operator
of eqn (\ref{d5Weinberg}) can be  obtained reliably by
using LSZ reduction or Wick's theorem, which 
gives the  signs for fermion interchange.
The fermion  fields are expanded  as \cite{P+S}
$$\psi(x) =  \sum_s\int \frac{d^3 k}{(2\pi)^3} \frac{1}{\sqrt{2E}}
(a_k^s u_s(k) e^{-ik\cdot x}  +  b_k^{s\,\dagger} v_s(k) e^{+ik\cdot x})$$
so the amplitude $\mathcal{M}_{fi}$ is
\begin{equation}
  \label{eq:8}
  \begin{split}
\langle \ell_{\a j} {H}_I| i \frac{C_5^{ \s \r}}{2\Lambda} (\overline{\ell_{ \s n }}\varepsilon_{nN} {H^{N*})(\ell^c_{ \r m} \varepsilon_{mM} {H^{M*}}}) |\ell^c_{\b i} H_J^{*} \rangle
=&
 (-i) i \frac{C_5^{ \a \b }}{2\Lambda} \left( \overline{u}_{ \a j} P_R u_{ \b  i } + \overline{u}_{\beta i} P_R u_{\alpha  j} \right) (\varepsilon_{iI} \varepsilon_{jJ} + \varepsilon_{iJ} \varepsilon_{jI} ) 
 \\
= (-i) i \frac{C_5^{ \a \b } + C_5^{ \b \a }}{2 \Lambda} \overline{u}_{ \a j} P_R u_{ \b  i } (\varepsilon_{iI} \varepsilon_{jJ} + \varepsilon_{iJ} \varepsilon_{jI})
=& (-i) i \frac{C_5^{ \a \b }}{\Lambda} \overline{u}_{ \a j} P_R u_{ \b  i } (\varepsilon_{iI} \varepsilon_{jJ} + \varepsilon_{iJ} \varepsilon_{jI} ) \,,
  \end{split}
\end{equation}
where the SU(2) lepton indices are lower case, Higgs
indices are upper case, $\ell_{\a j}$ and $\ell^c_{\a j}$ represent a final state lepton and an initial state anti-lepton respectively.
The factor $i$ is the usual factor for Feynman rules and the factor $(-i)$ is due to the calculation of $\mathcal{M}_{fi}$. This expression agrees with Feynman rule of Reference~\cite{BGJ}.

A Feynman-rule to attach a $W$-boson to the $\ell^c$ line  also will be needed.
With the following identities \cite{Denner}
\bea
\ell^c =  C\overline{\ell}^T ~~~,~~~ C= i\g_0 \g_2~~~,~
C^{-1} = C^\dagger ~~~,~~~ C^\dagger \gamma^{\mu T} C = -\g^\mu
\\
~~ \overline{\ell^c} = [ C \g_0^T \ell^* ]^\dagger \g_0=
\ell^T \g_0 C^\dagger \g_0 = \ell^T C^\dagger C \g_0 C^\dagger \g_0 =
-\ell^T C^\dagger \g_0\g_0 = -\ell^T C^{-1}
\eea
one obtains (where the  (-1) is for   interchanging fermions)
\bea
{\Big [} \overline{\ell_i} \tau_{ij} \Wslash P_L \ell_j {\Big ]}^T
&=&(-1){\Big [} -\overline{\ell_j^c} C \tau^{a*}_{ji} P_L^T W_\mu^a\g_\mu^TC^{-1} \ell^c {\Big ]}\\
&=& \overline{\ell^c} \tau^{a*} W_\mu^aC \g_\mu^T C^{-1}P_R \ell^c
\\
&=&- \overline{\ell^c} \tau^{a*} W_\mu^a  \g_\mu P_R \ell^c
\eea
and recall that $\tau = \tau^\dagger$, so $\tau^* = \tau^T$. 



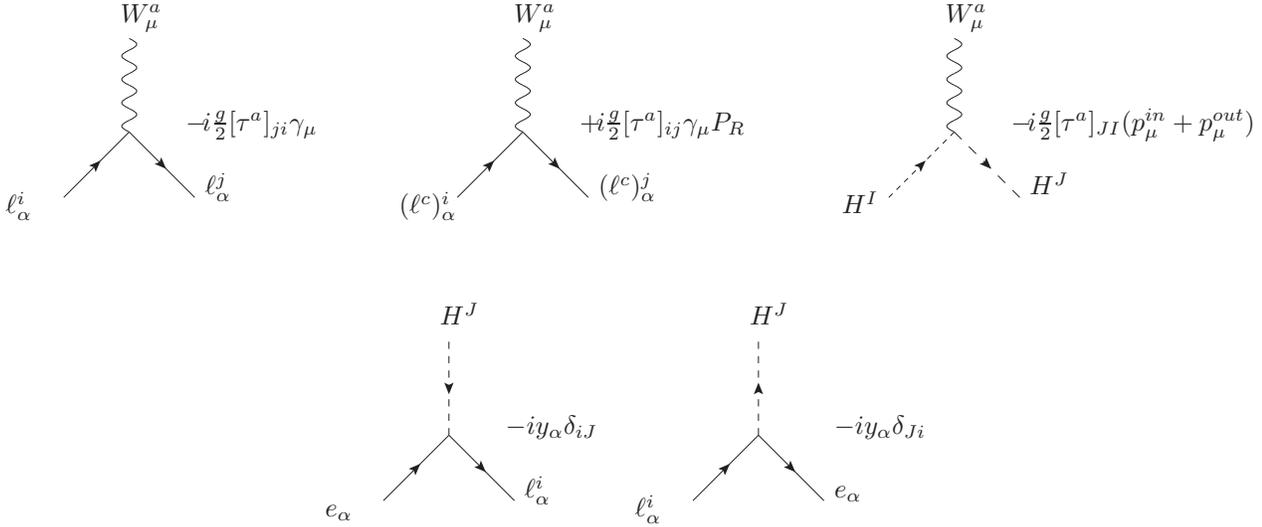
\begin{figure}[ht]
\unitlength.25mm
\SetScale{.7080}
 \begin{center}
$~$\hspace{-2cm}
\begin{picture}(180,200)(0,0)
 \Photon(60,210)(60,160){4}{4} 
\ArrowLine(25,125)(60,160)
\ArrowLine(60,160)(95,125)
\put(55,218){$W^a_\mu$}
\put(-6,116){$\ell^{i}_{ \alpha}$} 
\put(100,126){$\ell^{j}_{\alpha}$}
\put(90,160){$ {  - \! i  \frac{g}{2} [\tau^a]_{ji} \gamma_\mu } $}
\end{picture}
\hspace{.5cm}
\begin{picture}(180,200)(0,0)
 \Photon(60,210)(60,160){4}{4} 
\ArrowLine(25,125)(60,160)
\ArrowLine(60,160)(95,125)
\put(55,218){$W^a_\mu$}
\put(-6,116){$(\ell^c)^{i}_{ \alpha}$} 
\put(100,126){$(\ell^c)^{j}_{\alpha}$}
\put(90,160){$ {  + \! i  \frac{g}{2} [\tau^a]_{ij} \gamma_\mu P_R} $}
\end{picture}
\hspace{1cm}
\begin{picture}(180,220)
\Photon(60,210)(60,160){4}{4} 
\DashArrowLine(25,125)(60,160){3}
\DashArrowLine(60,160)(95,125){5}
\put(55,218){$W^a_\mu$}
\put(0,116){$H^I$} 
\put(100,126){$H^J$}
\put(90,160){$ {  -  \!   i  \frac{ g}{2} [\tau^a]_{JI} (p_\mu^{in} + p_\mu^{out})} $}
\end{picture}
\hspace{1cm}
\begin{picture}(160,160)(0,0)
 \DashArrowLine(60,210)(60,160){4}  
\ArrowLine(25,125)(60,160)
\ArrowLine(60,160)(95,125)
\put(55,218){$H^J$}
\put(-6,116){$e_ \alpha$} 
\put(100,126){$\ell^{i}_{ \alpha}$}
\put(90,160){$ { -  i  y_\a  \delta_{iJ}} $}
\end{picture}
\begin{picture}(180,160)(0,0)
\DashArrowLine(60,160)(60,210){4}
\ArrowLine(25,125)(60,160)
\ArrowLine(60,160)(95,125)
\put(55,218){$H^J$}
\put(-6,116){$\ell^{i}_{ \alpha}$} 
\put(100,126){$e_ \alpha$}
\put(100,160){$ {   -i  y_\a  \delta_{Ji}} $}
\end{picture}
\vspace{-2.5cm}
 \end{center}
\caption{ Feynman rules for dimension-four interactions \label{fig:1}}
\end{figure}

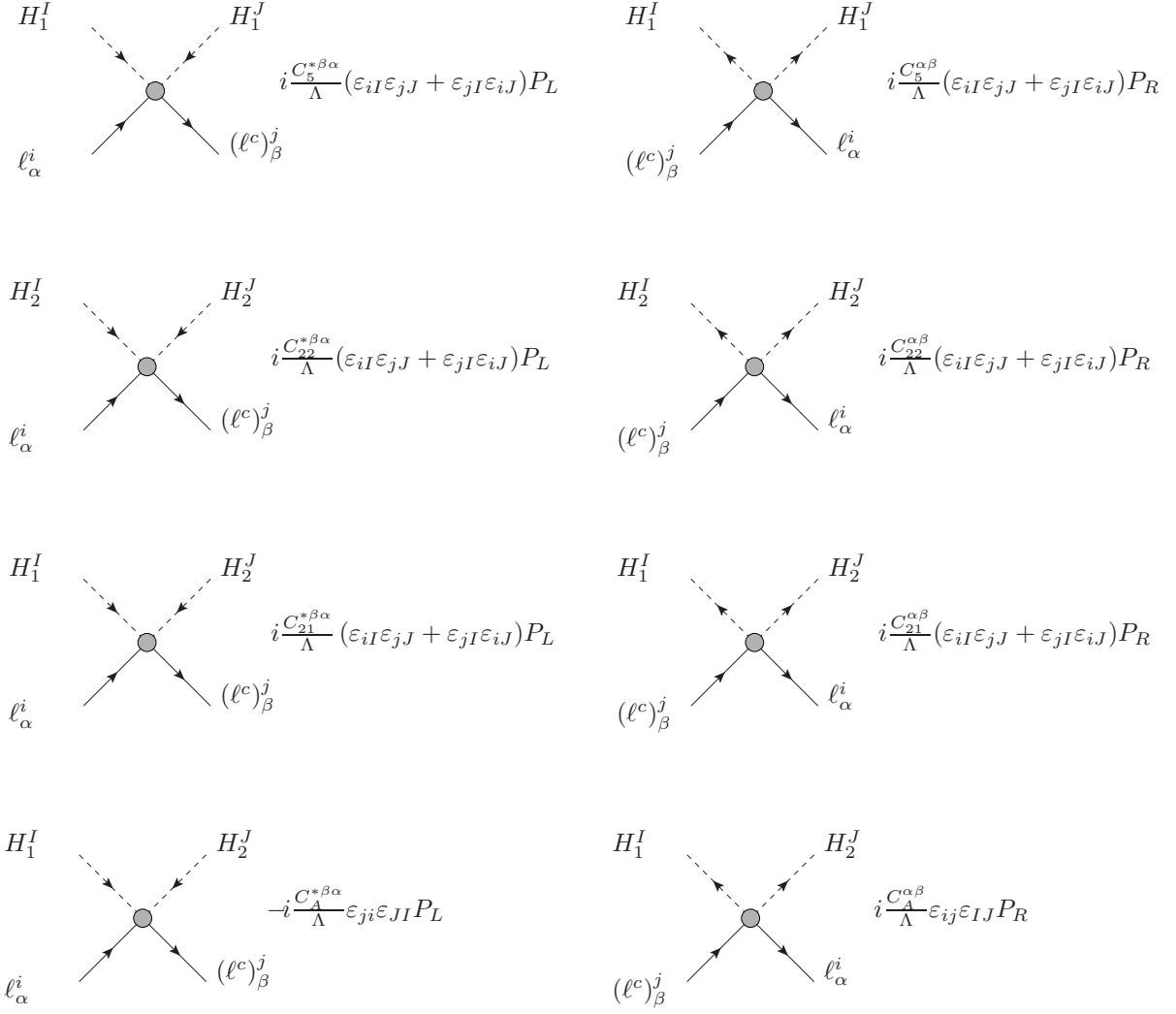
\begin{figure}[ht]
\unitlength.25mm
\SetScale{.7080}
 \begin{center}
$~$
\begin{picture}(244,150)(0,0)
\ArrowLine(25,125)(58,158)
\ArrowLine(62,158)(95,125)
\DashArrowLine(25,195)(58,162){3}
\DashArrowLine(95,195)(62,162){3}
\GCirc(60,160){5}{.7}
\put(-16,116){$\ell^{i}_ {\alpha}$} 
\put(100,126){$(\ell^{c})^{j}_{\beta}$}
\put(-16,196){$H_{1}^{I}$} 
\put(100,196){$H_{1 }^{J}$}
\put(130,160){$ {   \! i  \frac{C_5^{*\b\a}}{\Lambda}
(\varepsilon_{iI } \varepsilon_{jJ} +\varepsilon_{jI } \varepsilon_{iJ}) P_{L} } $}
\end{picture}
\hspace{2cm}
\begin{picture}(244,150)(0,0)
 \ArrowLine(25,125)(58,158)
\ArrowLine(62,158)(95,125) 
\DashArrowLine(58,162)(25,195){3}
\DashArrowLine(62,162)(95,195){3}
\GCirc(60,160){5}{.7}
\put(-16,116){$(\ell^{c})^{j}_{\beta}$}
\put(100,126){$\ell^{i}_{\alpha}$} 
\put(-16,196){$H_{1}^{I}$} 
\put(100,196){$H_{1}^{J}$}
\put(130,160){$ {   \! i  \frac{C_5^{\a\b}}{\Lambda} (\varepsilon_{iI } \varepsilon_{jJ}+\varepsilon_{jI } \varepsilon_{iJ}) } P_{R} $}
\end{picture}
\begin{picture}(244,150)(0,0)
  \ArrowLine(25,125)(58,158)
\ArrowLine(62,158)(95,125)
\DashArrowLine(25,195)(58,162){3}
\DashArrowLine(95,195)(62,162){3}
\GCirc(60,160){5}{.7}
\put(-16,116){$\ell^{i}_{\alpha}$} 
\put(100,126){$(\ell^{c})^{j}_{\beta}$}
\put(-16,196){$H_{2}^{I}$} 
\put(100,196){$H_{2}^{J}$}
\put(130,160){$ {   \! i  \frac{C_{22}^{*\b\a}}{\Lambda}
(\varepsilon_{iI } \varepsilon_{jJ} +\varepsilon_{jI } \varepsilon_{iJ}) } P_{L} $}
\end{picture}
\hspace{2cm}
\begin{picture}(244,150)(0,0)
  \ArrowLine(25,125)(58,158)
\ArrowLine(62,158)(95,125)
\DashArrowLine(58,162)(25,195){3}
\DashArrowLine(62,162)(95,195){3}
\GCirc(60,160){5}{.7}
\put(-16,116){$(\ell^{c})^{j}_{\beta}$}
\put(100,126){$\ell^{i}_{\alpha}$} 
\put(-16,196){$H_{2}^{I}$} 
\put(100,196){$H_{2}^{J}$}
\put(130,160){$ {   \! i  \frac{C_{22}^{\a\b}}{\Lambda} (\varepsilon_{iI } \varepsilon_{jJ}+\varepsilon_{jI } \varepsilon_{iJ}) } P_{R} $}
\end{picture}
%
%
\begin{picture}(244,150)(0,0)
   \ArrowLine(25,125)(58,158)
\ArrowLine(62,158)(95,125)
\DashArrowLine(25,195)(58,162){3}
\DashArrowLine(95,195)(62,162){3}
\GCirc(60,160){5}{.7}
\put(-16,116){$\ell^{i}_{\alpha}$} 
\put(100,126){$(\ell^{c})^{j}_{\beta}$}
\put(-16,196){$H_{1}^{I}$} 
\put(100,196){$H_{2}^{J}$}
\put(130,160){$ {   \! i \frac{C_{21}^{*\b\a}}{\Lambda} \left(
\varepsilon_{iI } \varepsilon_{jJ} + \varepsilon_{jI } \varepsilon_{iJ} \right) } P_{L}$}
\end{picture}
\hspace{2cm}
\begin{picture}(244,150)(0,0)
    \ArrowLine(25,125)(58,158)
\ArrowLine(62,158)(95,125)
\DashArrowLine(58,162)(25,195){3}
\DashArrowLine(62,162)(95,195){3}
\GCirc(60,160){5}{.7}
\put(-16,116) {$(\ell^{c})^{j}_{\beta}$}
\put(100,126){$\ell^{i}_{\alpha}$}
\put(-16,196){$H_{1}^{I}$} 
\put(100,196){$H_{2}^{J}$}
\put(130,160){$ {   \! i \frac{C_{21}^{\a\b}}{\Lambda} ( \varepsilon_{iI } \varepsilon_{jJ}+     \varepsilon_{jI } \varepsilon_{iJ} }) P_{R} $}
\end{picture}
%
%
\begin{picture}(244,150)(0,0)
      \ArrowLine(25,125)(58,158)
\ArrowLine(62,158)(95,125)
\DashArrowLine(25,195)(58,162){3}
\DashArrowLine(95,195)(62,162){3}
\GCirc(60,160){5}{.7}
\put(-16,116){$\ell^{i}_{\alpha}$} 
\put(100,126){$(\ell^{c})^{j}_{\beta}$}
\put(-16,196){$H_{1}^{I}$} 
\put(100,196){$H_{2}^{J}$}
\put(130,160){$    \!
- \! i  \frac{C_A^{*\b\a}}{\Lambda} \varepsilon_{ji } \varepsilon_{JI}  P_{L}  $}
\end{picture}
\hspace{2cm}
\begin{picture}(244,150)(0,0)
    \ArrowLine(25,125)(58,158)
\ArrowLine(62,158)(95,125)
\DashArrowLine(58,162)(25,195){3}
\DashArrowLine(62,162)(95,195){3}
\GCirc(60,160){5}{.7}
\put(-16,116) {$(\ell^{c})^{j}_{\beta}$}
\put(100,126){$\ell^{i}_{\alpha}$}
\put(-16,196){$H_{1}^{I}$} 
\put(100,196){$H_{2}^{J}$}
\put(130,160){$ {  \! i  \frac{C_A^{\a\b}}{\Lambda} \varepsilon_{ij } \varepsilon_{IJ} }  P_{R}$}
\end{picture}
\vspace{-1.5cm}
 \end{center}
\caption{   Feynman rules for dimension-five interactions,  in the
  single and two Higgs Doublet Models. $H_1$ is the SM Higgs.
  $H_2$ is the second Higgs of the 2HDM, with the same
  hypercharge as the SM Higgs, opposite to the lepton doublet.
  \label{fig:Fr2HDM}}
\end{figure}

\begin{figure}[h!]
\unitlength.25mm
\SetScale{.7080}
 \begin{center}
$~$
\begin{picture}(244,150)(0,0)
  \ArrowLine(25,125)(58,158)
\ArrowLine(62,158)(95,125)
\DashArrowLine(25,195)(58,162){3}
\DashArrowLine(62,162)(95,195){3}
\GCirc(60,160){5}{.7}
\put(-16,116){$\ell^{n}_{\alpha}(p_{i})$} 
\put(100,126){$\ell^{i}_{\beta}(p_{f})$}
\put(-20,196){$H_{1}^{M}(q_{i})$} 
\put(100,196){$H_{1}^{J}(q_{f})$}
\put(130,160){$ \! i  \frac{C_{H\ell(1)}^{*\b\a}}{\Lambda^{2}} 
\big( \slashed{q}_{i}+\slashed{q}_{f}\big ) P_{L}  \delta_{in}\delta_{JM}$}
\end{picture}
\hspace{2cm}
\begin{picture}(244,150)(0,0)
\ArrowLine(25,125)(58,158)
\ArrowLine(62,158)(95,125)
\DashArrowLine(25,195)(58,162){3}
\DashArrowLine(62,162)(95,195){3}
\Photon(58,158)(58, 210){3}{6}
\GCirc(60,160){5}{.7}
\put(0,116){$\ell^{n}_{\alpha}$} 
\put(100,126){$\ell^{i}_{\beta}$}
\put(-10,196){$H_{1}^{M}$} 
\put(100,196){$H_{1}^{J}$}
\put(50, 220){$B_{\mu}$}
\put(130,160){$ -2ig_{1} y_{H} \frac{C_{H\ell(1)}^{\beta \alpha}}{\Lambda^{2}} \gamma^{\mu}P_{L}\delta_{in}\delta_{JM}$}
\end{picture}
%
%
\begin{picture}(244,150)(0,0)
\ArrowLine(25,125)(58,158)
\ArrowLine(62,158)(95,125)
\DashArrowLine(25,195)(58,162){3}
\DashArrowLine(62,162)(95,195){3}
\Photon(58,158)(58, 210){3}{6}
\GCirc(60,160){5}{.7}
\put(0,116){$\ell^{n}_ {\alpha}$} 
\put(100,126){$\ell^{i}_{\beta}$}
\put(-10,196){$H_{1}^{I}$} 
\put(100,196){$H_{1 }^{J}$}
\put(50, 220){$W_{\mu}^{a}$}
\put(130,160){$  -2ig_{2} \frac{C_{H\ell(1)}^{*\b\a}}{\Lambda^{2}} \gamma^{\mu}P_{L} \delta_{in}S^{a}_{JM}$}
\end{picture}
\hspace{2cm}
\begin{picture}(244,150)(0,0)
  \ArrowLine(25,125)(58,158)
\ArrowLine(62,158)(95,125)
\DashArrowLine(25,195)(58,162){3}
\DashArrowLine(62,162)(95,195){3}
\GCirc(60,160){5}{.7}
\put(-16,116){$\ell^{n}_{\alpha}(p_{i})$} 
\put(100,126){$\ell^{i}_{\beta}(p_{f})$}
\put(-20,196){$H_{1}^{M}(q_{i})$} 
\put(100,196){$H_{1}^{J}(q_{f})$}
\put(130,160){$ 4 i  \frac{C_{H\ell(3)}^{*\b\a}}{\Lambda^{2}} 
\big( \slashed{q}_{i}+\slashed{q}_{f}  \big ) P_{L} S^{a}_{in}S^{a}_{JM}$}
\end{picture}
\begin{picture}(244,150)(0,0)
\ArrowLine(25,125)(58,158)
\ArrowLine(62,158)(95,125)
\DashArrowLine(25,195)(58,162){3}
\DashArrowLine(62,162)(95,195){3}
\Photon(58,158)(58, 210){3}{6}
\GCirc(60,160){5}{.7}
\put(0,116){$\ell^{n}_{\alpha}$} 
\put(100,126){$\ell^{i}_{\beta}$}
\put(-10,196){$H_{1}^{M}$} 
\put(100,196){$H_{1}^{J}$}
\put(50, 220){$B_{\mu}$}
\put(130,160){$ -8ig_{1} y_{H} \frac{C_{H\ell(3)}^{\beta \alpha}}{\Lambda^{2}} \gamma^{\mu}P_{L}S^{a}_{in}S^{a}_{JM}$}
\end{picture}
\hspace{2cm}
\begin{picture}(244,150)(0,0)
\ArrowLine(25,125)(58,158)
\ArrowLine(62,158)(95,125)
\DashArrowLine(25,195)(58,162){3}
\DashArrowLine(62,162)(95,195){3}
\Photon(58,158)(58, 210){3}{6}
\GCirc(60,160){5}{.7}
\put(0,116){$\ell^{n}_ {\alpha}$} 
\put(100,126){$\ell^{i}_{\beta}$}
\put(-10,196){$H_{1}^{I}$} 
\put(100,196){$H_{1 }^{J}$}
\put(50, 220){$W_{\mu}^{a}$}
\put(130,170){$  -4ig_{2} \frac{C_{H\ell(3)}^{*\b\a}}{\Lambda^{2}} \gamma^{\mu}P_{L}  \big(S^{b}_{JK}S^{a}_{KM}$}
\put(150, 150){$+S^{a}_{JK}S^{b}_{KM} \big)S^{b}_{in}$}
\end{picture}
%

\begin{picture}(244,150)(0,0)
\DashArrowLine(25,125)(58,158){3}
\DashArrowLine(95,125)(62,158){3}
\ArrowLine(25,195)(58,162)
\ArrowLine(62,162)(95,195)
\DashArrowLine(60,158)(60, 210){3}
\GCirc(60,160){5}{.7}
\put(-10,116){$H_{1}^{K}$} 
\put(100,126){$H_{1}^{J}$}
\put(0,196){$e_{\alpha}$} 
\put(100,196){$\ell_{\beta}^{n}$}
\put(50, 220){$H_{1}^{I}$}
\put(130,160){$i\frac{C_{eH}^{\beta \alpha}}{\Lambda^{2}} \big ( \delta_{JK}\delta_{In} + \delta_{IJ}\delta_{Kn} \big )P_{R}$}
\end{picture}
\hspace{2cm}
\begin{picture}(244,150)(0,0)
  \ArrowLine(25,125)(58,158)
\ArrowLine(95,125)(62,158)
\ArrowLine(58,162)(25,195)
\ArrowLine(62,162)(95,195)
\GCirc(60,160){5}{.7}
\put(0,116){$\ell^{j}_{\alpha}$} 
\put(100,126){$\ell^{l}_{\rho}$}
\put(0,196){$\ell^{i}_{\sigma}$} 
\put(100,196){$\ell^{k}_{\beta}$}
\put(130,160){$ \!2 i  \frac{C_{\ell \ell}^{*\beta\sigma\rho \alpha}}{\Lambda^{2}} 
( \gamma_{\mu}P_{L}\otimes\gamma^{\mu}P_{L})\delta_{ik}\delta_{jl}$}
\end{picture}
\vspace{-3cm}
 \end{center}
\caption{   Feynman rules for dimension-six operators of the SMEFT using the ``Warsaw''-basis \cite{polonais}. $H_1$ is the SM Higgs.
  \label{fig:Fr2HDM}}
\end{figure}
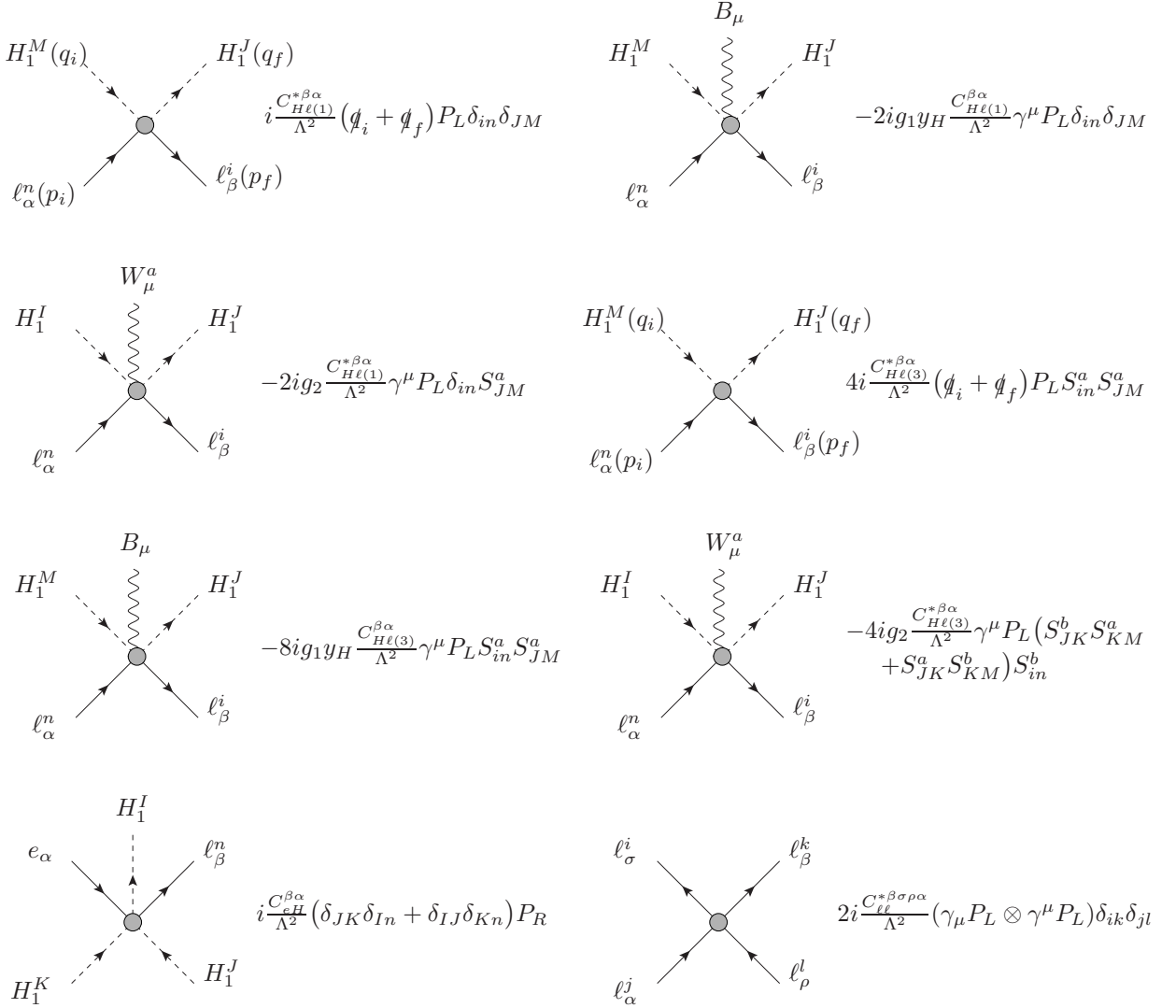
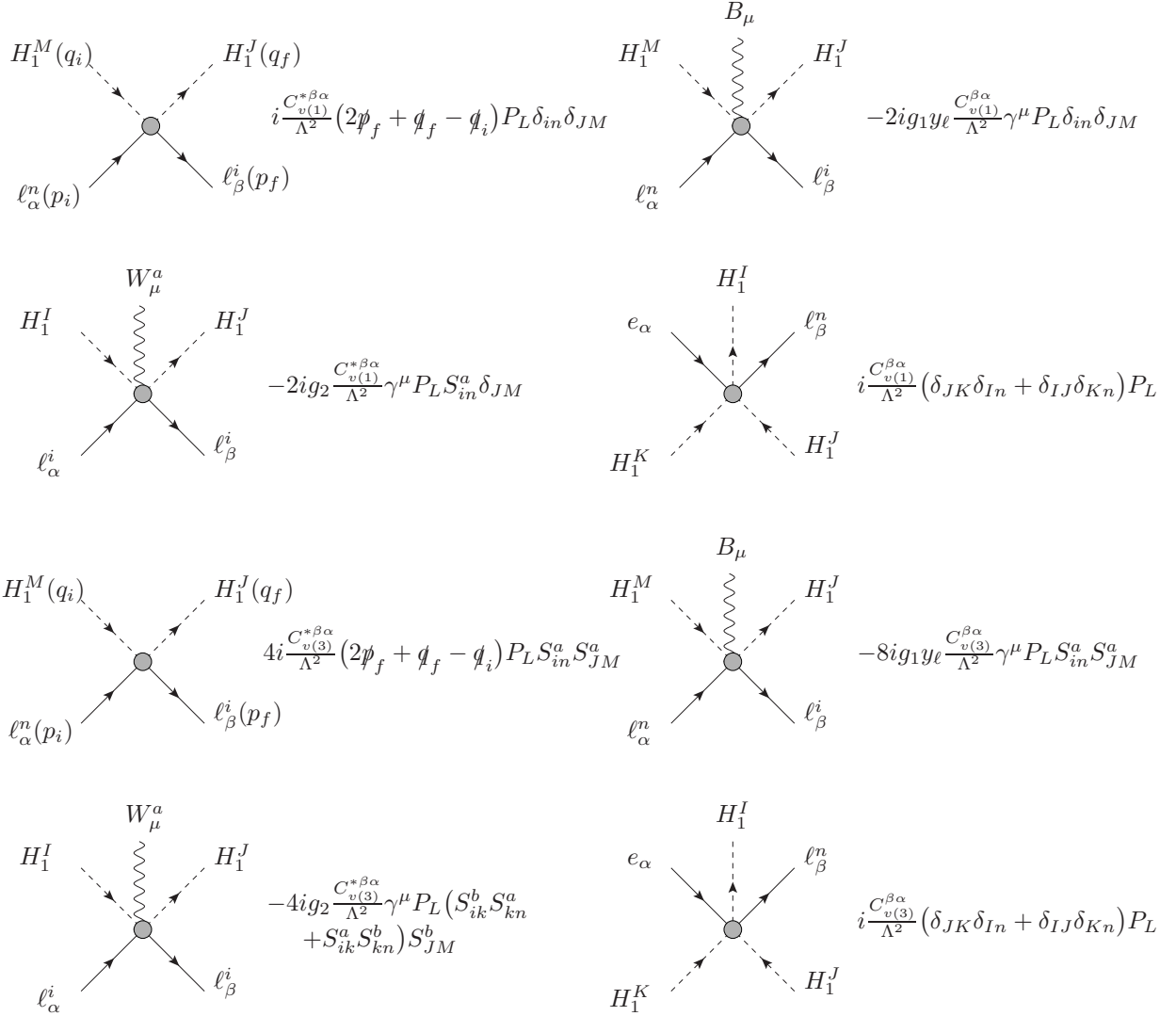
\begin{figure}[h!]
\unitlength.25mm
\SetScale{.7080}
 \begin{center}
$~$
\begin{picture}(244,150)(0,0)
  \ArrowLine(25,125)(58,158)
\ArrowLine(62,158)(95,125)
\DashArrowLine(25,195)(58,162){3}
\DashArrowLine(62,162)(95,195){3}
\GCirc(60,160){5}{.7}
\put(-16,116){$\ell^{n}_{\alpha}(p_{i})$} 
\put(100,126){$\ell^{i}_{\beta}(p_{f})$}
\put(-20,196){$H_{1}^{M}(q_{i})$} 
\put(100,196){$H_{1}^{J}(q_{f})$}
\put(130,160){$ \! i  \frac{C_{v(1)}^{*\b\a}}{\Lambda^{2}} 
\big( 2\slashed{p}_{f}+\slashed{q}_{f} -\slashed{q}_{i}  \big ) P_{L} \delta_{in}\delta_{JM}$}
\end{picture}
\hspace{2cm}
\begin{picture}(244,150)(0,0)
\ArrowLine(25,125)(58,158)
\ArrowLine(62,158)(95,125)
\DashArrowLine(25,195)(58,162){3}
\DashArrowLine(62,162)(95,195){3}
\Photon(58,158)(58, 210){3}{6}
\GCirc(60,160){5}{.7}
\put(0,116){$\ell^{n}_{\alpha}$} 
\put(100,126){$\ell^{i}_{\beta}$}
\put(-10,196){$H_{1}^{M}$} 
\put(100,196){$H_{1}^{J}$}
\put(50, 220){$B_{\mu}$}
\put(130,160){$ -2ig_{1} y_{\ell} \frac{C_{v(1)}^{\beta \alpha}}{\Lambda^{2}} \gamma^{\mu}P_{L}\delta_{in}\delta_{JM}$}
\end{picture}
\begin{picture}(244,150)(0,0)
\ArrowLine(25,125)(58,158)
\ArrowLine(62,158)(95,125)
\DashArrowLine(25,195)(58,162){3}
\DashArrowLine(62,162)(95,195){3}
\Photon(58,158)(58, 210){3}{6}
\GCirc(60,160){5}{.7}
\put(0,116){$\ell^{i}_ {\alpha}$} 
\put(100,126){$\ell^{i}_{\beta}$}
\put(-10,196){$H_{1}^{I}$} 
\put(100,196){$H_{1 }^{J}$}
\put(50, 220){$W_{\mu}^{a}$}
\put(130,160){$  -2ig_{2} \frac{C_{v(1)}^{*\b\a}}{\Lambda^{2}} \gamma^{\mu}P_{L} S^{a}_{in}\delta_{JM}$}
\end{picture}
\hspace{2cm}
\begin{picture}(244,150)(0,0)
\DashArrowLine(25,125)(58,158){3}
\DashArrowLine(95,125)(62,158){3}
\ArrowLine(25,195)(58,162)
\ArrowLine(62,162)(95,195)
\DashArrowLine(60,158)(60, 210){3}
\GCirc(60,160){5}{.7}
\put(-10,116){$H_{1}^{K}$} 
\put(100,126){$H_{1}^{J}$}
\put(0,196){$e_{\alpha}$} 
\put(100,196){$\ell_{\beta}^{n}$}
\put(50, 220){$H_{1}^{I}$}
\put(130,160){$i\frac{C_{v(1)}^{\beta \alpha}}{\Lambda^{2}} \big ( \delta_{JK}\delta_{In} + \delta_{IJ}\delta_{Kn} \big )P_{L}$}
\end{picture}
\begin{picture}(244,150)(0,0)
  \ArrowLine(25,125)(58,158)
\ArrowLine(62,158)(95,125)
\DashArrowLine(25,195)(58,162){3}
\DashArrowLine(62,162)(95,195){3}
\GCirc(60,160){5}{.7}
\put(-16,116){$\ell^{n}_{\alpha}(p_{i})$} 
\put(100,126){$\ell^{i}_{\beta}(p_{f})$}
\put(-20,196){$H_{1}^{M}(q_{i})$} 
\put(100,196){$H_{1}^{J}(q_{f})$}
\put(130,160){$ \!4 i  \frac{C_{v(3)}^{*\b\a}}{\Lambda^{2}} 
\big( 2\slashed{p}_{f}+\slashed{q}_{f} -\slashed{q}_{i}  \big )P_{L} S^{a}_{in}S^{a}_{JM}$}
\end{picture}
\hspace{2cm}
\begin{picture}(244,150)(0,0)
\ArrowLine(25,125)(58,158)
\ArrowLine(62,158)(95,125)
\DashArrowLine(25,195)(58,162){3}
\DashArrowLine(62,162)(95,195){3}
\Photon(58,158)(58, 210){3}{6}
\GCirc(60,160){5}{.7}
\put(0,116){$\ell^{n}_{\alpha}$} 
\put(100,126){$\ell^{i}_{\beta}$}
\put(-10,196){$H_{1}^{M}$} 
\put(100,196){$H_{1}^{J}$}
\put(50, 220){$B_{\mu}$}
\put(130,160){$ -8ig_{1} y_{\ell} \frac{C_{v(3)}^{\beta \alpha}}{\Lambda^{2}} \gamma^{\mu}P_{L}S^{a}_{in}S^{a}_{JM}$}
\end{picture}
\begin{picture}(244,150)(0,0)
\ArrowLine(25,125)(58,158)
\ArrowLine(62,158)(95,125)
\DashArrowLine(25,195)(58,162){3}
\DashArrowLine(62,162)(95,195){3}
\Photon(58,158)(58, 210){3}{6}
\GCirc(60,160){5}{.7}
\put(0,116){$\ell^{i}_ {\alpha}$} 
\put(100,126){$\ell^{i}_{\beta}$}
\put(-10,196){$H_{1}^{I}$} 
\put(100,196){$H_{1 }^{J}$}
\put(50, 220){$W_{\mu}^{a}$}
\put(130,170){$  -4ig_{2} \frac{C_{v(3)}^{*\b\a}}{\Lambda^{2}} \gamma^{\mu}P_{L} \big( S^{b}_{ik}S^{a}_{kn}$}
\put(150,150){$+S^{a}_{ik}S^{b}_{kn} \big) S^{b}_{JM}$}
\end{picture}
\hspace{2cm}
\begin{picture}(244,150)(0,0)
\DashArrowLine(25,125)(58,158){3}
\DashArrowLine(95,125)(62,158){3}
\ArrowLine(25,195)(58,162)
\ArrowLine(62,162)(95,195)
\DashArrowLine(60,158)(60, 210){3}
\GCirc(60,160){5}{.7}
\put(-10,116){$H_{1}^{K}$} 
\put(100,126){$H_{1}^{J}$}
\put(0,196){$e_{\alpha}$} 
\put(100,196){$\ell_{\beta}^{n}$}
\put(50, 220){$H_{1}^{I}$}
\put(130,160){$i\frac{C_{v(3)}^{\beta \alpha}}{\Lambda^{2}} \big ( \delta_{JK}\delta_{In} + \delta_{IJ}\delta_{Kn} \big ) P_{L}$}
\end{picture}

\vspace{-3cm}
 \end{center}
\caption{   Feynman rules for dimension-six operators that are vanishing by the equations of motion, in the
  single Higgs Doublet Model (SMEFT). $H_1$ is the SM Higgs.
  \label{fig:Fr2HDM}}
\end{figure}

\newpage


Note that we have chosen a convention for our Feynman rules to eliminate any dependence on the momentum of the incoming lepton, $p_{i}$, since all momenta are not independent. 

\subsection{Identities}

The following identities  are useful:
\bea
2 \varepsilon_{iI} \varepsilon_{jJ} &= &\d_{ij} \d_{IJ} -\tau_{ij}^a \tau_{a,IJ} ~~{\rm Fierz}\label{Fiertz}\\
\frac{1}{4}\tau_{ij}^a \tau_{a,kl} &=& \frac{1}{2}\d_{il} \d_{kj} - \frac{1}{4}\d_{ij} \d_{kl}  ~~~~{\rm SU(N)}
\label{SU2id}
\\
\varepsilon_{ab}\varepsilon_{cd} + \varepsilon_{bc}\varepsilon_{ad} + \varepsilon_{ac}\varepsilon_{bd}  &=& 0 
\label{anotherid}
\\
\varepsilon_{iJ}\varepsilon_{kJ}&=&\delta_{ik}
\\
\varepsilon_{ij}S^{a}_{jk}\varepsilon_{kl} &=& S^{a}_{li} 
\\
0 &=&\delta_{i j}S^{a}_{kl} -\delta_{jl}S^{a}_{ki} +\delta_{kl}S^{a}_{ji}-\delta_{ik}S^{a}_{jl} \\
\varepsilon_{ij}\varepsilon_{kl} &=& \delta_{ik}\delta_{jl}-\delta_{il}\delta_{jk},
\eea
where 
$$ \varepsilon  =  \left[
\begin{array}{cc}
0&1\\
-1 &0
\end{array}
\right] ~~,~~\vec{\tau} = \left(\left[
\begin{array}{cc}
0&1\\
1 &0
\end{array}
\right],  \left[
\begin{array}{cc}
0&-i\\
i &0
\end{array}
\right] ,  \left[
\begin{array}{cc}
1&0\\
0 &-1
\end{array}
\right]
\right)
$$
and the SU(2) generators are $S^a =\tau^a/2$. 

\section{The Loop Calculation}
\label{app:renormalisation}
\subsection{Flavour dependence}

We allow for multiple operators at both
  dimension-five and -six, 
  and denote a particular Wilson coefficient by $C_X^{\zeta}$, where $X$ and $\zeta$
  are the operator and flavour labels respectively.
  Then the bare Wilson coefficients of the dimension-six standard
  model effective theory Lagrangian can be written as
\begin{equation}
  \sum_{\zeta, X} C^{\zeta}_{X,\mathrm{bare}} Q^{\zeta}_{X,\mathrm{bare}} = \mu^{2\epsilon} \sum_{\theta, Y}
  \left( \sum_{\zeta,X} C_X^{\zeta} Z_{XY}^{\zeta\theta} + 
    \sum_{\zeta, \eta} C_5^{\zeta} \big[C_5^{\eta}\big]^{\dagger} Z_{5\bar{5},Y}^{\zeta\eta\theta}\right) Q^{\theta}_{Y,\mathrm{bare}} \,,
\end{equation}
where $\zeta$, $\eta$ and $\theta$ represent generation indices of an operator, and  the renormalisation constants $Z_{XY}^{\zeta\theta}$ encode the mixing of dimension-six Wilson coefficients amongst themselves, which can be extracted from the anomalous dimensions of reference~\cite{JMT}. 
In the standard model, the mixing of two dimension-five Wilson coefficients into a dimension-six coefficient is given by $Z_{5\bar{5},Y}^{\zeta\eta\theta}$. They are induced by the double-insertions of dimension-five operators, as shown in figure \ref{fig:2}.
In the case of a 2HDM effective field theory we extend the summation of the dimension-five flavour indices to a sum over all dimension-five operators and their respective flavour components.


The renormalisation constants can be expanded in the number of loops and powers of
epsilon. At one-loop in the $\overline{\mathrm{MS}}$ scheme the counterterms of the physical and EOM-vanishing operators are pure $1/\epsilon$ poles, and the renormalisation of evanescent operators does not play a role. Hence we can expand
\begin{equation}
\label{eq:zij-expansion}
Z_{5\bar{5},j}^{\zeta\eta\theta} = \frac{1}{16 \pi^2}\frac{1}{\epsilon} \delta Z_{5\bar{5},j}^{\zeta\eta\theta}
\end{equation}
and write the generation summation in the case of an operator involving four fermions explicitly as: 
\begin{equation}
\label{eq:flavour-summation-tensor}
C_5^{\zeta} C_5^{\eta\dagger} \delta Z_{5\bar{5},X}^{\zeta \eta \theta} Q_X^\theta = 
C_5^{\alpha\beta} C_5^{\delta\gamma *} 
\delta Z_{5\bar{5},X}^{\alpha\beta\, \gamma\delta, \rho\sigma\tau\upsilon} 
Q_X^{\rho \sigma \tau \upsilon} .
\end{equation}
The sum over generation indices reduces trivially for operators that
involve less fermions.
The corresponding renormalisation equation ensures that the pole of
the one-loop off-shell matrix element of an insertion of two
dimension-five operators is cancelled by its counterterm. Factoring out the common overall factor $C_5^{\alpha\beta} C_5^{\delta \gamma*}$ we write:
\begin{equation}
  \label{eq:3}
  \langle f | Q_5^{\alpha\beta} (Q_5^{\gamma\delta})^{\dagger} | i \rangle|_{1/\epsilon}^{(1)} + 
  \left( \delta Z_{5\bar{5},X}^{\alpha\beta\, \gamma\delta, \rho\sigma\tau\upsilon} \langle f | Q_{X}^{\rho\sigma\tau\upsilon} | i \rangle + \mathrm{h.c.} \right) = 0 \,,
\end{equation}
where $|_{1/\epsilon}^{(1)}$ denotes the $1/\epsilon$ pole of a one-loop diagram and $\langle f|$ and $|i \rangle$ are arbitrary off-shell final and initial states.

In calculations of the loop diagrams the following generation structures arose:
\begin{equation}
  \label{eq:2}
  \begin{split}
      T_1^{\alpha\beta\, \gamma\delta, \rho\sigma} &= 
  \frac{1}{2} \left(
  \delta _{\delta \sigma } \delta _{\alpha  \rho } \delta_{\beta  \gamma }
 +\delta _{\alpha  \delta} \delta _{\beta  \rho } \delta_{\gamma  \sigma }
\right) \,, \\
  T_{1A}^{\alpha\beta\, \gamma\delta, \rho\sigma} &= 
  \frac{1}{2} \left(
  \delta _{\delta \sigma } \delta _{\alpha  \rho } \delta_{\beta  \gamma }
 -\delta _{\alpha  \delta} \delta _{\beta  \rho } \delta_{\gamma  \sigma }
\right) \,, \\
  T_2^{\alpha\beta\, \gamma\delta, \rho\sigma} &= 
  \frac{1}{2} \left(
    \delta _{\alpha  \rho } \delta _{\beta  \gamma } Y_{\delta \sigma }
   +\delta _{\alpha  \delta } \delta _{\beta  \rho } Y_{\gamma \sigma}
  \right) \,, \\
  T_{2A}^{\alpha\beta\, \gamma\delta, \rho\sigma} &= 
  \frac{1}{2} \left(
  \delta _{\alpha  \rho } \delta _{\beta  \gamma }
        Y^{(2)}_{\delta \sigma }
       -\delta _{\alpha  \delta } \delta _{\beta\rho }
        Y^{(2)}_{\gamma \sigma }
   \right) \,, \\
  T_3^{\alpha\beta\, \gamma\delta, \rho\sigma\tau\upsilon} &= 
\frac{1}{4} \left(\delta _{\delta  \sigma } \delta _{\gamma  \upsilon }+\delta _{\delta  \upsilon } \delta _{\gamma  \sigma
   }\right) \left(\delta _{\alpha  \tau } \delta _{\beta  \rho }+\delta _{\alpha  \rho } \delta _{\beta  \tau }\right) \,, \\
  T_{3A}^{\alpha\beta\, \gamma\delta, \rho\sigma\tau\upsilon} &= 
-\frac{1}{4} \left(
   \delta _{\delta  \sigma } \delta _{\gamma  \upsilon }-\delta _{\delta \upsilon } \delta _{\gamma  \sigma }\right) 
   \left(\delta _{\alpha  \tau } \delta _{\beta  \rho }-\delta _{\alpha  \rho } \delta _{\beta  \tau }\right)  \,. \\
  \end{split}
\end{equation}
These were matched onto the generation structures of the dimension-six operators (the matching is more subtle for the four-lepton operator $\mathcal{O}_{\ell \ell}^{\alpha \beta \gamma \delta}$, where the matching is done via a Fierz-evanescent dimension-six operator $\mathcal{O}_{e}^{\alpha \beta \gamma \delta}$), and the generation structure therefore extracted from the renormalisation constants, which can then be written as a generation structure multiplied by a numerical factor.  

At one-loop we find the following non-vanishing mixing into the physical dimension-six operators
\begin{align}
\label{eq:ztensor-phys}
\delta Z_{5\overline{5},H\ell(1)}^{\alpha\beta\, \gamma\delta, \rho\sigma}
 &= - \frac{3}{4} T_1^{\alpha\beta\, \gamma\delta, \rho\sigma}  \,,
 & 
\delta Z_{21\overline{21},H\ell(1)}^{\alpha\beta\, \gamma\delta, \rho\sigma}
 &= - \frac{3}{4} T_1^{\alpha\beta\, \gamma\delta, \rho\sigma} \,,
\nonumber \\
\delta Z_{A\overline{A},H\ell(1)}^{\alpha\beta\, \gamma\delta, \rho\sigma}
 &= - \frac{1}{4} T_1^{\alpha\beta\, \gamma\delta, \rho\sigma} \,,
&
\delta Z_{5\overline{5},H\ell(3)}^{\alpha\beta\, \gamma\delta, \rho\sigma}
 &= \frac{1}{2} T_1^{\alpha\beta\, \gamma\delta, \rho\sigma} \,,
\nonumber \\
\delta Z_{21\overline{21},H\ell(3)}^{\alpha\beta\, \gamma\delta, \rho\sigma}
 &= \frac{1}{2} T_1^{\alpha\beta\, \gamma\delta, \rho\sigma}  \,,
&
\delta Z_{A\overline{21},H\ell(3)}^{\alpha\beta\, \gamma\delta, \rho\sigma}
 &= \frac{1}{4} T_{1A}^{\alpha\beta\, \gamma\delta, \rho\sigma}  \,,
\nonumber\\
\delta Z_{21\overline{A},H\ell(3)}^{\alpha\beta\, \gamma\delta, \rho\sigma}
 &= \frac{1}{4} T_{1A}^{\alpha\beta\, \gamma\delta, \rho\sigma}  \,,
&
\delta Z_{5\overline{5},eH}^{\alpha\beta\, \gamma\delta, \rho\sigma}
&= \frac{3}{4} T_2^{\alpha\beta\, \gamma\delta, \rho\sigma} \,,
\nonumber\\
\delta Z_{21\overline{5},eH}^{\alpha\beta\, \gamma\delta, \rho\sigma}
&= T_2^{\alpha\beta\, \gamma\delta, \rho\sigma} \,,
&
\delta Z_{A\overline{5},eH}^{\alpha\beta\, \gamma\delta, \rho\sigma}
&= - T_{2A}^{\alpha\beta\, \gamma\delta, \rho\sigma} \,,
\\
\delta Z_{A\overline{A},eH}^{\alpha\beta\, \gamma\delta, \rho\sigma}
&= - \frac{1}{4} T_2^{\alpha\beta\, \gamma\delta, \rho\sigma} \,,
&
\delta Z_{A\overline{21},eH}^{\alpha\beta\, \gamma\delta, \rho\sigma}
&= \frac{1}{4} T_{2A}^{\alpha\beta\, \gamma\delta, \rho\sigma} \,,
\nonumber\\
\delta Z_{21\overline{A},eH}^{\alpha\beta\, \gamma\delta, \rho\sigma}
&= \frac{1}{4} T_{2A}^{\alpha\beta\, \gamma\delta, \rho\sigma} \,,
&
\delta Z_{21\overline{21},eH}^{\alpha\beta\, \gamma\delta, \rho\sigma}
&= - \frac{1}{4} T_{2}^{\alpha\beta\, \gamma\delta, \rho\sigma} \,,
\nonumber\\
 \delta Z_{5\overline{5},\ell \ell}^{\alpha\beta\, \gamma\delta, \rho\sigma\tau\upsilon}
&= - \frac{1}{4} T_3^{\alpha\beta\, \gamma\delta, \rho\sigma\tau\upsilon} \,,
 & 
\delta Z_{22\overline{22},\ell \ell}^{\alpha\beta\, \gamma\delta, \rho\sigma\tau\upsilon}
 &= - \frac{1}{4} T_3^{\alpha\beta\, \gamma\delta, \rho\sigma\tau\upsilon} \,,
\nonumber\\
\delta Z_{21\overline{21},\ell \ell}^{\alpha\beta\, \gamma\delta, \rho\sigma\tau\upsilon}
 &= - \frac{1}{2} T_3^{\alpha\beta\, \gamma\delta, \rho\sigma\tau\upsilon} \,,
&
\delta Z_{A\overline{A},\ell \ell}^{\alpha\beta\, \gamma\delta, \rho\sigma\tau\upsilon}
 &=  \frac{1}{2} T_{3A}^{\alpha\beta\, \gamma\delta, \rho\sigma\tau\upsilon} \,.
\nonumber\end{align}

\subsection{Four-lepton Green's function}

In the following we will explicitly present the renormalisation of a Green's function involving four lepton doublets.
When we consider double-insertions of dimension-five operators one additional operator that vanishes in the limit $d \to 4$, a so-called evanescent operator, appears in our calculation.
The exact definition of the evanescent operator in $d$ dimensions is not important, but will induce a scheme dependence beyond one-loop. We use
\begin{equation}
  \label{eq:eva-operator}
  \mathcal{O}^{\alpha\beta\gamma\delta}_{\mathrm{eva}} = 
\frac{1}{2} \delta_{ij}\delta_{kl} (\overline{\ell}_{i\alpha} \ell_{k\gamma}^c) (\overline{\ell^c}_{l\delta} \ell_{j\beta}) - 
\frac{1}{2} \mathcal{O}^{\alpha\beta\gamma\delta}_{\ell\ell} \,,
\end{equation}
where the first term has a left-right chirality structure and $i$, $j$, $k$,  $l$ are SU(2) indices. 

Denoting the flavour and SU(2) component of the final state $\langle f| = \langle \ell_{k,\phi} \ell_{l,\chi}|$ and the initial state $| i\rangle = | \ell_{i,\psi} \ell_{j,\omega} \rangle$ by $\phi$, $\chi$, $\psi$, $\omega$, and $i$, $j$, $k$, $l$ respectively, we find for the third diagram of figure~\ref{fig:2}
\begin{equation}
  \label{eq:4}
  \langle f | Q_5^{\alpha\beta} (Q_5^{\gamma\delta})^{\dagger} | i \rangle|_{1/\epsilon}^{(1)} = 
\frac{(\bar{u}_{\psi i} P_L v_{\omega j})(\bar{v}_{\phi k} P_R u_{\chi l})}{64 \pi^2 }
\left(\delta _{\psi\delta } \delta _{\omega\gamma }+\delta
   _{\omega\delta } \delta _{\psi\gamma }\right) \left(\delta
   _{\chi\alpha } \delta _{\phi\beta }+\delta _{\phi\alpha }
   \delta _{\chi\beta }\right) \left(\delta _{il} \delta
   _{jk}+\delta _{ik} \delta _{jl}\right)  ,
\end{equation}
which exactly matches the scalar contribution of the evanescent operator $\mathcal{O}_{\mathrm{eva}}$ at tree level
\begin{equation}
\label{eq:5}
\begin{split}
  \langle f |& \left( \delta Z_{5\bar{5},e}^{\alpha\beta\, \gamma\delta, \rho\sigma\tau\upsilon} Q_\mathrm{eva,scalar}^{\rho\sigma\tau\upsilon} + \mathrm{h.c.} \right) | i \rangle_{LR} = \delta Z_{5\bar{5},e}^{\alpha\beta\, \gamma\delta, \rho\sigma\tau\upsilon} 
(\bar{u}_{\psi i} P_L v_{\omega j})(\bar{v}_{\phi k} P_R u_{\chi l}) \times \\ &
  [\delta _{il} \delta _{jk} 
  \left( \delta _{\psi\sigma } \delta_{\omega\upsilon } \delta _{\chi\rho } \delta _{\phi\tau} +
         \delta _{\omega\sigma } \delta _{\psi\upsilon } \delta_{\phi\rho } \delta _{\chi\tau }
  \right)+ 
 \delta _{ik} \delta_{jl} 
  \left( \delta _{\omega\sigma } \delta _{\psi\upsilon }\delta _{\chi\rho } \delta _{\phi\tau } +
         \delta _{\psi\sigma } \delta _{\omega\upsilon } \delta _{\phi\rho } \delta_{\chi\tau }
  \right) ]\,,
\end{split}
\end{equation}
where we have used the hermiticity condition of the renormalisation constants\footnote{The four-lepton renormalisation constants fulfil the hermiticity condition $Z^{\alpha\beta\,\gamma\delta,\rho\sigma\tau\upsilon}_{5\bar{5},\ell \ell} = (Z^{\delta\gamma\,\beta\alpha,\sigma\rho\upsilon\tau}_{5\bar{5},\ell \ell})^{*}$.}. 
The one-loop contribution to the $L \times R$ part is then renormalised by the renormalisation constant $\delta Z_{5\bar{5},\mathrm{eva}}^{\alpha\beta\, \gamma\delta, \rho\sigma\tau\upsilon} = - \frac{1}{2} T_3^{\alpha\beta\, \gamma\delta, \rho\sigma\tau\upsilon}$. As there is no $(V-A)\times(V-A)$ contribution to the Green's function, the $(V-A)\times(V-A)$ parts have to cancel between the counterterms of $\mathcal{O}_{\mathrm{eva}}$ and $\mathcal{O}_{\mathrm{\ell\ell}}$, i.e. $\delta Z_{5\bar{5},\mathrm{\ell \ell}}^{\alpha\beta\, \gamma\delta, \rho\sigma\tau\upsilon} = (1/2) \delta Z_{5\bar{5},\mathrm{eva}}^{\alpha\beta\, \gamma\delta, \rho\sigma\tau\upsilon} =
- \frac{1}{4} T_3^{\alpha\beta\, \gamma\delta, \rho\sigma\tau\upsilon}$.


\subsection{$W$ emission}
\label{app:W}
The values of renormalisation constants may be checked by renormalising other loop processes involving a double-insertion of dimension-five operators, and matching them to the same operator basis $\mathcal{O}_{H\ell(1)}^{\beta \alpha},  \mathcal{O}_{H\ell(3)}^{\beta \alpha}, \mathcal{O}_{v(1)}^{\beta \alpha}$ and $\mathcal{O}_{v(3)}^{\beta \alpha}$. The internal Higgs and lepton lines of the loop diagram may couple to $B_{\mu}$ or $W_{\mu}^{a}$ bosons of the $\rm U(1)_{Y}$ and $\rm SU(2)_{L}$ groups respectively. Since the group structure of $\rm U(1)_{Y}$ is trivial, we concentrate here on the calculation resulting from emission of a $W_{\mu}^{a}$ boson. The results for emission of $B_{\mu}$ emission may be retrieved from these results by replacing the $\rm SU(2)_{L}$ generators everywhere by $\rm U(1)_{Y}$ generators, $\frac{1}{2}\tau_{ij}^{a} \to y_{H,\ell} \delta_{ij}$ at the beginning of the calculation.

The renormalisation equation for the process $H^{M}\ell^{n}_{\alpha} \to H^{J}\ell_{\beta}^{i} W_{\mu}^{a}$ in $\rm \overline{MS}$ is 
\begin{align*}
&0= \bra{\ell_{\beta}^{i}H^{J} W_{\mu}^{a}}\mathcal{O}_{5}^{\gamma \delta}(\mathcal{O}_{5}^{\eta \kappa})^{\dagger}\ket{\ell_{\alpha}^{n} H^{M}} \Big |_{\frac{1}{\epsilon}}^{(1)} \\
&+ {Z_{5\bar{5},H \ell (1)}^{\gamma\delta\eta \kappa, \beta \alpha}}  \brackets{- g_{2} } \sbrackets{\bar{u}_{ \beta i} \gamma^{\mu} P_{L}u_{\alpha n}}\tau^{a}_{JM} \delta_{in}  
+  {Z_{5\bar{5},H \ell (3)}^{\gamma \delta \eta \kappa, \beta \alpha}} \brackets{-g_{2}} \sbrackets{\bar{u}_{\beta i} \gamma^{\mu} P_{L}u_{\alpha n}} \brackets{\delta_{JM}\tau^{a}_{in} }  \\
&+  {Z_{5\bar{5},v (1)}^{\gamma \delta \eta \kappa, \beta \alpha}}\brackets{-g_{2}} \sbrackets{\bar{u}_{\beta i}\gamma^{\mu} P_{L} u_{\alpha n}} \delta_{JM}\tau^{a}_{in}  
+ {Z_{5\bar{5},v (3)}^{\gamma \delta \eta \kappa, \beta \alpha}}\brackets{-g_{2}} \sbrackets{\bar{u}_{\beta i}\gamma^{\mu} P_{L} u_{\alpha n}} \brackets{ \delta_{in}\tau^{a}_{JM}}\,.
\end{align*}
where the tree-level matrix elements are replaced by their respective amplitudes and the SU(2) algebra has been simplified.

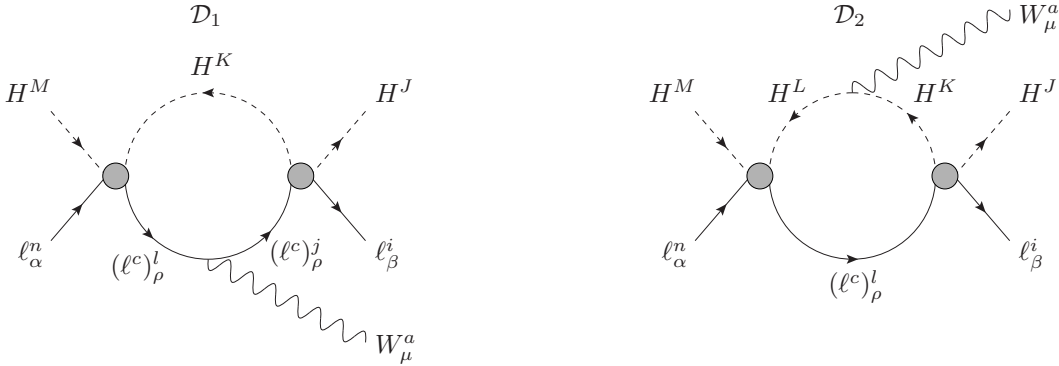
\begin{figure}[h]
\centering
\SetScale{0.7}
\begin{picture}(180,180)(00,0)
\centering
\ArrowLine(25,125)(55,160)
\DashArrowLine(25,195)(55,160){3}
\DashArrowArc(110,160)(45,0,180){3}
\Arc[arrow](110,160)(45, 180, 270)
\Arc[arrow](110,160)(45, 270, 0)
\ArrowLine(165,160)(195,125)
\DashArrowLine(165,160)(195,195){3}
\Photon(110, 115)(195, 70){-5}{7}
\GCirc(160,160){7}{0.7}
\GCirc(60,160){7}{0.7}
\put(5,80){$\ell_{\a}^{n}$} 
\put(40,75){$(\ell^c)^{l}_ {\r}$}
\put(100,80){$(\ell^{c})^{j}_{\rho}$}
\put(140,80){$\ell^{i}_ {\b}$}
\put(0,140){$H^{M }$} 
\put(140,140){$H^{J }$}
\put(70,150){$H^{K }$}
\put(140, 45){$W_{\mu}^{a}$}
\put(70, 170){$\mathcal{D}_{1}$}
\end{picture}
\hspace{2cm}
\begin{picture}(180,180)(0,0)
\centering
\ArrowLine(25,125)(55,160)
\DashArrowLine(25,195)(55,160){3}
\DashArrowArc(110,160)(45,0,90){3}
\DashArrowArc(110,160)(45,90,180){3}
\Photon(110, 205)(195, 250){5}{7}
\ArrowArc(110,160)(45,180,0)
\ArrowLine(165,160)(195,125)
\DashArrowLine(165,160)(195,195){3}
\GCirc(160,160){7}{.7}
\GCirc(60,160){7}{.7}
\put(5,80){$\ell_{\a}^{n}$} 
\put(68,68){$(\ell^c)^{l}_ {\r}$}
\put(140,80){$\ell^{i}_ {\b}$}
\put(0,140){$H^{M }$} 
\put(140,140){$H^{J }$}
\put(45,140){$H^{L}$}
\put(100,140){$H^{K}$}
\put(140,170){$W_{\mu}^{a}$}
\put(70, 170){$\mathcal{D}_{2}$}
\end{picture}
\vspace{-1.4cm}
\caption{Double insertions of dimension-five operators with associated emission of $W_{\mu}^{a}$ that mix into dimension-six operators.}
\label{fig:w}
\end{figure}

Two diagrams must be evaluated for the double insertion of dimension-five operators with associated emission of a $W_{\mu}^{a}$ boson, which can couple to either the internal Higgs or internal lepton. These diagrams are denoted by $\mathcal{D}_{1}$ and $\mathcal{D}_{2}$, and are shown in figure \ref{fig:w}.

Calculating the diagrams and isolating the $1/\epsilon$ poles gives
\begin{align}
\mathcal{D}_{1} \big |_{\frac{1}{\epsilon}}&=\frac{1}{\epsilon}\frac{ g_{2}}{64 \pi^{2} } \sbrackets{ \bar{u}_{\beta i} \gamma^{\mu} P_{L}u_{\alpha n} }\brackets{\delta_{\alpha \kappa} \delta_{\beta \gamma} \delta_{\delta \eta} }
\left( 2\delta_{JM}\tau^{a}_{in} - \delta_{Jn}\tau^{a}_{iM} - \delta_{iM} \tau^{a}_{Jn}-\delta_{in}\tau^{a}_{JM} \right) \,,  \\
\mathcal{D}_{2} \big |_{\frac{1}{\epsilon}}&=\frac{1}{\epsilon}\frac{g_{2}}{64 \pi^{2} } \sbrackets{\bar{u}_{\beta i} \gamma^{\mu} P_{L}u_{\alpha n} }\brackets{\delta_{\alpha \kappa} \delta_{\beta \gamma} \delta_{\delta \eta} } 
\left( \delta_{iM}\tau^{a}_{Jn}+\delta_{Jn}\tau^{a}_{iM}-3\delta_{JM}\tau^{a}_{in} \right) \,,
\end{align}
we find the total amplitude of the double-insertion of dimension-five operators:
\begin{align}
\bra{\ell_{\beta}^{i}H^{J} W_{\mu}^{a}}\mathcal{O}_{5}^{\gamma \delta}(\mathcal{O}_{5}^{\eta \kappa})^{\dagger}\ket{\ell_{\alpha}^{n} H^{M}} \Big |_{\frac{1}{\epsilon}}^{(1)}  &= 
- \frac{1}{\epsilon} \frac{g_{2}}{64\pi^{2}} \sbrackets{ \bar{u}_{\beta i} \gamma^{\mu} P_{L}u_{\alpha n} } 
 \brackets{\delta_{\alpha \kappa} \delta_{\beta \gamma} \delta_{\delta \eta} }\brackets{\delta_{JM}\tau^{a}_{in}+\delta_{in}\tau^{a}_{JM}} \,.
\end{align}
In this form it is simple to set up simultaneous equations for the renormalisation condition by comparing the loop and tree amplitudes, 
\begin{align}
&{Z_{5\bar{5},H \ell (1)}^{\gamma\delta\eta \kappa, \beta \alpha}}  \brackets{- g_{2} } +  {Z_{5\bar{5},v (3)}^{\gamma \delta \eta \kappa, \beta \alpha}}\brackets{- g_{2}} - \frac{1}{\epsilon} \frac{g_{2}}{64\pi^{2}}  \brackets{\delta_{\alpha \kappa} \delta_{\beta \gamma} \delta_{\delta \eta} } = 0, \\
& {Z_{5\bar{5},v (1)}^{\gamma \delta \eta \kappa, \beta \alpha}}\brackets{-g_{2}} +  {Z_{5\bar{5},H \ell (3)}^{\gamma \delta \eta \kappa, \beta \alpha}} \brackets{-g_{2}} -\frac{1}{\epsilon} \frac{g_{2}}{64\pi^{2}} \brackets{\delta_{\alpha \kappa} \delta_{\beta \gamma} \delta_{\delta \eta} } = 0.
\end{align}
This underconstrained set of equations may be constrained by substituting in solutions for ${Z_{5\bar{5},v (1)}^{\gamma \delta \eta \kappa, \beta \alpha}}$ and $ {Z_{5\bar{5},v (3)}^{\gamma \delta \eta \kappa, \beta \alpha}}$ from the momentum-dependent calculation, to verify the solutions 
\begin{equation}
{\delta Z_{5\bar{5},H \ell (1)}^{\gamma\delta\eta \kappa, \beta \alpha}}  = -  \frac{3}{4}T_{1}^{\kappa\beta\gamma\delta,\alpha\eta}\,, \hspace{10mm} {\delta Z_{5\bar{5},H \ell (3)}^{\gamma\delta\eta \kappa, \beta \alpha}}  = \frac{1}{2}T_{1}^{\kappa\beta\gamma\delta,\alpha\eta}.
\end{equation}

\section{ Renormalisation Group Equations}
\label{app:RGEs}

The bare  Wilson coefficients of dimension-five operators can be written as
\begin{equation}
\vec{C}_{X, \rm bare}^{\eta} =\mu^{2 \epsilon}
\vec{C}_{Y}^{\theta}(\mu)  Z^{\theta \eta}_{YX}(\mu),
\end{equation}
where
$\vec{C}_{Y}^{\theta}(\mu)$ is the renormalised Wilson coefficient, $Z_{YX}^{\theta\eta}(\mu)$ is the renormalisation matrix, and $\mu$ is the renormalisation scale. The $\mu^{2 \epsilon}$ introduces an additional term proportional to $\epsilon$ into the $d$-dimensional renormalisation group equation
\begin{equation}
\mu \frac{d}{d \mu} \vec{C}_{X}^{\eta} = 
- \vec{C}_{Y}^{\theta} \brackets{\mu \frac{d}{d \mu}Z_{YZ}^{\theta\zeta}} \left[ Z^{-1} \right]^{\zeta\eta}_{ZX} - 2 \epsilon \vec{C}_{X}^{\eta}.
\label{eq:g6d}
\end{equation}
This reduces to the renormalisation group equation in $d=4$ dimensions
\begin{equation}
(16 \pi^2) \mu \frac{d}{d \mu} \vec{C}_{X}^{\eta} \overset{d=4}{=} 
\vec{C}_{Y}^{\theta} \gamma_{YX}^{\theta\eta},
\end{equation}
where the 4-dimensional anomalous dimension matrix
\begin{equation}
  \label{eq:7}
  \gamma_{YX}^{\theta\eta} = 
- (16 \pi^2) \left( \mu \frac{d}{d \mu}Z_{YZ}^{\theta\zeta} \right) 
\left[ Z^{-1} \right]^{\zeta\eta}_{ZX}
\end{equation}
is independent of the choice of the overall factor $\mu^{2 \epsilon}$.
Therefore the $\mu^{2 \epsilon}$ term can be neglected when only considering mixing amongst operators of equal dimensions. 
In the case of mixing between operators of different dimensions a more careful treatment is required.


At loop level, operators of different dimensions can mix via multiple operator insertions \cite{Herrlich:1994kh}. Consider the specific case of loop diagrams involving two dimension-five operators mixing into diagrams with a single dimension-six operator insertion. We denote dimension-six quantities with a tilde, quantities that mix dimension-five and -six with a hat, and dimension-five quantities without a tilde or hat. The bare dimension-six Wilson coefficient is
\begin{equation}
\label{eq:cbare6-flavour}
\tilde{C}_{X, \rm bare}^{\eta}=
\mu^{2 \epsilon}\tilde{C}_{Y}^{\theta}(\mu)\hat{Z}_{YX}^{\theta \eta} (\mu) +
\mu^{2 \epsilon}C_{A}^{\zeta}(\mu) \tilde{Z}^{\zeta \theta, \eta}_{AB,X}(\mu) \big [C_{B}^{\theta}\big]^{\dagger}(\mu),
\end{equation}
where $\tilde{C}_{\rm bare}$ is $\mu$-independent. Therefore the renormalisation group equation is
\begin{equation}
(16 \pi^2) \mu \frac{d}{d\mu}\tilde{C}_X^{\eta} =
\tilde{C}_Y^{\theta}\hat{\gamma}_{YX}^{\theta \eta} +
C_A^{\zeta} \tilde{\gamma}_{AB, X}^{\zeta \theta, \eta} \big[ C_B^{\theta} \big]^{\dagger} ,
\end{equation}
where $\hat{\gamma}^{\theta \eta}_{YX}$ is  defined analogously to  equation (\ref{eq:g6d}), and 
\begin{align}
\tilde{\gamma}_{AB, X}^{\zeta \theta, \eta}  =&
(16 \pi^2) \left ( 2 \epsilon \tilde{Z}_{AB, Y}^{\zeta \theta, \upsilon} - 
\mu \frac{d}{d\mu}\tilde{Z}_{AB, Y}^{\zeta \theta, \upsilon} \right ) 
\big[ \hat{Z}^{-1} \big]_{YX}^{\upsilon \eta} \nonumber \\
& - (16 \pi^2) \left ( \big[ \gamma_{BD}^{\theta \omega} \big]^{\dagger} \delta_{AC}^{\zeta \chi} + 
            \gamma_{AC}^{\zeta\chi} \delta_{BD}^{\theta \omega} \right ) 
\tilde{Z}_{CD, Y}^{\chi \omega, \upsilon} \big[ \hat{Z}^{-1} \big]_{YX}^{\upsilon \eta} \,
\end{align}
where the explicit form in terms of generation indices is $[ \gamma_{AB}^{\alpha\beta\, \gamma\delta} ]^{\dagger} = [ \gamma_{AB}^{\beta\alpha\, \delta\gamma} ]^{*}$ and $\delta_{AB}^{\alpha\beta \, \gamma\delta} = \delta_{AB} \delta_{\alpha\gamma} \delta_{\beta\delta}$. 
The terms in the second line of the above equation only contribute beyond one-loop.
Furthermore, the contribution to the renormalisation tensor $Z_{AB, Y}^{\zeta \theta, \upsilon}$ is $\mu$ independent at one-loop and only the term proportional to $2 \epsilon$ contributes in our calculation.
A comment regarding the sign of the $2 \epsilon$ contribution is in order. 
The factor in $\mu^2\epsilon$ in \eqref{eq:cbare6-flavour} generates a term proportional to $- 2 \epsilon$, while the derivative of the dimension-five Wilson coefficients generates a contribution proportional to $2\times2 \epsilon$ from \eqref{eq:g6d}.
Hence the one-loop anomalous dimension matrix reads
\begin{equation}
  \label{eq:6}
\tilde{\gamma}_{AB, C}^{\zeta \eta, \theta} = 2 \delta \tilde{Z}_{AB, C}^{\zeta \eta, \theta}
\end{equation}
in terms of the one-loop renormalisation constants defined in eq.~\eqref{eq:zij-expansion}. Correspondingly we find $[\tilde{\gamma}] = 2 (16\pi^2) \epsilon [ \tilde{Z} ]$.


\section{Operators}
\label{app:ops}

This Appendix lists dimension-six,  
SM-gauge invariant  operators 
that change lepton flavour.
The  operators are in the  Buchmuller-Wyler
basis, as pruned  in 
Grzadkowski {\it et.al.} \cite{polonais},
commonly refered to as the ``Warsaw'' basis.
All operators are  added to the Lagrangian $+ \mathrm{h.c.}$, as given in
eqn (\ref{L3}):
$$
\delta {\cal L}_6 = \sum_{X,\zeta}  \frac{C^\zeta_X}{\Lambda^2} {\cal O}^\zeta_X + \mathrm{h.c.} ~~~
$$
where the flavour indices  are represented by $\zeta$,
and are all summed over all generations. 
In  the conventions of  \cite{polonais} and
\cite{JMT}, the hermitian conjugate
is not added for ``self-conjugate'' operators, for which
$\sum_{\zeta} C^\zeta_X {\cal O}^\zeta_X =
[\sum_{\zeta} C^\zeta_X {\cal O}^\zeta_X]^\dagger$.
(For instance, ${\cal O}^{\a\b\r\s}_{\ell\ell}$ of eqn (\ref{OLL})
is hermitian, because
 $[(\overline{e} \gamma^\m   \mu ) (\overline{\tau} \gamma_\m  \tau)]^\dagger
 = (\overline{\mu} \gamma^\m   e ) (\overline{\tau} \gamma_\m   \tau)$).
So we define such operators 
 with a factor 1/2  to avoid this double-counting.

The four-fermion operators involving 
$\beta \leftrightarrow \alpha$ flavour change and two quarks are:
\bea
{\cal O}_{\ell q}^{(1) \alpha \beta nm} &=&  \frac{1}{2}(\overline{\ell}_\alpha \gamma^\mu \ell_\beta ) 
(\overline{q}_n \gamma_\mu q_m ) \label{OLQ1} \\
{\cal O}_{\ell q}^{(3)\alpha \beta nm} &=&  \frac{1}{2}(\overline{\ell}_\alpha \gamma^\mu \tau^a\ell_\beta ) 
(\overline{q}_n \gamma_\mu \tau^a q_m ) \label{OLQ3} \\
{\cal O}^{\alpha \beta nm}_{eq} &=&  \frac{1}{2}(\overline{e}_\alpha \gamma^\mu e_\beta ) 
(\overline{q}_n \gamma_\mu q_m ) \label{OEQ} \\
{\cal O}^{\alpha \beta nm}_{\ell u} &=& \frac{1}{2} (\overline{\ell}_\alpha \gamma^\mu \ell_\beta ) 
(\overline{u}_n \gamma_\mu u_m ) 
~~~~~~~~~~~~~~~~~~~~~~~~ \label{OLU}\\
{\cal O}^{\alpha \beta nm}_{\ell d} &=& \frac{1}{2} (\overline{\ell}_\alpha \gamma^\mu \ell_\beta ) 
(\overline{d}_n \gamma_\mu d_m ) 
~~~~~~~~~~~~~~~~~~~~~~~~ \label{OLD}\\
{\cal O}^{\alpha \beta nm}_{eu} &=& \frac{1}{2} (\overline{e}_\alpha \gamma^\mu e_\beta ) 
(\overline{u}_n \gamma_\mu u_m )
~~~~~~~~~~~~~~~~~~~~~~~~ \label{OEU}\\
{\cal O}^{\alpha \beta nm}_{ed} &=& \frac{1}{2} (\overline{e}_\alpha \gamma^\mu e_\beta ) 
(\overline{d}_n \gamma_\mu d_m )
~~~~~~~~~~~~~~~~~~~~~~~~ \label{OED}\\
{\cal O}^{\alpha \beta nm}_{\ell e q u}&=&  (\overline{\ell}_\alpha^A  e_\beta ) \varepsilon_{AB}
(\overline{q}^B_n u_m )
~~~~~~~~~~~~~~~~~~~~~~~~
\label{scalaremu}\\
{\cal O}^{\alpha \beta nm}_{\ell e  d q}&=& (\overline{\ell}_\alpha  e_\beta ) 
(\overline{d}_n q_m )
~~~~~~~~~~~~~~~~~~~~~~~~
\label{scalarDemu}\\
{\cal O}^{\alpha \beta nm}_{T, \ell equ} &=&  (\overline{\ell}_\alpha^A \sigma^{\beta\nu} e_\beta ) \varepsilon_{AB}
(\overline{q}^B_n \sigma_{\beta\nu} u_m )
\label{tensoremu}
\eea
where $\ell,q$ are doublets and $e,u$ are singlets,
$n,m$ are possibly equal  quark family indices, and
  $A,B$ are SU(2) indices.
The operator names are as in \cite{polonais} 
with $\varphi \to H$; the flavour indices are in
 superscript. 

 In the case of  four-lepton operators, the flavour change can be
 by one or two units. Notice that in the case of ${\cal O}_{ee}$ and
 ${\cal O}_{\ell\ell}$, which are symmetric
 under interchange of the two bilinears 
 $(\overline{e} \gamma^\mu   \mu ) (\overline{\tau} \gamma_\mu  \tau)
 = (\overline{\tau} \gamma^\mu   \tau) (\overline{e} \gamma_\mu  \mu )$,
 there will be two equal coefficients that
 contribute to the Feynman rule:
\bea
{\cal O}^{\alpha \beta \r \s}_{\ell \ell} &=& \frac{1}{2}(\overline{\ell}_\alpha \gamma^\mu \ell_\beta ) 
(\overline{\ell}_\r \gamma_\mu \ell_\s) 
\label{OLL}\\
{\cal O}^{\alpha \beta \r\s}_{\ell e} &=& \frac{1}{2}(\overline{\ell}_\alpha \gamma^\mu \ell_\beta ) 
(\overline{e}_\r \gamma_\mu e_\s) 
\label{EL} \\
{\cal O}^{\alpha \beta \r\s}_{ee} &=& \frac{1}{2}(\overline{e}_\alpha \gamma^\mu e_\beta ) 
(\overline{e}_\r\gamma_\mu e_\s ) \,.
~~~~~~~~~~~~~~~~~~~~~~~~
\label{OEE}
\eea

Then there are the operators  allowing interactions
with gauge bosons and Higgses.  This includes the dipoles,
which are normalised with  the muon Yukawa coupling so
as to match onto the normalisation of Kuno-Okada \cite{KO}:
  \begin{eqnarray}
    {\cal O}^{\alpha \beta}_{eH } &=&  (H^\dagger H) (\overline{\ell}_\alpha H  e_\beta )
 \label{yuk6L} \\
{\cal O}^{\alpha \beta}_{eW } &=&  Y_\beta (\overline{\ell}_\alpha  \tau^a H \sigma^{\mu \nu} e_\beta ) W^a_{\mu \nu}
 \label{magmo2L} \\
{\cal O}^{\alpha \beta}_{eB } &=&   Y_\beta(\overline{\ell}_\alpha H \sigma^{\mu \nu}  e_\beta ) B_{\mu \nu}
 \label{magmo1L} \\
{\cal O}^{ \alpha\beta}_{H\ell (1)} &=& \frac{i}{2}(H^\dagger \stackrel{ \leftrightarrow}{  D_\mu } H)(\overline{\ell}_\alpha \gamma^\mu \ell_\beta ) 
\label{LLpenguin} \\
{\cal O}^{\alpha\beta}_{H\ell (3) } &=& \frac{i}{2} (H^\dagger  \stackrel{ \leftrightarrow}{  D_\mu^{a} }  H) (\overline{\ell}_\alpha \gamma^\mu \tau^{a}\ell_\beta ) 
\label{LL3penguin} \\
{\cal O}^{\alpha\beta}_{He } &=& \frac{i}{2}(H^\dagger \stackrel{ \leftrightarrow}{  D_\mu } H)(\overline{e}_\alpha \gamma^\mu e_\beta ) \,,
\label{EEpenguin} 
  \end{eqnarray}
  where $Y_\beta$ denotes the Yukawa coupling of a charged lepton $e_{\beta}$ in the mass basis, the double derivatives are defined in eqn
    (\ref{doubleD}), and we include factors of 1/2 for hermitian operators
    as discussed above eqn (\ref{OLQ1}).

\section{  Experimental bounds on coefficients}
\label{app:bds}

The aim of this appendix is to  obtain experimental constraints  on
the coefficients of the LFV operators of eqn (\ref{LFVWarsaw}),
evaluated at the weak scale $m_W$.  We are
interested in this subset of operators  because
they are generated at one loop by double-insertions
of dimension-five, lepton number changing (LNV) operators.
Such  constraints will 
allow an estimation of the sensitivity of LFV processes to  the coefficients of LNV  operators. 

Recall that  constraints
and  sensitivities are different. A constraint is an exclusion,
which tells the range of values a coefficient cannot have. 
For instance, the dipole coefficient (evaluated
at the muon mass scale) $ C_{D,R}^{e\mu} (m_\mu)$,
cannot be larger than $1.05 \times 10^{-8}$ 
because the branching ratio searched for by the MEG experiment \cite{TheMEG:2016wtm} is 
$$
BR(\meg) = 384\pi^2 \frac{v^2}{\Lambda^2}
(|C^{e \mu}_{D,L}|^2 +|C^{e \mu}_{D,R}|^2) ~~~,
 $$
and the current experimental search
  imposes this constraint.
Sensitivity is often discussed when an observable depends on many
coefficients, and gives  the range of values where
a coefficient could have been seen.
For instance,  among the many
loop processes that contribute to $\mu \to e \gamma$,
there are  two-loop  diagrams involving  flavour-changing
Higgs coupling $C_{eH}^{e\mu} (m_W)$. Calculating these diagrams
and imposing that they saturate the  current experimental
bound gives
$$
\left| \frac{e\alpha_e y_t}{8 \pi^3 y_\mu}C_{eH}^{e\mu}(m_W)\right| = 1.05 \times 10^{-8} ~~~.
$$
Smaller  values of  $C_{eH}^{e\mu}$  are allowed(the experiment could not have
seen them),  but larger values are not excluded
by MEG,
because many other operator coefficients could
contribute to the rate, with possibly cancellations. 

\begin{figure}[ht]
\begin{center}
\epsfig{file=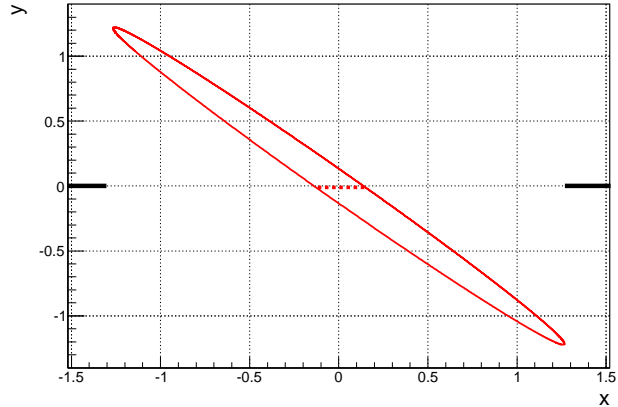, height=6cm,width=9cm}
\end{center}
\caption{An illustration of  constraints  vs sensitivities: the red ellipse represents an experimentally allowed region of parameter space. Parameter $x$ is excluded outside the projection of the ellipse onto the axis (thick black line). The experiment is insensitive to $x$ inside the ellipse. 
  \label{fig:ellipse}}
\end{figure}

The difference between an exclusion and
a sensitivity is illustrated in figure \ref{fig:ellipse},
where the allowed region is the diagonal ellipse. 
The horizontal variable $x$ is excluded outside the
projection of the ellipse onto the $x$-axis
(where the axis is thickened).   But the experiment  is only
insensitive to
$x$   inside the  intersection of the axis 
with the ellipse (dashed red line). Values of $x$ between these two
regions are allowed, provided that $y$ has the appropriately
correlated value. 

Three ways to relate low-energy experimental bounds  to  the
coefficients of operators at a higher scale  are:
\ben
\item to calculate the sensitivity of
an experimental process to a  particular operator coefficient.
This is usually simple.
\item
To express  an experimental rate as a function of
high-scale coefficients. This  is slightly more difficult,
because   more coefficients are involved: each
coefficient that contributes at the experimental scale
will become a linear combination  of
high scale coefficients due the renormalisation group
mixing.
\item To obtain constraints on coefficients at the high
  scale. This is more involved, because  a sufficient number
  of experimental constraints must be combined, in order
  to  obtain a finite allowed region  in  coefficient
  space (no ``flat directions'').  Then the allowed
  region must be projected  onto the various axes,
  in order to obtain  constraints. 
  \een
The third option is  the most useful, but
beyond the scope of this work.  Instead here, 
we partially follow the second option, as  a contribution
to the third: we  consider experimental bounds on  the dimension-six
 operators   which are generated in RGE
evolution by double-insertions of dimension-five  operators
that change lepton number.  We aim to quote
these bounds at $m_W$. The processes in question
are LFV Higgs and $Z$ decays (which occur at
the weak scale),     and flavour-changing lepton
decays at low energy (these bounds must be translated
to the weak scale via the RGEs of QED and QCD).
So we will not  succeed in our aim of setting
constraints on coefficients at $m_W$, because the
low-energy experimental bounds depend on many
coefficients at the weak scale, and we do not
include enough experimental bounds.

In the following sections, we outline the calculations
of the various rates, and summarise the experimental constraints
on coefficients at $m_W$ in table \ref{tab:bds}.

\subsection{Rates and calculations}

\begin{center}
 \renewcommand{\arraystretch}{1.3} 
$\begin{array}{|c|c|c|}
  \hline
{\rm process}& {\rm BR} <& \frac{v^2}{\Lambda^2}|\sum C| <  \\
\hline
Z\to e^\pm \mu^\mp &  7.5\times 10^{-7} \cite{Aad:2014bca}
& |C^{e\mu}_{H\ell (1)} + C^{e\mu}_{H\ell (3)}| <  1.2\times 10^{-3} \\
Z\to \tau^\pm \mu^\mp & 1.2\times 10^{-5}\cite{Abreu:1996mj}
& | C^{\mu\tau}_{H\ell (1)} + C^{\mu \tau}_{H\ell (3)}| <  4.6 \times 10^{-3} \\
Z\to e^\pm \tau^\mp & 9.8\times 10^{-6}\cite{Akers:1995gz}
& | C^{e\tau}_{H\ell (1)} + C^{e\tau}_{H\ell (3)} | <  4.1 \times 10^{-3} \\
h\to e^\pm \mu^\mp & 3.5\times 10^{-4}\cite{Khachatryan:2016rke}&
|C^{\mu e}_{eH}| , |C^{e\mu}_{eH}| <  2.5\times 10^{-4}
\\
h\to \tau^\pm \mu^\mp & 1.5\times 10^{-2}\cite{Khachatryan:2015kon}&
|C^{\mu \tau}_{eH}| , |C^{\tau \mu}_{eH}| <  1.6\times 10^{-3}
\\
h\to e^\pm \tau^\mp & 6.9\times 10^{-3}\cite{Khachatryan:2016rke} &
|C^{e \tau}_{eH}| , |C^{\tau e}_{eH}| <  1.1\times 10^{-3}
\\
\hline
\tau \to e e \overline{e} & 2.7\times 10^{-8}\cite{Hayasaka:2010np}
&| {C}^{e \tau ee}_{\ell \ell}
+ {C}^{ee e \tau }_{\ell \ell}    +g_L^e[C^{e\tau}_{H\ell(1)}
  + C^{e \tau}_{H\ell(3)}]-\delta C_{penguin}^{e \tau}| <   2.8\times 10^{-4}
\\
&&| {C}^{e \tau ee}_{\ell e}   +  g_R^e [C^{e \tau}_{H\ell(1)}
  + C^{e \tau}_{H\ell(3)}]-\delta C_{penguin}^{e \tau}| <   4.0\times 10^{-4} \\
\tau \to e \mu \overline{\mu} & 2.7\times 10^{-8}\cite{Hayasaka:2010np}
&
| {C}^{e \tau \mu \mu}_{\ell \ell}
+ {C}^{\mu \mu e\tau }_{\ell \ell}
+ {C}^{e \mu \mu \tau }_{\ell \ell}
+ {C}^{\mu \tau e\mu }_{\ell \ell} 
+g_L^e[C^{e \tau}_{H\ell(1)}
  + C^{e \tau}_{H\ell(3)}]-\delta C_{penguin}^{e \tau}| <   4.0\times 10^{-4}
\\
&&| {C}^{e \tau \mu \mu}_{\ell e}    +g_R^e [C^{\mu \tau}_{H\ell(1)}
  + C^{e \tau}_{H\ell(3)}] -\delta C_{penguin}^{e \tau}| <   4.0\times 10^{-4} 
\\
\tau \to \mu e \overline{e} & 1.8\times 10^{-8}\cite{Hayasaka:2010np}
&
| {C}^{\mu \tau ee}_{\ell \ell}
+ {C}^{ee \mu \tau }_{\ell \ell}  {C}^{\mu ee \tau }_{\ell \ell}
+ {C}^{e \tau \mu e}_{\ell \ell}   +g_L^e[C^{\mu \tau}_{H\ell(1)}
  + C^{\mu \tau}_{H\ell(3)}] -\delta C_{penguin}^{\mu \tau}| <   3.2\times 10^{-4}
\\
&&| {C}^{\mu \tau ee}_{\ell e}    +g_R^e [C^{\mu \tau}_{H\ell(1)}
  + C^{\mu \tau}_{H\ell(3)}] -\delta C_{penguin}^{\mu \tau}| <   3.2\times 10^{-4} \\
\tau \to \mu \mu \overline{\mu} & 2.1\times 10^{-8}\cite{Hayasaka:2010np}
&| {C}^{\mu \tau  \mu \mu}_{\ell \ell}
+ {C}^{ \mu \mu \mu \tau }_{\ell \ell}   +g_L^e[C^{\mu \tau}_{H\ell(1)}
  + C^{\mu \tau}_{H\ell(3)}]  -\delta C_{penguin}^{\mu \tau}| <   2.5\times 10^{-4}
\\
&&| {C}^{\mu \tau  \mu \mu}_{\ell e}    +g_R^e [C^{\mu \tau}_{H\ell(1)}
  + C^{\mu \tau}_{H\ell(3)}]  -\delta C_{penguin}^{\mu \tau}| <   3.5\times 10^{-4} \\
\tau \to e e \overline{\mu} & 1.5\times 10^{-8}\cite{Hayasaka:2010np}&
| {C}^{e \tau e\mu }_{\ell \ell} + {C}^{e \mu e\tau }_{\ell \ell}|<  3.2\times 10^{-4}
\\
\tau \to \mu \mu \overline{e} & 1.7\times 10^{-8}\cite{Hayasaka:2010np}
&
| {C}^{\mu \tau \mu e}_{\ell \ell} + {C}^{\mu e \mu \tau }_{\ell \ell}|<  3.2\times 10^{-4}
\\
\mu \to 3e & 1\times 10^{-12}\cite{Bellgardt:1987du}
&
| {C}^{e \mu ee}_{\ell \ell}
+ {C}^{ee e \mu }_{\ell \ell}   +g_L^e[C^{e\mu}_{H\ell(1)}
  + C^{e \mu}_{H\ell(3)}]  -\delta C_{penguin}^{e\mu}| <   7.1\times 10^{-7}
\\
&&| {C}^{e \mu ee}_{\ell e}  +g_R^e [C^{e \mu}_{H\ell(1)}
  + C^{e \mu}_{H\ell(3)}] - \delta C_{penguin}^{e\mu}| <   1.0\times 10^{-6}
\\
\tau\to e \g & 3.3\times 10^{-8}\cite{Aubert:2009ag}
&|C^{\tau e *}_{e \g} + \frac{e\alpha_e y_t}{8\pi^3 y_\mu} C^{\tau e *}_{e H} + \frac{e g_L^e }{16\pi^2 } C^{ e \tau}_{ H e} |< 7.3\times 10^{-6} \\
&& |C^{ e \tau}_{e \g} + \frac{e\alpha_e y_t}{8\pi^3 y_\mu} C^{e \tau }_{e H}
+ \frac{e g_R^e }{16\pi^2 } [ C^{ e \tau}_{ H \ell (1)} 
+  C^{ e \tau}_{ H \ell (3)} ] |< 7.3\times 10^{-6} \\
\tau \to  \mu \g & 4.4\times 10^{-8}\cite{Aubert:2009ag,Hayasaka:2007vc}
& |C^{\tau \mu *}_{e \g} + \frac{e\alpha_e y_t}{8\pi^3 y_\mu} C^{\tau \mu *}_{e H} + \frac{e g_L^e }{16\pi^2 } C^{ \mu \tau}_{ H e} |< 8.1\times 10^{-6} \\
&& |C^{ \mu \tau}_{e \g} + \frac{e\alpha_e y_t}{8\pi^3 y_\mu} C^{\mu \tau }_{e H}
+ \frac{e g_R^e }{16\pi^2 } [ C^{ \mu \tau}_{ H \ell (1)} 
+  C^{ \mu \tau}_{ H \ell (3)} ]|< 8.1\times 10^{-6} \\
\mu \to e \g & 4.2\times 10^{-13}\cite{TheMEG:2016wtm}
&
|C^{\mu e *}_{e \g} + \frac{e\alpha_e y_t}{8\pi^3 y_\mu} C^{\mu e *}_{e H} +\frac{e g_L^e }{16\pi^2 } C^{ e \mu}_{ H e} |< 1.05\times 10^{-8} \\
&& |C^{ e \mu}_{e \g} + \frac{e\alpha_e y_t}{8\pi^3 y_\mu} C^{e \mu }_{e H}
+ \frac{e g_R^e }{16\pi^2 } [ C^{ e \mu}_{ H \ell (1)} 
+  C^{ e \mu}_{ H \ell (3)} ] |< 1.05\times 10^{-8} \\
\hline
\end{array}
$
 \renewcommand{\arraystretch}{1} 
 \captionof{table}{ \label{tab:bds}
   Bound on operator coefficients of the SMEFT, evaluated at $m_W$,
   from the  bounds listed in column 2 on the processes of
   column 1. 
   The  bounds on coefficients of hermitian operators (${\cal O}_{H\ell(1)}$,  $ {\cal O}_{H\ell(3)}$, $ {\cal O}_{\ell\ell}, {\cal O}_{\ell e}$)
   also apply
  to   the conjugate coefficient.  All the bounds apply to running coefficients
  evaluated at $m_W$, and are for $\Lambda = v\simeq m_t$. The combination
  of coefficients $C_{penguin}$ is defined in eqn(\ref{Cpingouin}) and before
  eqn (\ref{last}), $\delta$ is defined after eqn  (\ref{last}),
  and 
 $g_R^e
  = 2s_W^2$,   $g_L^e = -1+2s_W^2$.}
\end{center}

\subsubsection{ $\bm{Z\to l_\a\bar{l}_\b}$ decay}

When  the Higgs gets a vev, the ``penguin'' operators 
${\cal O}_{H\ell(1)}$ and   $ {\cal O}_{H\ell(3)}$   generate
a vertex involving the $Z$ and  two charged leptons. 
If the flavour-changing  $Z$-fermion vertex is written in
a   SM-like  form : $-\overline{l_\a} Z^\mu \frac{g}{2c_W}
\g_\mu (g_V - g_A \g_5) l_\b$, then
\beq
g_V = g_A = - (C_{H\ell (1)} + C_{H\ell (3)}) \frac{v^2}{\Lambda^2}
\eeq
(for $v\sim m_t$).

The branching ratio can be written
\beq
BR(Z\to l_\a \overline{l_\b}) =
\frac{M_Z}{2.5{\rm GeV}} \frac{g^2}{48\pi c_W^2} (|g_V|^2 + |g_A|^2)
\eeq
where 2.5 GeV is the $Z$ width in the SM.  Since
${\cal O}_{H\ell (1)}$ and ${\cal O}_{H\ell (3)}$
are hermitian,  the conjugate process 
$Z\to l_\b \overline{l_\a}$ neccessarily occurs at the same rate,
so the BR to the experimental final state  is
\beq
BR(Z\to l^\pm_\a l^\mp_\b) =
BR(Z\to l_\a \overline{l_\b}) + BR(Z\to l_\b \overline{l_\a}) =
\frac{M_Z}{2.5{\rm GeV}} \frac{g^2}{12\pi c_W^2} |(C^{\a\b}_{H\ell (1)} + C^{\a\b}_{H\ell (3)}|^2 \frac{v^4}{\Lambda^4} 
\eeq
and the bounds we obtain on the operator coefficients,
evaluated at $\sim m_W$, are given in table \ref{tab:bds}.

\subsubsection{ $\bm{h \to \ell_\a^+ e_\b^-,  e_\a^+ \ell_\b^-}$ decays}

The flavour-changing Higgs decays
occur via the   non-hermitian operator ${\cal O}_{eH}$. 
When the Higgs has a vev,  it induces  the   Feynman rules   for
a  flavour-changing Higgs vertex with two fermions:
\bea
C^{\a\b}_{eH}{\cal O}^{\a\b}_{eH}  \longrightarrow i\frac{3C^{\a\b}_{eH} v^2}{\sqrt{2}\Lambda^2} P_R  ~~~~~,~~~~~
C^{\b\a *}_{eH}{\cal O}^{\b \a *}_{eH}  \longrightarrow i\frac{3C^{\b\a *}_{eH} v^2}{\sqrt{2}\Lambda^2} P_L \,.
\eea
We calculate the  flavour-changing  branching ratio
by comparing to
$BR (h\to  b\bar{b}) = 0.575 \pm 0.32 $
(from  the  Appendix of the Higgs Working Group  Report \cite{HWG},
for $m_h = 125.1 $ GeV), assuming the Feynman rule
for $h  b\bar{b}$  is  $-\frac{i}{\sqrt{2}}y_b(m_h) P_{L,R}$.
We use a one-loop approximation \cite{burashouches} for the running
$b$ mass 
\beq
 y_b(m_h) v = m_b(m_b) \left[\frac{\alpha(m_h)}{\alpha(m_b)}\right]^{\g_m^{(0)}/2\b^{(0)}} \simeq 3.0 ~{\rm GeV}
\eeq
where   $\alpha(m_h) \simeq 0.12,
\alpha(m_b) \simeq .23$,$\g_m^{(0)}= 8$,
$\b^{(0)}= 23/3$ and $m_b(m_b) = 4.2$ GeV.

The  operator ${\cal O}_{eH}$ is not hermitian,
but is always  included in the Lagrangian $+\mathrm{h.c.}$.
So $C^{e\mu}_{eH}{\cal O}^{e \mu}_{eH} + \mathrm{h.c.}$
will induce  both $h\to e_L \overline{\mu_R}$
and  $h\to \mu_R \overline{ e_L}$ at the same
rate:
\beq
\frac{BR(h\to \overline{e_L}  \mu_R)}{BR(h\to b \bar{b})}
=  \frac{9 |C^{e \mu}_{eH}|^2 v^2}{6 y_b^2  \Lambda^4} ~~~,
\eeq
where downstairs there is a 3 for quark
colour sums, and a 2 from the chiral projectors
in the lepton decay.
The experimental  search  sums
the  $ e_L \overline{\mu_R}$
and  $ \mu_R \overline{ e_L}$   final states, so we obtain 
\bea
3 v^2\frac{|C^{ \a \b}_{eH}|^2 }{ \Lambda^2} ~,~
3 v^2\frac{ |C^{\b\a}_{eH}|^2}{ \Lambda^2}
\leq
y_b^2 (m_h) \frac{BR(h\to l_\a^\pm l_\b^\mp)}{BR(h\to b \bar{b})}
\eea
and the resulting contraints are given in table \ref{tab:bds}.

\subsubsection{ Including the low energy decays}

The flavour-changing $\tau$ and $\mu$ decays listed
in table \ref{tab:bds} occur at
 energies $\sim m_\mu,m_\tau$, so
the decay rates are usually written
in terms of the coefficients of dimension-six operators
from the QCD$\times$QED invariant  basis appropriate
at low energies. 
These  ``low energy'' coefficients,
which we denote with a tilde $\widetilde{C}$,
 can be expressed in terms of
SMEFT coefficients at $m_W$ by
running them up to $m_W$, then matching the
QCD$\times$QED-invariant operator basis onto
the SMEFT. This was performed in
\cite{megmW}  for $\meg$, so we use
the results  of \cite{megmW} for the radiative decays  of
section \ref{sssec:meg}.
Reference \cite{PSI} studied the  Renormalisation
Group evolution,  below the weak scale,
of the coefficients who mediate $\meee$
(as well those for  as $\meg$ and $\mec$);
we use these results, combined with the
weak-scale matching conditions of \cite{megmW},
for  the discussion in section \ref{sssec:meee} 
of three body leptonic decays of $\tau$s and
$\mu$s. The minor differences between
$\mu$ and $\tau$ decays are discussed
in section \ref{sssec:meee}.


 In the EFT below $m_W$,
 we use the basis of  lepton-flavour-changing four-fermion  operators
 introduced  in \cite{KO,megmW}  for $\mu\leftrightarrow e$
 flavour change \footnote{In this basis,  the flavour indices
 are written explicitly,  so the 2 discussed above
eqn (\ref{OLL}) is absent, and Fierz transformations
are used to put the flavour change in one bilinear in
the case of $\Delta L = 1$ four-fermion operators.}.
The operators and coefficients
have as subscript their Lorentz structure ($V,S,T$) and
the chiral projection operators  of the two   fermion
bilinears, and the flavour indices of the four
fermions  as superscript. They wear tildes 
to distinguish them from the  coefficients  of SMEFT
operators. We restrict to the dipole and vector
operators,  and neglect the scalars and tensors,
which will turn out to be irrelevant  for
our study of LFV  operators generated by double-insertions
of LNV operators. So the  four-fermion operator basis below $m_W$ is
  \bea
  \delta {\cal L}_{4f} &=& \sum_{\a\b} \sum_f{\Big [}
    \widetilde{C}^{ \a\b ff}_{V,LL} (\overline{e_\a} \g^\omega P_L e_\b)(\overline{f} \g_\omega P_L f) +   \widetilde{C}^{ \a\b ff}_{V,LR} (\overline{e_\a} \g^\omega P_L e_\b)(\overline{f} \g_\omega P_R f)  {\Big ]}  + h.c. \nonumber \\
    && + \sum_{\a\b\s\r} {\Big [}  \widetilde{C}^{ \a\b \s\r}_{V,LL} (\overline{e_\a} \g^\omega P_L e_\b)
(\overline{e_\s} \g_\omega P_L e_\r)
 {\Big ]} + h.c.
  \eea
  where $\a\b \in\{ e\mu, \mu\tau,e\tau\}$, $f\in \{e,\mu,\tau,u,d,s,c,b\}$,
  and $\a\b\s\r \in\{ e\tau e\mu, \mu\tau \mu e\}$.
  In addition,  below $m_W$ we consider the  photon dipole operators
\beq
\delta {\cal L}_{dipole} =  \frac{ m_\b}{\Lambda^2} \left(
C^{\a\b}_{D,L} \overline{e^\a_R} \sigma^{\r\s} e^\b_L F_{\r\s} +
C^{\a\b}_{D,R} \overline{e^\a_L} \sigma^{\r\s} e^\b_R F_{\r\s}\right) + h.c.
\label{Lmeg}
\eeq
because the SMEFT  operators ${\cal O}_{Hl(1)}$, ${\cal O}_{Hl(3)}$
and ${\cal O}_{eH}$ match onto the dipole at $m_W$. 
The current bounds on $\meg$, $\teg$ and $\tmg$ will give
the best sensitivity to the coefficients $C_{H\ell(1)}$, $C_{H\ell(3)}$
and $C_{eH}$ .

\subsubsection{$\bm{\tau \to 3l}$ and $\bm{\mu \to 3e}$ }
\label{sssec:meee}

The first step is  to translate the experimental bounds into
constraints on  operator coefficients
at the experimental scale.  
For the three-body leptonic decays of the $\tau$, it is convenient
to define
\beq
\widetilde{BR}(\tau \to 3l) \equiv \frac{BR(\tau \to 3l)}{BR(\tau \to \mu\bar{\nu}\nu)}
\eeq
(where  $BR(\tau \to \mu\bar{\nu}\nu) = 0.174$ \cite{PDB}). Then
$\widetilde{BR}(\tau \to 3l)$  can be directly compared to
the branching ratio for
  $\mu\to 3e$  \cite{KO}:
  \bea
  BR(\meee)\frac{\Lambda^2}{v^2}  & =&  \frac{|\widetilde{C}_{S,LL}|^2+
    |\widetilde{C}_{S,RR}|^2}{8} +2 |\widetilde{C}_{V,RR} + 4e\widetilde{C}_{D, R}|^2
 +2 |\widetilde{C}_{V,LL} + 4e\widetilde{C}_{D ,L}|^2  \label{BRmeee}\\
&&+ (64 \ln\frac{m_\mu}{m_e} -88) 
 (|e\widetilde{C}_{D ,R}|^2 +|e\widetilde{C}_{D ,L}|^2) +
 |\widetilde{C}_{V,RL} + 4e\widetilde{C}_{D ,R}|^2
 + |\widetilde{C}_{V,LR} + 4e\widetilde{C}_{D, L}|^2 ~~,
\nonumber
\eea
where $2 \sqrt{2} G_F = 1/v^2$ and  the generalisation to $\tau$ decays is straightforward, after
accounting for  2s as we now discuss.

We calculate the decay rates   in
the  approximation that all final state fermions
are massless. Factors of 2   can arise
when there are two identically-flavoured fermions in the final state:
there   will be 2 diagrams,  and a factor of 1/2 in the final-state phase space. 
  Then there are two cases:\\
  a) if the identical fermions have the same chirality, there is constructive
  interference between the two diagrams (despite the fact that they have
  relative minus signs due to Fermi statistics), which doubles the rate.
  (This is consistent with $\mu \to 3e$ rate of Kuno and
  Okada \cite{KO} given above.)\\
  b) if the fermions have different chirality, the interference
  is suppressed by final state masses (which are neglected),
  so  the two for
  two diagrams cancels the 1/2 from  phase space.

 We set
  the dipole coefficients  to zero, because they are
  better constrained by the  radiative decays discussed
  in the next subsection (see table \ref{tab:bds}). Then  it is clear that
  each upper bounds on  a three-body leptonic
  decay  of the $\tau$  or $\mu$, implies six independent
  constraints on operator coefficients (evaluated at the experimental scale),
  those of interest to us are given in table \ref{tab:taumtau}.
  \begin{center}
  \renewcommand{\arraystretch}{1.5}  
$\begin{array}{|c|c|c|}
  \hline
{\rm process}& {\rm \widetilde{ BR}} <& \frac{v^2}{\Lambda^2}| C| <  \\
\hline
\tau \to e e \overline{e} & 1.6\times 10^{-7}
& \widetilde{C}^{e\tau ee}_{V,LL} < 2.8\times 10^{-4} ,  \widetilde{C}^{e\tau ee}_{V,LR} < 4\times 10^{-4} \\
\tau \to e \mu \overline{\mu} & 1.6\times 10^{-7}
& \widetilde{C}^{e\tau \mu\mu}_{V,LR}  , 
\widetilde{C}^{e\tau \mu\mu}_{V,LL}
< 4\times 10^{-4}  \\
\tau \to \mu e \overline{e} & 1.0\times 10^{-7}
&
\widetilde{C}^{ \mu\tau ee}_{V,LR}  ,   
\widetilde{C}^{\mu \tau ee}_{V,LL}
< 3.2\times 10^{-4} 
\\
\tau \to \mu \mu \overline{\mu} & 1.2\times 10^{-7}
& \widetilde{C}^{e\tau \mu \mu}_{V,LL} < 2.5\times 10^{-4} ,  \widetilde{C}^{e\tau \mu \mu}_{V,LR} < 3.5\times 10^{-4} \\
\tau \to e e \overline{\mu} & 8.6\times 10^{-8}
&
\widetilde{C}^{e \tau e \mu}_{V,LL}
  < 3.2\times 10^{-4} , \\
\tau \to \mu \mu \overline{e} & 1.0\times 10^{-7}
&
\widetilde{C}^{\mu \tau \mu e}_{V,LL}
  < 3.2\times 10^{-4}  \\
\hline
\mu \to e e \overline{e} & 1.0\times 10^{-12}
& \widetilde{C}^{e\mu ee}_{V,LL} < 7.1\times 10^{-7} ,
\widetilde{C}^{e\mu ee}_{V,LR} <  10^{-6} \\
\hline
\end{array}
$ 
\renewcommand{\arraystretch}{1} 
 \captionof{table}{ \label{tab:taumtau}
   Bounds on  some operator coefficients from three-body
   lepton decays, evaluated at the experimental scale. }
\end{center}

The operator coefficients  $\widetilde{C}_X(m_\tau)$ given in   table
\ref{tab:taumtau} can be
expressed in terms of coefficients  at $m_W$ using the one-loop RGEs~\cite{megmW,PSI}:
$$
\mu \frac{d}{d \mu } \widetilde{C}_I = \frac{\alpha_e}{4 \pi}\widetilde{C}_J[\gamma_e]_{JI}
\Rightarrow \widetilde{C}_I (m_\tau) = \widetilde{C}_J(m_W)
[\delta_{JI} - \frac{\alpha_e}{4 \pi}\ln \frac{m_W}{m_\tau}[\gamma_e]_{JI} + ...]
$$
where $[\gamma_e]$ is the one-loop anomalous dimension matrix of QED, $\ln \frac{m_W}{m_\tau} = 3.85$,
 $\ln \frac{m_W}{m_\mu} = 6.64$ and the approximate solution neglects the running
of $\alpha_e$. The one-loop QED corrections involve photon exchange between two
legs of the operator, which does not change the flavour or chiral indices, and
also ``penguin'' diagrams, where  two legs of the operator are closed in
a loop,  and a photon is attached, which turns into two external
leg fermions. The ``penguins'' can change the chirality and flavour,
and  allow 2-lepton-2-quark operators to mix with the four-lepton operators.
We therefore need a recipe for dealing with the quark-sector threshholds
$m_b$, $m_c$ and $\Lambda_{QCD}$. We  make the simplest approximation,
which is to  have a single low-energy threshhold at $m_\tau$, and
run from $m_W \to m_\tau$ with  five flavours of quark,  and
we use this low-energy  scale also for the decays
of the $\mu$. In this
approximation, it is convenient to define the combination
of operator coefficients
\bea
\widetilde{C}^{\a\b}_{penguin} &=&-\frac{4N_c}{3} \sum_{q}Q_q (\widetilde{C}^{\a\b qq}_{V,LL} + \widetilde{C}^{\a\b qq}_{V,LR} )
+ \frac{4}{3} \sum_{l}  ([1+\delta_{\a l} + \delta_{\b l}]
\widetilde{C}^{\a\b ll}_{V,LL} + \widetilde{C}^{\a\b ll}_{V,LR} )
\label{Cpingouin}
\eea
where  $l \in \{e,\mu,\tau\}$, $q \in \{u,d,s,c,b\}$,  and $Q_q$ is the electric charge of the quark.
Then the coefficients constrained in table \ref{tab:taumtau} can be written
\bea
\widetilde{C}^{e \mu ee}_{V,LR}(m_\tau) &=& [
 1+12 \frac{\alpha_e}{4 \pi}\ln \frac{m_W}{m_\tau}] \widetilde{C}^{e \mu ee}_{V,LR}(m_W)  - 
\frac{\alpha_e}{4 \pi}\ln \frac{m_W}{m_\tau}  \widetilde{C}^{e\mu}_{penguin} (m_W)  \\
\widetilde{C}^{e \mu ee}_{V,LL}(m_\tau) &=& 
[ 1-12 \frac{\alpha_e}{4 \pi}\ln \frac{m_W}{m_\tau}]
\widetilde{C}^{e \mu ee}_{V,LL}(m_W) - 
\frac{\alpha_e}{4 \pi}\ln \frac{m_W}{m_\tau} \widetilde{C}^{e\mu}_{penguin} (m_W)  \\
\widetilde{C}^{e \tau ll}_{V,LR}(m_\tau) &=& 
[ 1+12 \frac{\alpha_e}{4 \pi}\ln \frac{m_W}{m_\tau}]
\widetilde{C}^{e \tau ll}_{V,LR}(m_W)- 
\frac{\alpha_e}{4 \pi}\ln \frac{m_W}{m_\tau}  \widetilde{C}^{e\tau}_{penguin} (m_W) \\
\widetilde{C}^{e \tau ll}_{V,LL}(m_\tau) &=& 
[ 1-12 \frac{\alpha_e}{4 \pi}\ln \frac{m_W}{m_\tau}]
\widetilde{C}^{e \tau ll}_{V,LL}(m_W)- 
\frac{\alpha_e}{4 \pi}\ln \frac{m_W}{m_\tau} \widetilde{C}^{e\tau}_{penguin} (m_W)  \\
\widetilde{C}^{\mu \tau ll}_{V,LR}(m_\tau) &=& 
[ 1+12 \frac{\alpha_e}{4 \pi}\ln \frac{m_W}{m_\tau}]
\widetilde{C}^{\mu \tau ll}_{V,LR}(m_W)- 
\frac{\alpha_e}{4 \pi}\ln \frac{m_W}{m_\tau}  \widetilde{C}^{\mu\tau}_{penguin} (m_W)  \\
\widetilde{C}^{\mu \tau ll}_{V,LL}(m_\tau) &=&
[ 1-12 \frac{\alpha_e}{4 \pi}\ln \frac{m_W}{m_\tau}]
 \widetilde{C}^{\mu \tau ll}_{V,LL}(m_W)- 
\frac{\alpha_e}{4 \pi}\ln \frac{m_W}{m_\tau} \widetilde{C}^{\mu\tau}_{penguin} (m_W)  \\
\widetilde{C}^{\mu \tau \mu e}_{V,LL}(m_\tau) &=& [1 -
12 \frac{\alpha_e}{4 \pi}\ln \frac{m_W}{m_\tau}  ]\widetilde{C}^{\mu \tau \mu e}_{V,LL}(m_W) \\
\widetilde{C}^{e \tau e\mu }_{V,LL}(m_\tau) &=&[1- 12
\frac{\alpha_e}{4 \pi}\ln \frac{m_W}{m_\tau} ]\widetilde{C}^{e \tau e\mu}_{V,LL}(m_W) \,.
\eea

Finally, the  combinations of coefficients that are constrained by
data can be matched at $m_W$ onto coefficients of SMEFT operators
\cite{megmW}\footnote{These equations differ  from
  \cite{megmW}   due to different conventions
  for operator normalisation and signs, and also due to some errors
  in \cite{megmW}. 
 The  SMEFT basis used here  is normalised according to \cite{polonais},
 where there are  ``redundant'' flavour-changing
  four-fermion operators,
  which are absent from the basis used  below $m_W$ in \cite{megmW}, compare $e.g.$ the left and right hand sides of eqn.~(\ref{4fmWmatch}). Then, the sign convention  used here for the $g^f_{L,R}$
  and the $Z$-vertex Feynman rule  agrees with the PDG but
  is opposite to that of   \cite{megmW}. Finally, in  \cite{megmW}, 
  there is an incorrect factor of 2 mutiplying  the penguin coefficients
  which generate $s$ and $t$ channel diagrams; this 2 should not appear,
  because the four-fermion operator generates the same
  $s$ and $t$ channel diagrams.}:
\bea
\widetilde{C}^{e \tau e\mu }_{V,LL}(m_W)& =& {C}^{e \tau e\mu }_{\ell \ell}(m_W) +  {C}^{ e\mu e \tau }_{\ell \ell}(m_W) \label{4fmWmatch}\\
\widetilde{C}^{\mu \tau \mu e}_{V,LL}(m_W) &=& {C}^{\mu \tau \mu e}_{\ell \ell}(m_W)
+ {C}^{\mu e \mu \tau }_{\ell \ell}(m_W) \nonumber \\
\widetilde{C}^{\mu \tau ee}_{V,LL}(m_W) &=& {C}^{\mu \tau ee}_{\ell \ell}(m_W)
+ {C}^{ee \mu \tau }_{\ell \ell}(m_W) +{C}^{\mu ee \tau }_{\ell \ell}(m_W)
+ {C}^{e \tau \mu e}_{\ell \ell}(m_W) 
+ g_L^e[C^{\mu \tau}_{H\ell(1)}(m_W)
  + C^{\mu \tau}_{H\ell(3)}(m_W)]  \nonumber\\
\widetilde{C}^{\mu \tau \mu \mu}_{V,LL}(m_W) &=& {C}^{\mu \tau \mu \mu}_{\ell \ell}(m_W)
+ {C}^{\mu \mu \mu \tau }_{\ell \ell}(m_W)  
+ g_L^e[C^{\mu \tau}_{H\ell(1)}(m_W)
  + C^{\mu \tau}_{H\ell(3)}(m_W)]  \nonumber\\
\widetilde{C}^{\mu \tau ll}_{V,LR}(m_W) &=& {C}^{\mu \tau ll}_{\ell e}(m_W)
  +  g_R^e [C^{\mu \tau}_{H\ell(1)}(m_W)  + C^{\mu \tau}_{H\ell(3)}(m_W)] \nonumber \\
\widetilde{C}^{e \tau \mu \mu}_{V,LL}(m_W) &=& {C}^{e \tau \mu \mu}_{\ell \ell}(m_W)
+ {C}^{\mu \mu e \tau }_{\ell \ell}(m_W)   +{C}^{e \mu \mu \tau }_{\ell \ell}(m_W)
+ {C}^{\mu \tau e \mu }_{\ell \ell}(m_W) 
+ g_L^e[C^{e \tau}_{H\ell(1)}(m_W)
  + C^{e \tau}_{H\ell(3)}(m_W)]  \nonumber\\
\widetilde{C}^{e \tau ee}_{V,LL}(m_W) &=& {C}^{e \tau ee}_{\ell \ell}(m_W)
+ {C}^{ee e \tau }_{\ell \ell}(m_W)  
+ g_L^e[C^{e \tau}_{H\ell(1)}(m_W)
  + C^{e \tau}_{H\ell(3)}(m_W)]  \nonumber\\
\widetilde{C}^{e \tau ll}_{V,LR}(m_W) &=& {C}^{e \tau ll}_{\ell e}(m_W)
   +  g_R^e [C^{e\tau}_{H\ell(1)}(m_W)
  + C^{e \tau}_{H\ell(3)}(m_W)]  \nonumber\\
\widetilde{C}^{e \mu ee}_{V,LL}(m_W) &=& {C}^{e \mu ee}_{\ell \ell}(m_W)
+ {C}^{ee e \mu }_{\ell \ell}(m_W)  
+ g_L^e[C^{e \mu}_{H\ell(1)}(m_W)
  + C^{e \mu}_{H\ell(3)}(m_W)]  \nonumber\\
\widetilde{C}^{e \mu ee}_{V,LR}(m_W) &=& {C}^{e \mu ee}_{\ell e}(m_W)
   +  g_R^e[C^{e \mu}_{H\ell(1)}(m_W)
  + C^{e \mu}_{H\ell(3)}(m_W)]  \nonumber
\eea
where  $l \in \{ e,\mu\}$ in the above equations,
and $g_R^e = 2\sin^2\theta_W$,
 $g_L^e = -1 + 2\sin^2\theta_W$.
In order to match the ``penguin'' coefficient
of eqn (\ref{Cpingouin}) onto coefficients
of the SMEFT, matching conditions for operators
with a quark bilinear are also required:
\bea
\widetilde{C}^{\a\b uu}_{V,LL}(m_W) &=& {C}^{\a\b uu}_{\ell q(1)}(m_W)
- {C}^{\a\b uu}_{\ell q(3)}(m_W)
+ g_L^u[C^{\a\b}_{H\ell(1)}(m_W)
  + C^{\a\b}_{H\ell(3)}(m_W)]  \nonumber\\
\widetilde{C}^{\a\b dd}_{V,LL}(m_W) &=& {C}^{\a\b dd}_{\ell q(1)}(m_W)
+ {C}^{\a\b dd}_{\ell q(3)}(m_W)
+ g_L^d[C^{\a\b}_{H\ell(1)}(m_W)
  + C^{\a\b}_{H\ell(3)}(m_W)]  \nonumber\\
\widetilde{C}^{\a\b uu}_{V,LR}(m_W) &=& {C}^{\a\b uu}_{\ell u}(m_W)
  +  g_R^u [C^{\a\b}_{H\ell(1)}(m_W)  + C^{\a\b}_{H\ell(3)}(m_W)] \nonumber \\
\widetilde{C}^{\a\b dd}_{V,LR}(m_W) &=& {C}^{\a\b dd}_{\ell d}(m_W)
+  g_R^d [C^{\a\b}_{H\ell(1)}(m_W)  + C^{\a\b}_{H\ell(3)}(m_W)]
\label{D21}
\eea
where $\a\b \in \{ \mu\tau, e\tau ,e\mu \}$, 
$g_L^u = 1 - \frac{4}{3} \sin^2\theta_W$, 
$g_R^u = - \frac{4}{3} \sin^2\theta_W$, $g_L^d = -1 + \frac{2}{3} \sin^2\theta_W$ and, $g_R^d = \frac{2}{3}\sin^2\theta_W$. Combining
the definition  (\ref{Cpingouin}) with the matching conditions
of eqn (\ref{D21}) allows the definition of a combination of SMEFT
coefficients $C_{penguin}^{\a\b} (m_W)$. Then the experimental
constraint on, for instance $\widetilde{C}^{e\tau  \mu \mu}_{V,LL}(m_\tau)$,
gives
\beq
    {\Big |} {\Big [}1-12\delta {\Big ]}{\Big [}{C}^{e \tau \mu \mu}_{\ell \ell}
+ {C}^{\mu \mu e \tau }_{\ell \ell}   +{C}^{e \mu \mu \tau }_{\ell \ell}
+ {C}^{\mu \tau e \mu }_{\ell \ell} 
+ g_L^e[C^{e \tau}_{H\ell(1)}
  + C^{e \tau}_{H\ell(3)}] {\Big ]}
    -\delta C_{penguin}^{\a\b}   {\Big |}<4\times 10^{-4}
\label{last}
    \eeq
where all the coefficients are evaluated at $m_W$, and
$\delta = \frac{\alpha_e}{4 \pi} \ln\frac{m_W}{m_\tau}\sim 1/400$.
This and other constraints from 3-body $\tau$ decays
are given in table \ref{tab:bds}, where for compactness,
$ [1 \pm 12\delta ]$ is approximated as 1.

\subsubsection{ $\bm{l_\b\to l_\a \gamma}$}
\label{sssec:meg}

The radiative decays $l_\b\to l_\a \gamma$ provide some
of the most restrictive  bounds on lepton flavour violation.
The branching ratio at $m_{\b}$ can be written
\beq
\widetilde{BR}(l_\b\to l_\a \gamma) \equiv \frac{BR(l_\b\to l_\a \gamma)}
          {BR(l_\b\to l_\a \bar{\nu}\nu)} = 384\pi^2 \frac{v^2}{\Lambda^2}
          (|C^{\a\b}_{D,L}|^2 +|C^{\a\b}_{D,R}|^2)
\leq \left\{
\begin{array}{ll}
  4.2 \times 10^{-13} & \meg\\
  2.0 \times 10^{-7} & \teg\\
  2.5 \times 10^{-7} & \tmg\\
  \end{array}\right.
          \eeq
where the low energy dipole operators are added to
the Lagrangian as in eqn (\ref{Lmeg}).

The dipole coefficients evaluated at the experimental
scale can be expressed in terms of SMEFT coefficients at the
weak scale as \cite{megmW}
\bea
C^{\a\b}_{D,L}(m_\tau) &=& C^{\b \a *}_{e \g} (m_W) + \frac{e\alpha_e y_t}{8\pi^3 y_\mu} C^{\b \a *}_{e H}(m_W) + \frac{e g_L^e }{16\pi^2 } C^{ \a \b}_{ H e} (m_W) +... \\
C^{\a\b}_{D,R}(m_\tau) &=& C^{\a\b}_{e \g} (m_W) + \frac{e\alpha_e y_t}{8\pi^3 y_\mu} C^{\a\b}_{e H}(m_W) + \frac{e g_R^e }{16\pi^2 } [ C^{\a\b}_{ H \ell (1)} (m_W)
+  C^{\a\b}_{ H \ell (3)} (m_W)+...
\eea
where
the contributions of  scalar and  tensor  four-fermion operators were
neglected, $g_R^e$ and $g_L^e$  are defined after
eqn(\ref{4fmWmatch}), and 
\beq
C^{\a\b}_{e \g} = c_W C^{\a\b}_{eB} -s_W  C^{\a\b}_{eW}   ~~.
\eeq

\section{Comparison with the Literature}
\label{sec-1-3}

The standard model calculation has been performed in Reference~\cite{BGJ} in a different operator basis.
We disagree with their final results even after transforming our results to their basis.
To do this we specify our basis 
\begin{equation}
\tilde{\mathcal{O}} = \left( 
\mathcal{O}_{H \ell (1)}, \mathcal{O}_{H \ell (3)},
\mathcal{O}_{e H}, \mathcal{O}_{e H}^{\dagger}, 
\mathcal{O}_{v(1)}, \mathcal{O}_{v(1)}^{\dagger}, 
\mathcal{O}_{v(3)}, \mathcal{O}_{v(3)}^{\dagger} 
\right)^T \,,
\end{equation}
and the one used in Reference~\cite{BGJ}
\begin{equation}
\label{eq:basisBGJ}
\tilde{Q} = \left( Q_{\phi \ell}^{(-)}, Q_{\phi \ell}^{(+)}, Q_{e\phi}, Q_{e\phi}^{\dagger}, \mathcal{O}_{v(1)}, \mathcal{O}_{v(1)}^{\dagger}, \mathcal{O}_{v(3)}, \mathcal{O}_{v(3)}^{\dagger}
\right)^T \,,
\end{equation}
where the additional operators are defined as
\begin{eqnarray}
Q_{\phi \ell}^{(-)} &=& \frac{i}{2} \left[ (H^{\dagger} D_{\mu} H)(\bar{\ell} \gamma^{\mu} \ell) - (H^{\dagger} D_{\mu}^a H)(\bar{\ell} \tau^a \gamma^{\mu} \ell) \right]\\
Q_{\phi \ell}^{(+)} &=& \frac{i}{2} \left[ (H^{\dagger} D_{\mu} H)(\bar{\ell} \gamma^{\mu} \ell) + (H^{\dagger} D_{\mu}^a H)(\bar{\ell} \tau^a \gamma^{\mu} \ell) \right] \\
Q_{e\phi} &=& \mathcal{O}_{eH} \,.
\end{eqnarray}
Here we drop the generation indices and note that the operators $Q_{\phi \ell}^{(-)}$ and $Q_{\phi \ell}^{(+)}$ are not hermitian. 
For this reason we treat the operator $\mathcal{O}_{eH}$ and the EOM-vanishing operators independent from their hermitian conjugate in our basis transformation. 
Writing the resulting linear transformation as
\begin{equation*}
\tilde{\mathcal{O}} = \hat{R} \tilde{Q} \,,
\end{equation*}
only the first two rows of $\hat{R}$ have entries that are not proportional to an identity transformation.
These two rows are determined by the following linear transformation\footnote{To perform the change of basis we have to move covariant derivatives from one term to another. This can be done by noting that the total derivatives $D_{\mu} \left[ \left( H^{\dagger} H \right) (\bar{\ell} \gamma^{\mu} \ell) \right]$ and $D_{\mu} \left[ \left( H^{\dagger} \tau^a H \right) (\bar{\ell} \tau^a \gamma^{\mu} \ell) \right]$ are vanishing.}:
\begin{equation}
\label{eq:rnonhc}
\begin{pmatrix}
  \mathcal{O}_{H \ell(1)}\\ \mathcal{O}_{H \ell(3)}
\end{pmatrix}
= 
\begin{pmatrix}
 2 & 2 & Y & -Y^{\dagger} & 1 & -1 & 0 & 0 \\
 -2 & 2 & Y & -Y^{\dagger} & 0 & 0 & 1 & -1 \\  
\end{pmatrix}
\tilde{O} \,.
\end{equation}
The Wilson coefficients and renormalisation constants will consequently fulfil our hermiticity conditions in our basis, but not necessarily in the basis of Reference~\cite{BGJ}.
The counterterms of the Wilson coefficients transform in the same way as the respective Wilson coefficients under our change of basis, i.e. as
\begin{equation*}
\delta\tilde{c} = \hat{R}^T \delta \tilde{C} \,,
\end{equation*}
where $\delta \tilde{C} = (16 \pi^2) \epsilon \vec{C} \tilde{Z} \vec{C}^\dagger$  represent the counterterms multiplied with $(16 \pi^2) \epsilon$, while $\delta\tilde{c}$ correspond to the analogous expression in the $\tilde{Q}$ basis.

Using the counterterms presented in Equations~\eqref{ct0}, \eqref{ct1} and \eqref{ct2}, and the results of \eqref{eq:rnonhc} we obtain
\begin{equation}
  \label{eq:bwnohcct}
  \delta\tilde{c} = 
  \left( 
  - \frac{5}{2} \big[ C_5 C_5^* \big],
  - \frac{1}{2} \big[ C_5 C_5^* \big],
    \frac{1}{2} \big[ C_5 C_5^* Y \big],
    \big[ Y^{\dagger} C_5^* C_5 \big]
  \right)^T \,,
\end{equation}
which fulfil the hermiticity condition of the overall Lagrangian, even though this is not immediately apparent due to the choice of basis. 
These results are in disagreement with the final results quoted in Reference~\cite{BGJ}. Yet using the results quoted in the individual diagrams in Appendix B of Reference~\cite{BGJ} we find agreement with the expression of
eqn~\eqref{eq:bwnohcct} apart from a global minus sign, which suggests that a different projection was performed. 
We explicitly checked that the diagrams given in Appendices B.1, B.3 and B.4 of their calculation have the opposite sign compared with our calculation, while we agree with their lepton conserving contribution presented in Appendix B.2.
Following the explanations of the calculation it appears that part (the $\delta\delta$ part) of the diagram evaluated in Appendix B.1 of Reference~\cite{BGJ} is projected onto an operator basis where the operators $Q_{\phi \ell}^{(\pm)}$ are replaced by $Q_{\phi \ell}^{(\pm)\prime} = Q_{\phi \ell}^{(\pm)} + (Q_{\phi \ell}^{(\pm)})^{\dagger}$, while another part (the $\epsilon\epsilon$ part) is projected onto the basis presented in eqn~\eqref{eq:basisBGJ}.

Transforming now to the primed basis, where the hermitian conjugate is added to the first two operators of eqn~\eqref{eq:basisBGJ} we find that the non-trivial transformation matrix involves only the first two elements of our and the primed basis. Writing explicitly
\begin{equation}
\label{eq:rnonhc}
\begin{pmatrix}
  \mathcal{O}_{H \ell(1)} \\ \mathcal{O}_{H \ell(3)}
\end{pmatrix}
= 
\begin{pmatrix}
 1 & 1 \\
 -1 & 1 \\ 
\end{pmatrix}
\begin{pmatrix}
  \mathcal{O}_{H \ell}^{(-)\prime} \\ \mathcal{O}_{H \ell}^{(+)\prime}
\end{pmatrix} \,,
\end{equation}
we find 
\begin{equation}
  \label{eq:bwhcct}
  \delta\tilde{c}' = 
  \left( 
  - \frac{5}{4} \big[ C_5 C_5^* \big],
  - \frac{1}{4} \big[ C_5 C_5^* \big],
    \frac{3}{4} \big[ C_5 C_5^* Y \big],
    \frac{3}{4} \big[ Y^{\dagger} C_5^* C_5 \big]
  \right)^T \,.
\end{equation}
Again, this result does not agree with Reference~\cite{BGJ}.
Finally, note that projecting 
the results quoted for the individual diagrams in Appendix B of Reference~\cite{BGJ},
except the $\varepsilon\varepsilon$ part, would give 
\begin{equation}
  \label{eq:epsepshc}
 \delta\tilde{c}'_{\mathrm{not}\; \varepsilon\varepsilon} = 
  \left( 
  + \frac{1}{4} \big[ C_5 C_5^* \big],
  + \frac{1}{4} \big[ C_5 C_5^* \big],
  - \frac{3}{4} \big[ C_5 C_5^* Y \big],
  - \frac{3}{4} \big[ Y^{\dagger} C_5^* C_5 \big]
  \right)^T \,,
\end{equation}
while projecting only the $\varepsilon\varepsilon$ part on the non-hermitian basis yields
$\delta\tilde{c}_{\varepsilon\varepsilon} = 
  \left( 
  + 2 \big[ C_5 C_5^* \big], 0, 0, 0
  \right)^T$. Summing these two terms would reproduce the results of Reference~\cite{BGJ}.

\section{Flavour Conserving Contribution}
\label{sec:lept-cons-contr}

Even though the diagram in figure \ref{fig:2b} cannot induce lepton flavour violation it contributes to the renormalisation of $\mathcal{O}_{eH}$ and the corresponding operators that involve quarks. We also explicitly checked that the diagrams that involve six external Higgses vanish after summing over them.
Denoting the trace over the product of the two dimension-five Wilson coefficients by $\mathrm{Tr}[C_5C_5^{*}]$ we find
\begin{equation}
\label{eq:flavour-conserving}
\begin{split}
  (\vec{C} [\tilde{Z}] \vec{C}^\dagger)^{\beta\alpha}_{eH} &= 
- \frac{1}{16\pi^2 \epsilon} \mathrm{Tr}[C_5C_5^{*}] [Y_{e}]_{\beta\alpha} \,, \\
  (\vec{C} [\tilde{Z}] \vec{C}^\dagger)^{\beta\alpha}_{uH} &= 
- \frac{1}{16\pi^2 \epsilon} \mathrm{Tr}[C_5C_5^{*}] [Y_{u}]_{\beta\alpha} \,, \\
  (\vec{C} [\tilde{Z}] \vec{C}^\dagger)^{\beta\alpha}_{dH} &= 
- \frac{1}{16\pi^2 \epsilon} \mathrm{Tr}[C_5C_5^{*}] [Y_{d}]_{\beta\alpha} \,,
\end{split}
\end{equation}
where $Y_{u}$ and $Y_{d}$ are defined as $\Gamma_u$ and $\Gamma_d$ of reference~\cite{polonais}.
In addition we also generate the following mixing into operators that only comprise Higgs and gauge fields and write
\begin{equation}
\label{eq:flavour-conserving}
\begin{split}
  (\vec{C} [\tilde{Z}] \vec{C}^\dagger)_{H} &= 
- 2 \frac{1}{16\pi^2 \epsilon} \mathrm{Tr}[C_5C_5^{*}] \lambda \,, \\
  (\vec{C} [\tilde{Z}] \vec{C}^\dagger)_{HD} &= 
- 2 \frac{1}{16\pi^2 \epsilon} \mathrm{Tr}[C_5C_5^{*}] \,, \\
  (\vec{C} [\tilde{Z}] \vec{C}^\dagger)_{H \Box} &= 
- \frac{1}{16\pi^2 \epsilon} \mathrm{Tr}[C_5C_5^{*}] \,,
\end{split}
\end{equation}
where the additional operators are defined as:
\begin{equation}
  \label{eq:9}
  \begin{split}
    \mathcal{O}_{dH}^{\alpha \beta} &= (H^\dagger  H) \overline{q}_\alpha H d_\beta \\
    \mathcal{O}_{uH}^{\alpha \beta} &= (H^\dagger  H) \overline{q}_\alpha \epsilon H^{*} u_\beta \\
    \mathcal{O}_H &= \frac{1}{2} \left( H^{\dagger} H \right)^3 \\
    \mathcal{O}_{HD} &= \frac{1}{2} \left( H^{\dagger} D^{\mu} H \right)^{*} \left( H^{\dagger} D_{\mu} H \right) \\
    \mathcal{O}_{H\Box} &= \frac{1}{2} \left( H^{\dagger} H \right)
    \Box \left( H^{\dagger} H \right) \,.
  \end{split}
\end{equation}

\end{document}